\documentclass[journal]{IEEEtran}
\usepackage{cite}
\usepackage{tikz}
\usepackage{color}
\usepackage{caption}
\usepackage{subcaption}
\usepackage{pgf-pie}

\usepackage{url}
\usepackage{breakurl}
\usepackage[hidelinks]{hyperref}

\usepackage{makecell}
\usepackage{graphicx, rotating}
\usepackage{pifont, amssymb}
\usepackage{setspace}
\usepackage{colortbl}
\usepackage{amsmath}
\usepackage{longtable, afterpage}

\usepackage{tabularx,ragged2e}
\usepackage{multicol, supertabular}
\usepackage{pdflscape}
\usepackage{multirow,amssymb,array}

\newcolumntype{L}{>{\arraybackslash}m{3.3cm}}

\def\checkmark{\tikz\fill[scale=0.4](0,.35) -- (.25,0) -- (1,.7) -- (.25,.15) -- cycle;} 

\newcolumntype{Z}{>{\centering\arraybackslash}X}

\begin{document}
\title{Three Decades of Deception Techniques in Active Cyber Defense - Retrospect and Outlook}
\author{Li Zhang and  
	Vrizlynn~L.~L.~Thing
	\thanks{The authors are with Cybersecurity Strategic Technology Center, ST Engineering, Singapore 609602. (email: zhang.li@stengg.com; vrizlynn.thing@stengg.com)}
}


\IEEEtitleabstractindextext{
\begin{abstract}
Deception techniques have been widely seen as a game changer in cyber defense. In this paper, we review representative techniques in honeypots, honeytokens, and moving target defense, spanning from the late 1980s to the year 2021. Techniques from these three domains complement with each other and may be leveraged to build a holistic deception based defense. However, to the best of our knowledge, there has not been a work that provides a systematic retrospect of these three domains all together and investigates their integrated usage for orchestrated deceptions. Our paper aims to fill this gap. By utilizing a tailored cyber kill chain model which can reflect the current threat landscape and a four-layer deception stack, a two-dimensional taxonomy is developed, based on which the deception techniques are classified. The taxonomy literally answers which phases of a cyber attack campaign the techniques can disrupt and which layers of the deception stack they belong to. Cyber defenders may use the taxonomy as a reference to design an organized and comprehensive deception plan, or to prioritize deception efforts for a budget conscious solution. We also discuss two important points for achieving active and resilient cyber defense, namely deception in depth and deception lifecycle, where several notable proposals are illustrated. Finally, some outlooks on future research directions are presented, including dynamic integration of different deception techniques, quantified deception effects and deception operation cost, hardware-supported deception techniques, as well as techniques developed based on better understanding of the human element.    

\end{abstract}

}

\maketitle

\IEEEdisplaynontitleabstractindextext

\IEEEpeerreviewmaketitle

\section{Introduction}
\label{sec:intro}
\IEEEPARstart{I}{n} his book \textit{The Art of Deception}\cite{mitnickArtDeceptionControlling2007}, Kevin Mitnick, the world's most infamous hacker, asserted that the human element is security's weakest link. By attacking this link through various deception based social engineering techniques such as pretexting and phishing, cyber criminals have achieved wide success. For instance, according to 2019 Verizon data breach investigations report\cite{2019DataBreacha}, phishing attacks accounted for more than 80\% of reported security incidents. In the COVID-19 pandemic, we have also witnessed the enormous quantity of cases where hackers exploited coronavirus fears to deliver their phishing and malware attacks\cite{cimpanuCzechHospitalHit2020, palmerCoronavirusthemedPhishingAttacks2020, cimpanuThereNowCOVID192020}.   

Deception aims to manipulate humans' perception by exploiting their psychological vulnerabilities \cite{cybenkoCognitiveHackingBattle2002}, which has direct impact on their beliefs, decisions, and actions. It can be a powerful tool for both hackers and cyber defenders. In as early as the late 1980s, Clifford Stoll managed to set up an imaginary computer environment (now known as \textit{honeypot}), in which a fictitious account was created along with a number of fake documents with enticing names, to lure a hacker to reveal himself and his objectives \cite{stollCuckooEggTracking1989}. In the battle between the hacker and the cyber defender, a conventional wisdom is that the offense has the upper hand: cyber defenders have to make sure everything is properly maintained and prevent intrusions at every single point, whereas hackers may just need to take advantage of one vulnerability to breach the defense \cite{lynniiiDefendingNewDomain2010a}. At the same time, attackers can always gain knowledge about a target system or network through a variety of reconnaissance and discovery tactics, while defenders are usually short of intelligence about their adversaries. Such asymmetric disadvantage for cyber defenders is well promised to be re-balanced through the use of defensive deception, which is expected to deliver a game-changing impact on how threats are faced\cite{lawrencepingreeEmergingTechnologyAnalysis2015, ferguson-walterGameTheoryAdaptive2019a, shadeMoonrakerStudyExperimental2020}.

The perimeter-based defense strategy utilizing conventional security measures such as firewalls, authentication controls, and intrusion prevention systems (IPS) has been proven feeble against infiltration. Even with the defense-in-depth strategy \cite{u.s.departmentofhomelandsecurityRecommendedPracticeImproving2016}, where multiple layers of the conventional security controls are placed throughout the target network, cyber defenders still find it hard to prevent and detect sophisticated attacks like Advanced Persistent Threat (APT) based intrusions. Such targeted attacks typically exploit zero-day vulnerabilities to establish footholds on the target network and leave very few traces of their malicious activities behind for detection. Besides, conventional anomaly detection solutions such as intrusion detection systems (IDS) and behavior based malware scanners tend to raise an overwhelming number of false positive alerts, which plagues cyber defenders and hurts their efficacy in identifying and responding to the true attacks. Defensive deception, featured by its capability of detecting zero-day vulnerabilities and its low false alarm rates due to a clear line between legitimate user activities and malicious interactions, can act as an additional layer of defense to mitigate the issues. 

Instead of focusing on attackers' actions, defensive deception works on their perception by obfuscating the attack surface. The objective is to hide critical assets from attackers and confuse or mislead them, thereby increasing their risk of being detected, causing them to misdirect or waste resources, delaying the effect of attacks, and exposing the adversary tradecraft prematurely \cite{rossDevelopingCyberResilient2019}. In other words, defensive deception helps establish an \textit{active cyber defense} posture, wherein the key elements are to anticipate attacks before they happen, to increase the costs of the adversary, and to gather new threat intelligence for preventing similar attacks.  

Since Stoll's honeypot, there have been numerous honeypots of different flavors proposed. These honeypots can be classified from different perspectives, such as whether they are server-based or client-based, of low interaction or high interaction, and based on real systems or virtual machines (VMs). Despite of the various flavors, all the honeypots share the same definition of being a security resource whose value lies in being probed, attacked, or compromised\cite{spitznerHoneypotsTrackingHackers2002}. The term \textit{honeypot} typically refers to decoy computer systems. Multiple interconnected honeypots form a honeynet. For bait resources that are of other forms (e.g., accounts, user files, database entries, and passwords), they can be collectively termed as \textit{honeytokens} \cite{augustopaesdebarrosIDSRESProtocol2003, lancespitznerHoneytokensOtherHoneypot2003}. Take the honeyfiles proposed in \cite{yuillHoneyfilesDeceptiveFiles2004} as an example. These spurious files reside on a file server; once they are accessed, the server will send an alarm to alert a possible intrusion. Honeypots and honeytokens, when used in tandem, can introduce multi-tier fake attack surfaces for intruders. Unless the intruder can correctly select his target at every turn, his maneuver will be detected. 

Nevertheless, if the honeypots and honeytokens are left with static deployment and configurations, the adversary will have enough time to infer their existence, map out them, and in turn evade them. Even worse, honeypots, especially the high-interaction ones which offer the intruder a real Operating System (OS) environment to interact with, may be exploited by the intruder to gain privileged control and used as a pivot point to compromise other systems \cite{lancespitznerProblemsChallengesHoneypots2004}. This is where the moving target defense (MTD) comes into the picture, which was identified as a key cybersecurity R\&D theme by U.S. NITRD Program \cite{NITRDCSIAIWG2010}.  Specifically, MTD techniques accomplish defensive deception through randomization and reconfiguration of networks, assets, and defense tools\cite{pawlickGameTheoreticTaxonomySurvey2019}. By dynamically shifting both the real and fake attack surfaces, the attack surfaces of critical assets can be maximally obfuscated, with the attacker continuously confused and misled. For instance, Cohen reported in \cite{fredcohenMovingTargetDefenses2010} that a combination of MTD techniques and honeytokens (e.g., automated responses on all unused ports), can help achieve long-term effectiveness of deceptions. 

In this paper, we present a systematic review on the three aspects of defensive deception techniques (i.e., honeypots, honeytokens, and MTD) that have been proposed in the past three decades. The aim is to facilitate a better understanding of the advancement in each aspect and provide clues on how to better integrate them to build a holistic and resilient deception based defense. We limit our scope to techniques that can be directly applied to counter network intrusions. For example, the client-side honeypots \cite{seifertHoneyCLowInteractionClient2007, nazarioPhoneyCVirtualClient2009}, which manifest as vulnerable user agents and actively troll malicious servers to study the client-side attacks, will be excluded in our survey. Sophisticated cyber attacks usually involve phased progressions, and an effective defense should be designed to disrupt each phase of the attack lifecycle. In view of this, the surveyed methods are classified based on the attack phases where they can be applied as countermeasures. In particular, we will use our proposed cyber kill chain model, which is specifically developed to model the network intrusion end to end and can reflect the current threat landscape. In each unique phase of the kill chain model, we further categorize the defensive deception methods according to a four-layer deception stack~\cite{lawrencepingreeEmergingTechnologyAnalysis2015} composed of the network, system, software, and data layers. Such a two-dimensional taxonomy can be employed as a reference for deciding what techniques can be used to disrupt which attack stages and what techniques can complement with each other. 

There have been some excellent surveys on cyber defensive deception. However, most of them just focus on either honeypots and honeytokens \cite{christianseifertTaxonomyHoneypots2006, nawrockiSurveyHoneypotSoftware2016, scottsmithCatchingFliesGuide2016, hanDeceptionTechniquesComputer2018, efendiSurveyDeceptionTechniques2019} or MTD techniques \cite{okhraviSurveyCyberMoving2013,caiMovingTargetDefense2016, b.c.wardSurveyCyberMoving2018, leiMovingTargetDefense2018, senguptaSurveyMovingTarget2020}. Although the survey of deception technology in \cite{fraunholzDemystifyingDeceptionTechnology2018a} includes MTD techniques, its main focus is on honeypots and honeytokens; MTD techniques are just briefly mentioned and not systematically reviewed. In \cite{pawlickGameTheoreticTaxonomySurvey2019}, twenty-four articles that use game theory to model defensive deception, comprising honeypots, honeytokens, and MTD techniques, are surveyed. Despite of representing an important direction, these game-theoretic models are just a small part of the literature. To the best of our knowledge, there has not been a work that provides a systematic retrospect of honeypots, honeytokens, as well as MTD techniques and investigates their integrated usage for orchestrated deceptions. Our survey aims to fill this gap.  

The remainder of this paper is organized as follows. The proposed cyber kill chain model is illustrated in Section~II, while the survey method is presented in Section~III. Representative honeypots, honeytokens, and MTD techniques that can disrupt the adversary kill chain are reviewed in Section IV to VI, respectively. Section VII discusses how to use the deception techniques to achieve active and resilient cyber defense from two aspects, i.e., deception in depth and deception lifecycle. Finally, the paper is concluded in Section VIII with reflection and outlook of defensive deception research.

\section{Cyber Kill Chain Model}
\label{sec:threatModel}

\begin{figure*}
	\centering
	\includegraphics[width=0.8\linewidth]{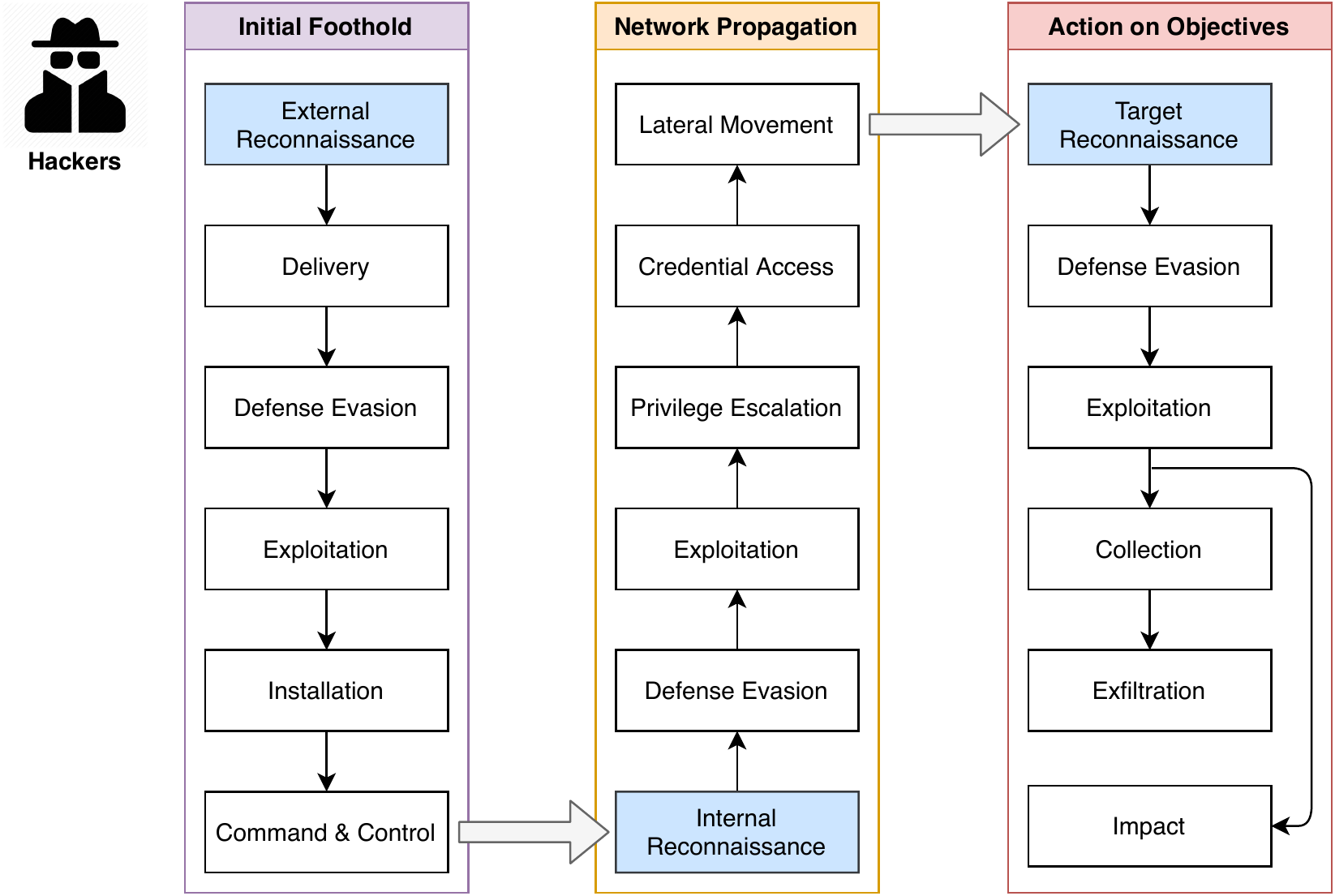}
	\caption[architecture]{The proposed cyber kill chain model} 
	\label{fig:zlkillchain}
\end{figure*}

A sophisticated cyber attack typically has to go through multiple consecutive phases before accomplishing its objective, be it stealthy collection and exfiltration of sensitive data or violation of critical assets' integrity or availability. The Lockheed Martin's intrusion kill chain \cite{hutchinsIntelligenceDrivenComputerNetwork2011} has been widely applied to assist structured analyses of the phased progressions. Nonetheless, this kill chain model, which consists of seven phases (i.e., \textit{reconnaissance}, \textit{weaponization}, \textit{delivery}, \textit{exploitation}, \textit{installation}, \textit{command \& control (C2)}, and \textit{actions on objectives}), is often criticized for reinforcing the perimeter-focused thinking and failing to cover the attack paths inside the network perimeter \cite{reidyCombatingInsiderThreat2013, gioraengelDeconstructingCyberKill2014}. There have been several efforts aimed at expanding it for improved coverage. For instance, to explicitly model the intruder's movement from an initially compromised system to the target system, Laliberte~\cite{marclaliberteTwistCyberKill2016} proposes to add a \textit{lateral movement} phase between the \textit{C2} and \textit{action on objectives} phase. Besides, Laliberte removes the \textit{weaponization} phase as it happens outside of the victim network and no security measure can directly defend against it. The removal of the \textit{weaponization} phase is in line with the kill chain model proposed in \cite{blaked.bryantNovelKillchainFramework2017}, which further introduces the \textit{privilege escalation} and \textit{exfiltration} phase. By contrast, the kill chain models proposed in \cite{maloneUsingExpandedCyber2016, polsUnifiedKillChain2017} significantly expands the Lockheed Martin's kill chain model. Both of them divide the adversary kill chain into three sub-kill chains, i.e., the external kill chain aiming to establish an initial foothold, the internal kill chain aiming to propagate inside the victim network, and the target manipulation kill chain aiming to manipulate the target system to achieve attack objectives. In particular, the kill chain model in \cite{polsUnifiedKillChain2017} includes four tactics from the MITRE ATT\&CK model\footnote{https://attack.mitre.org/}, namely \textit{defense evasion}, \textit{credential access}, \textit{execution}, and \textit{collection}, as additional attack phases. 

Our proposed cyber kill chain model, shown in Figure~\ref{fig:zlkillchain}, is developed based on \cite{maloneUsingExpandedCyber2016, polsUnifiedKillChain2017}. The purpose is to address some of their limitations. For example, there is only one attack objective in \cite{maloneUsingExpandedCyber2016}, represented by the final \textit{execution} phase (i.e., activating malware to subvert operations of the target system). This omits many other possible impacts that a threat actor may incur. Besides, defense evasion dominated the attack tactics in 2019 \cite{kellysheridanDefenseEvasionDominated2020}, which we believe should be explicitly modeled in the kill chain. However, it is missing in \cite{maloneUsingExpandedCyber2016}. Regarding the kill chain model in \cite{polsUnifiedKillChain2017}, we think its \textit{weaponization} and \textit{pivoting} phases are superfluous. For the former, we are in consonance with  \cite{marclaliberteTwistCyberKill2016, blaked.bryantNovelKillchainFramework2017} that it is not actionable to cyber defenders; for the latter, its actions and benefits actually have already been encompassed by the \textit{lateral movement} phase. In addition, the MITRE ATT\&CK model is recently supplemented with a new class of adversary tactics, i.e., \textit{impact}. The inclusion of an \textit{impact} phase in a kill chain model will enable it to model attack objectives more comprehensively. 

In the proposed kill chain model, there are also three sub-kill chains, whose intents have been described above. Each modeled sub-kill chain starts with a \textit{reconnaissance} phase as it provides the attacker with crucial information (such as network topology, vulnerabilities, and deployed security tools) to move further along the kill chain. All the three sub-kill chains also typically contain the \textit{defense evasion} phase due to the widely adopted defense-in-depth strategy. No matter where the intruder propagates, he has to ensure that his maneuver is not detected by defensive measures. The \textit{installation} phase in the external kill chain embodies both the \textit{persistence} and \textit{execution} tactics in the MITRE ATT\&CK model. The internal kill chain for network propagation may be repeated for several times before the attacker finally reaches the target system. In the target manipulation kill chain, a combination of the \textit{collection} and \textit{exfiltration} phases is used to model the attacker's action of covertly stealing sensitive data, while the \textit{impact} phase is used to represent the action of manipulating, interrupting, or destroying the critical assets.   
 
With the intrusion activities covered end to end, the proposed kill chain model can be used to guide attack analyses, threat intelligence extraction, as well as defensive measures selection and prioritization. Such an intelligence-driven, threat-focused approach is essential to establish the active cyber defense posture against threats from both external actors and malicious insiders. For example, synthesis of the remaining kill chain of a detected attack may reveal a zero-day exploit \cite{hutchinsIntelligenceDrivenComputerNetwork2011}. This yields insights into possible future attacks and thereby drive the defender to implement countermeasures beforehand. 

To facilitate the coordinated selection and deployment of deception techniques, the honeypot, honeytoken, and MTD techniques to be reviewed will be mapped to the proposed kill chain model's unique attack phases as listed in Table~\ref{tab_uniquePhases}. Based on the characteristics of each unique attack phase, the possible deception layers, in which the specific attack phase may be disrupted, are ticked accordingly.    

\begin{table*}[]
	\caption{Unique attack phases in the proposed cyber kill chain model}
	\label{tab_uniquePhases}
	\begin{tabular}{|l|L|c|c|c|c|}
		\hline
		\multicolumn{1}{|c|}{}                                        & \multicolumn{1}{c|}{}                                                                                                                                       & \multicolumn{4}{c|}{\textbf{Deception Layer}}      \\ \cline{3-6}  \\[-5.35ex]                                                                            \\
		\multicolumn{1}{|c|}{\multirow{-2}{*}{\textbf{Attack Phase}}} & \multicolumn{1}{c|}{\multirow{-2}{*}{\textbf{Description}}}                                                                                                 & \multicolumn{1}{c|}{Network} & \multicolumn{1}{c|}{System} & \multicolumn{1}{c|}{Software} & \multicolumn{1}{c|}{Data} \\ \hline
		Reconnaissance                                              & \multicolumn{1}{m{8.4cm}|}{Gather information from open-source intelligence, internet-facing/internal system probing, or network traffic sniffing}                 &    \checkmark                         &         \checkmark                      &      \checkmark                            &             \checkmark              \\ \hline
		Delivery                                                    & \multicolumn{1}{m{8.4cm}|}{Deliver the tailored malicious payload through entry vectors like email, websites, and removable media}               &                 \checkmark               &              \checkmark                 &        \checkmark                         &                          \\ \hline
		Defense Evasion                                             & \multicolumn{1}{m{8.4cm}|}{Evade defense by disabling security tools, abusing trusted process to hide malware, obfuscation/encryption, etc.}                      &             \checkmark                   &            \checkmark                   &                    \checkmark                &                          \\ \hline
		Exploitation                                                & \multicolumn{1}{m{8.4cm}|}{Exploit identified vulnerabilities or simply user negligence to facilitate the following kill chain phases}                              &       \checkmark                         &         \checkmark                      &          \checkmark                          &                          \\ \hline
		Installation                                                & \multicolumn{1}{m{8.4cm}|}{Drop the payload (e.g., install a remote access Trojan or backdoor to maintain the foothold)}    &                             &            \checkmark                   &                                 &                          \\ \hline
		C2                                     & \multicolumn{1}{m{8.4cm}|}{Beacon outbound to establish a C2 channel with attacker for additional commands or payloads}                                                                &           \checkmark                     &          \checkmark                     &                               &                          \\ \hline
		Privilege Escalation                                        & \multicolumn{1}{m{8.4cm}|}{Horizontally or vertically elevate privileges to gain access to sensitive assets}                                                                          &                        &            \checkmark                   &   \checkmark                              &                          \\ \hline
		Credential Access                                           & \multicolumn{1}{m{8.4cm}|}{Steal credentials like account names and passwords}                                                                                                         &                             &       \checkmark                        &          \checkmark                       &          \checkmark                \\ \hline
		Lateral Movement                                            & \multicolumn{1}{m{8.4cm}|}{Move through the network by using installed remote access tools or stolen credentials}                           &  \checkmark                              &   \checkmark                            &         \checkmark                         &          \checkmark                 \\ \hline
		Collection                                                  & \multicolumn{1}{m{8.4cm}|}{Collect data of interest from local databases/file system, screen/user input capture, shared network drives, etc.} &        \checkmark                        &     \checkmark                          &      \checkmark                              &      \checkmark                       \\ \hline
		Exfiltration                                                & \multicolumn{1}{m{8.4cm}|}{Process collected data and send it out through C2 channel or other channels}                    &     \checkmark                           &          \checkmark                     &                                   &                      \\ \hline
		Impact                                                      & \multicolumn{1}{m{8.4cm}|}{Disrupt availability or compromise integrity of critical data or system/network services}                                 &             \checkmark                   &         \checkmark                      &        \checkmark                            &        \checkmark                    \\ \hline
	\end{tabular}
	
\end{table*}


\section{Survey Method}
\label{sec:surveyMethod}
To build the paper repository for this survey, we firstly searched in two leading research databases (i.e., IEEE Xplore and ACM Digital Library) with the keyword \textit{cyber deception}, which returned a list of 253 research articles. Then we utilized the title and abstract of each paper to determine its relevance to our survey. The selection criteria is whether the paper is on defensive cyber deception (including evasion techniques of defensive deception). For example, papers on cyber deception attacks \cite{aminCyberSecurityWater2013, houRobustPartialNodesBasedState2020, zhangOptimalStealthyDeception2020, meira-goesSynthesisSupervisorsRobust2021} were removed. This reduced the list to 87 research works. By including relevant papers cited in these works, we managed to collect additional articles. Together with some writings that we saw in other venues or mediums and think important to our survey (e.g., Ph.D. dissertations available online), the final repository contains 192 research works, which are referred to as primary studies in a systematic survey \cite{kitchenhamProceduresPerformingSystematic2004}. Although our paper repository does not cover all the related papers that were published in the past three decades, we are confident that most representative deception techniques have been included, through which the overall development trends in this field are accurately pictured.   

\section{Honeypots}
\label{sec:honeypots}

Honeypots can be broadly classified into two categories: research and production\cite{lancespitznerValueHoneypotsPart2001}. Although research honeypots play an important role in gathering intelligence on the threat landscape, they do not directly benefit a specific organization. In contrast, production honeypots are placed in an organization's environment for attack detection and risk mitigation. They may be deployed as {sacrificial lamb}, {hacker zoo}, {minefield}, {proximity decoys}, {redirection shield}, and {deception ports (on production systems)} \cite{scottbergInternetHoneypotsProtection2002}, as described in Table~\ref{tab_commonDeploy}. 

Stoll's honeypot \cite{stollCuckooEggTracking1989} is a good example of the sacrificial lamb, which is the oldest and maybe also the most intuitive strategy. Being usually isolated from production systems, the sacrificial lamb honeypot may be easily identified and bypassed by attackers. The same limitation is shared by the hacker zoo. The minefield honeypots are commonly placed near the network perimeter, which will sound alarms upon attacker probing. This strategy helps enhance the perimeter based defense, but cannot handle attackers already inside the network. Both the proximity decoys and the redirection shield aim to lead the attacker astray and away from production systems. Their difference lies in that the redirection shield strategy, through the use of traffic rerouting or port redirection, does not require honeypots to be in the production network and hence has more flexibility. Among the five honeypot deployment strategies, the deception ports on production systems can be seen as the final defense. The various simulated vulnerable services on well-known ports can be used to detect and delay the attack even if the adversary reaches the production system. For example, the Deception Toolkit \cite{fredcohenDeceptionToolKit1998}, which is the first open source honeypot, can set up the deception services. 

Featured by deceiving to detect, derail and/or delay attacks, honeypots may be used to disrupt a number of attack phases in the cyber kill chain model, namely the {reconnaissance}, {delivery}, {exploitation}, {installation}, {C2}, {lateral movement}, and {impact} phase. On the other hand, in the {defense evasion} phase, the attacker may be able to identify honeypots and evade them. The remaining part of this section will be on these two aspects. 

\begin{table*}[]
	\caption{Common Deployment Strategies for Honeypots \cite{scottbergInternetHoneypotsProtection2002}}
	\label{tab_commonDeploy}
	\begin{center}
			\begin{tabular}{|l|L|}
			\hline
			\multicolumn{1}{|c|}{\textbf{Strategy}}                              & \multicolumn{1}{c|}{\textbf{Description}}                                                                                                                            \\ \hline
			Sacrificial Lamb                      & \multicolumn{1}{m{11cm}|}{An isolated system that has no entry point to production systems}                                                                 \\ \hline
			Hacker Zoo  & \multicolumn{1}{m{11cm}|}{An entire subnet of honeypots with varied platforms, services, vulnerabilities, and configurations, which are isolated from production systems}                                                       \\ \hline
			Minefield                      & \multicolumn{1}{m{11cm}|}{A number of honeypots placed in forefront to serve as first attack targets}                                                      \\ \hline
			Proximity Decoys                     & \multicolumn{1}{m{11cm}|}{Honeypots deployed in close proximity to production systems}           \\ \hline
			Redirection Shield                             & \multicolumn{1}{m{11cm}|}{External honeypots that appear on production systems through port redirection}                                                 \\ \hline
			Deception Ports                         & \multicolumn{1}{m{11cm}|}{Simulated services (e.g., SMTP, DNS, FTP) on production systems} \\ \hline
		\end{tabular}
	\end{center}

\end{table*}

\subsection{Disrupting the cyber kill chain}

The intruder relies on successful reconnaissance to achieve tactical advantage in the campaign. Sticky honeypots can be used to mitigate the threat from network scans. For example, LaBrea \cite{tomlistonLaBreaStickyHoneypot2001} can take over unused IP addresses in the network and create virtual hosts to attract worms and hackers; connection attempts to the impersonated hosts will then be tarpitted. Greasy \cite{leslieshingImprovedTarpitNetwork2016} further improves the sticky connection parameters to generate more realistic traffic. Besides slowing down the scanning activities, both LaBrea and Greasy are able to produce false network topologies and hence get adversaries confused. To dissimulate the network topology, many other honeypot techniques can also be used. For instance, HoneyD~\cite{provosVirtualHoneypotFramework2004} can simulate a large number of virtual systems with configurable fingerprints and provide arbitrary services and routing topologies. These honeypot techniques are typically of low interaction and virtually adopt the minefield or proximity decoys deployment strategy.

\begin{figure}
	\centering
	\includegraphics[width=\linewidth]{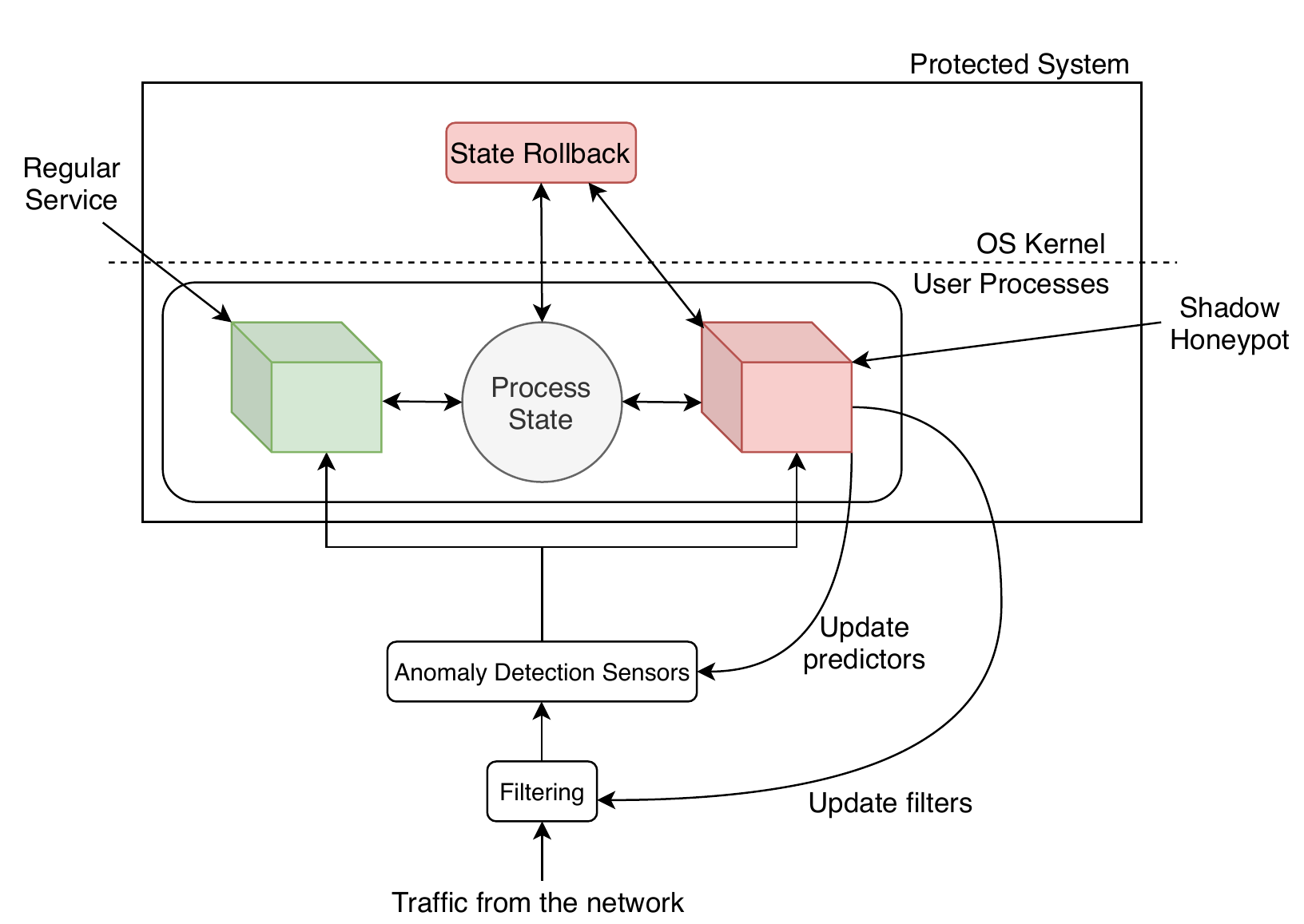}
	\caption[architecture]{The Shadow Honeypot architecture in \cite{anagnostakisDetectingTargetedAttacks2005}} 
	\label{fig:shadowHP}
\end{figure}

To disrupt the attack phases such as delivery, C2, and lateral movement, the key is to direct the malicious traffic to high-interaction honeypots. By providing high-fidelity forged environment to interact with attackers, the exploitation, installation, and impact phases may also be broken. In addition, attackers' time and resources will be wasted and their tactics, techniques, and procedures (TTPs) may be revealed. As these high-interaction honeypots typically monitor one IP address each and have the problem of limited field of view, the redirection shield strategy is often adopted. In \cite{anagnostakisDetectingTargetedAttacks2005}, it is proposed to handle anomalous traffic identified by IDS by a shadow honeypot, as shown in Figure~\ref{fig:shadowHP}. The shadow honeypot is an instrumented instance of the application (e.g., transactional applications) in protected system and share all internal states. Attacks mounted in the shadow honeypot will be caught and the induced state changes will be discarded, while legitimate traffic misclassified by IDS will be validated in the shadow honeypot and transparently handled. 
OpenFire~\cite{bordersOpenFireUsingDeception2007} presents additional false targets by appearing to attackers that all IPs and ports of an organization network are open. Suspicious traffic will then be forwarded to a cluster of decoy machines. In the cloud environment, Biedermann \emph{et al.} \cite{biedermannFastDynamicExtracted2012} propose to redirect potential attacks against an operational VM to a honeypot VM created through a live cloning process. The honeypot VM has exactly the same configuration as the original VM, but without the sensitive data. This way, the impact of the attack can be analyzed without risking the integrity of the original target VM. Similarly, in \cite{uriasGatheringThreatIntelligence2016}, the endpoint VM suspected of malicious activities will be cloned and forked in a deception environment with the same network and system configurations of the real network environment. If the suspicions for the VM are not found, the VM may be migrated back to the operational environment; otherwise, all artifacts related to the attack in the deception environment can be documented for further scrutiny.  

Besides standard IT systems, the above honeypot concepts are also applicable to industrial control systems (ICS). In \cite{dissoPlausibleSolutionSCADA2013}, after analysing the threat landscape and unique security requirements of supervisory control and data acquisition (SCADA) systems, a plausible honeypot system is built, which is composed of both a low-interaction HoneyD honeypot emulating the programmable logic controller (PLC) and a high-interaction honeypot using a genuine PLC. In \cite{winnConstructingCosteffectiveTargetable2015}, HoneyD is extended to address the authenticity flaw of emulated PLCs; together with the proxy technology, multiple high-interaction honeypots can be distributed at the cost of a single actual PLC. A number of recent honeypot applications in ICS are based on Conpot \cite{lukasristConpot2015}, which is a low-interaction virtual ICS honeypot designed for easy deployment, modification and extension and supports a range of common industrial control protocols such as Modbus TCP, SNMP, and BACnet. In \cite{zhaoResearchHighInteractive2017}, a high-interaction honeypot is created by improving Conpot in the aspects of control protocol, human-machine interface (HMI) and equipment simulation. In \cite{kumanExperimentUsingIMUNES2017}, with the use of Conpot and the IMUNES network simulator, a complex high-interaction ICS is emulated. 

\subsection{Evading the honeypot}
\label{sec_evadeHP}

Despite of being a powerful tool to trap, delay, and even gather information about intruders, honeypots have their own weakness. At best, they are counterfeits of the real target. If intruders are able to identify honeypots, they will circumvent them or keep the malicious payload dormant, making honeypots useless. To some extent, attackers are highly motivated to push the detection of honeypots to early phases of their kill chain, so that their intrusion efforts are not rendered in vain and their TTPs are not disclosed.

Honeypots may be fingerprinted based on timing or behavior discrepancies in probing responses. After the introduction of the seminal HoneyD, it was soon found that it can be remotely fingerprinted based on its response to bad packets \cite{josephcoreyAdvancedHoneyPot2003} or the latency of its emulated network links \cite{fuRecognizingVirtualHoneypots2006}. Degreaser in \cite{altUncoveringNetworkTarpits2014} can efficiently fingerprint sticky honeypots like LaBrea by sending a series of specially crafted probe packets; real hosts can then be discerned from tarpits based on the response. In \cite{vetterlBitterHarvestSystematically2018, vetterlHoneypotsAgeUniversal2019}, by leveraging the flaw of many honeypots' reliance on off-the-shelf libraries to implement the transport layer, distinguishing probes constructed at this layer is able to systematically fingerprint honeypots. 

Other unique features of honeypots may also be taken advantage of by attackers. Honeypot evader \cite{rrushiHoneypotEvaderActivityguided2019} exploits honeypots' innate characteristic of not initiating any network traffic and attacks only the hosts with obvious network activity. In \cite{wangHoneypotDetectionAdvanced2010}, by exploiting the liability constraint that cyber defenders cannot allow their honeypots to participate in real attacks that could cause damage to other entities, an attacker can detect honeypots by checking whether his compromised machines can successfully send out unmodified malicious traffic. A more specific example of this concept is given in \cite{krawetzAntihoneypotTechnology2004}, where spammers can simply check if an open proxy relay is a honeypot based on whether emails can be sent to themselves.  

\begin{figure}
	\centering
	\includegraphics[width=\linewidth]{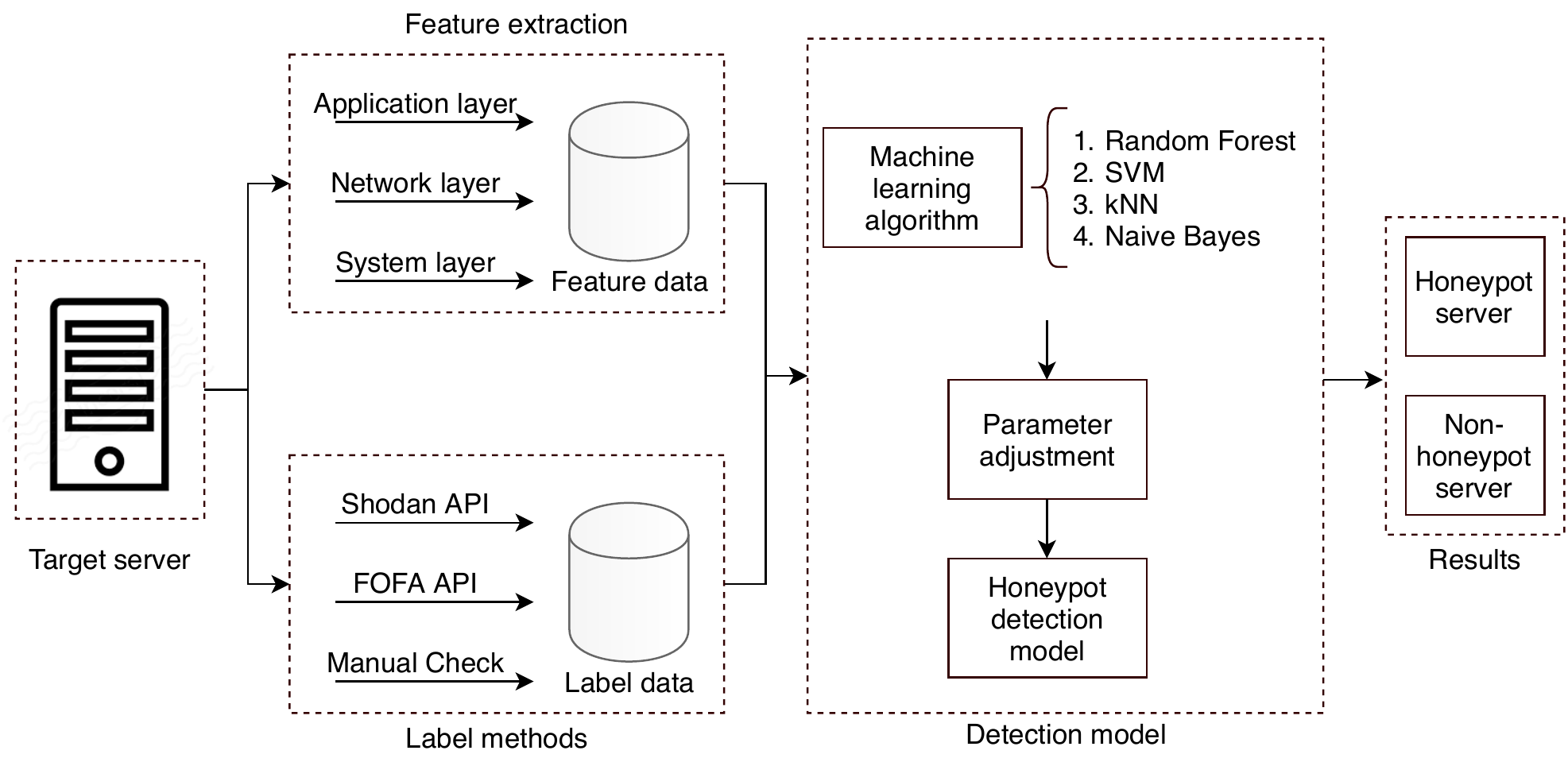}
	\caption[architecture]{The machine learning based Honeypot server identification method in \cite{huangAutomaticIdentificationHoneypot2019}} 
	\label{fig:mlHP}
\end{figure}

Besides exploiting a single factor to tell whether a target is honeypot, information collected from different factors may be combined to reach a more accurate decision. In \cite{hayatleDempsterShaferEvidenceCombining2012}, such combination is performed with Dempster-Shafer theory\cite{sentzCombinationEvidenceDempsterShafer2002}, while in \cite{huangAutomaticIdentificationHoneypot2019}, machine learning (ML) techniques are used. For the latter, the design is depicted in Figure~\ref{fig:mlHP}. 

\section{Honeytokens}
\label{sec:honeytokens}

Honeytokens share the same concept of honeypots, whose value lies in being used illicitly. In fact, the history of honeytokens is as long as that of honeypots. Besides the fictitious files with tempting names and contents in Stoll's honeypot in the late 1980s \cite{stollCuckooEggTracking1989}, Spafford built files with Unix sparse file structure in 1990s \cite{eugeneh.spaffordMorePassiveDefense2011}, which are of small size on disk but will result in ``endless" transfer for attacker's copy attempt. 

Honeytokens have remarkable flexibility. They can be in the form of any digital entity and placed anywhere across an organization's environment. The polymorphism and omnipresence bring two benefits for the defense: even though attackers are able to evade some forms of honeytokens, they may still be trapped by others; the uncertainty of whether and where honeytokens are placed will slow down attackers and may even turn them away (i.e., the deterrent effect). Typically, the honeytoken is simple to deploy and cost effective, making it considered as an exciting new dimension for honeypot \cite{lancespitznerHoneytokensOtherHoneypot2003}.  

Unlike honeypots which usually can only disrupt specific attack phases in the kill chain from the network and system layer, the various forms of honeytokens can be applied to thwart almost all the attack phases through all the four layers of the deception stack.   

In the external reconnaissance phase, before engaging the target network, attackers will actively gather information from open-source intelligence. For example, Project Spacecrab creates credential honeytokens in the form of Amazon Web Services (AWS) keys, and found that the average time for a hacker to exploit the honeytoken is just thirty minutes after it is posted on GitHub \cite{bourkeBreachDetectionScale2018}. There is a wealth of personal information on social network platforms, from which attackers might find valuable tips to drive targeted phishing campaigns. If bogus profiles are disseminated on these platforms \cite{stringhiniDetectingSpammersSocial2010}, attackers may be misdirected in the delivery attack phase. To increase the authenticity, Virvilis \emph{et al.} \cite{virvilisChangingGameArt2014} suggest that the created fake personas should have positions of interest to attackers, connections with people from both inside and outside the organization, valid email addresses, as well as real, but closely monitored, organization accounts. To facilitate the creation of the bogus profiles, a method for automatically generating realistic personally identifiable information (PII) based honeytokens is proposed in \cite{whiteCreatingPersonallyIdentifiable2010}. 

The internet-facing web servers of an organization is another important intelligence source in external reconnaissance. To confuse and misdirect malicious website visitors, decoy hyperlinks embedded in web pages are used in \cite{gavrilisFlashCrowdDetection2007}. These hyperlinks are invisible to legitimate human users, but can be detected by automated programs. An algorithm is also proposed for optimal placement of the decoys in website pages. Brewer \emph{et al.} \cite{douglasbrewerLinkObfuscationService2010} propose to add the decoy links under two design principles: the multiple-link principle where multiple decoy links are positioned off the visible page and valid links remain in their original; the shadow-link principle where multiple, invisible decoy links are stacked at the same coordinates as the valid link. In \cite{virvilisChangingGameArt2014}, three types of honey tokens are proposed for public web servers: fake entries in \texttt{robots.txt} files (used to tell crawlers which web pages to crawl and which ones not to), invisible decoy links as described above (e.g., white links with white font), and fake credentials in HTML comments. 

\begin{figure}
	\centering
	\includegraphics[width=\linewidth]{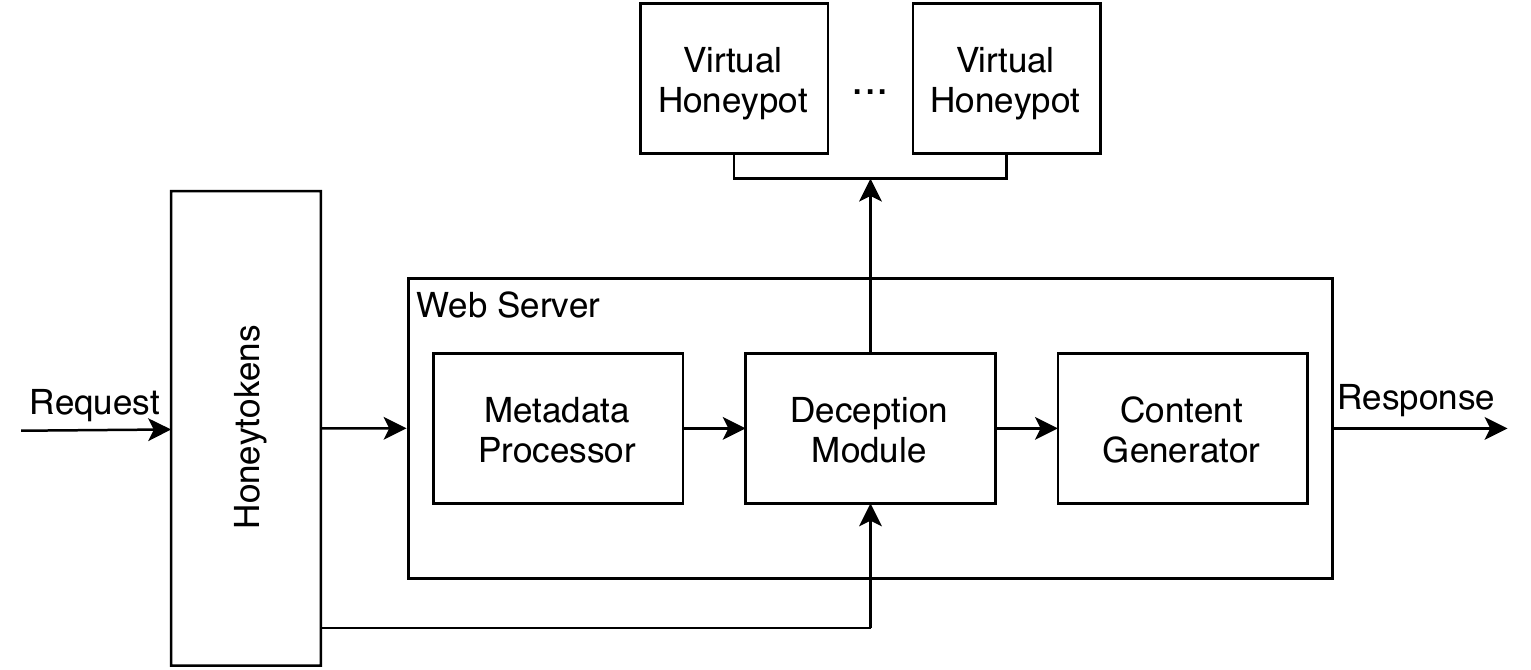}
	\caption[architecture]{The deceptive web server in Apache architecture in~\cite{katsinisFrameworkIntrusionDeception2013}} 
	\label{fig:deceptiveWebServer}
\end{figure}

When attackers probe the target network for more information, deceptive responses can be utilized to confuse them and delay their progress. To conceal the operating system (OS) related information that may be retrieved by attackers via OS fingerprinting, host-based OS obfuscation is suggested in \cite{murphyApplicationDeceptionCyberspace2010} as a deception technique. With the attacker being unsure of the OS or even assuming the wrong OS, his penetration will be impeded. In \cite{trassareTechniquePresentingDeceptive2013}, the defender controls the fake routes to be presented to attackers who use traceroute to map the target network's topology. Instead of directly rejecting the connection after an attack is suspected, which either is a false positive or will inform the adversary of being detected, deceptive delays are suggested in \cite{julianDelayingtypeResponsesUse2002, neilc.roweDeceptionDefenseComputer2007}. The defender can use excuses (e.g., a computation requires a long time) to keep the suspect waiting, and use the time to collect more evidence or reorganize the defense. Katsinis and Kumar~\cite{katsinisSecurityMechanismWeb2012,katsinisFrameworkIntrusionDeception2013} propose to deploy honeytokens such as fake form fields, fake parameters, and fake files in the web server. Alarms from these honeytokens will be sent to a deception module, which is responsible for redirecting the attacker traffic to a honeypot and supplying the attacker with misinformation that his attack is successful. By leveraging the modular design of Apache web server, the deception module can be conveniently inserted between the metadata processor and the content generator, as shown in Figure~\ref{fig:deceptiveWebServer}. A similar framework for achieving deceptive response is proposed in \cite{hanEvaluationDeceptionBasedWeb2017}, where the deception module is deployed as a transparent reverse proxy.  

A vital part of the attacker kill chain is to bypass the defense in the target network. Taking advantage of attackers' fear of having their TTPs exposed and resources wasted, Rowe \emph{et al.} \cite{roweDefendingCyberspaceFake2007} propose to plant clues in systems such that they appear as honeypots (i.e., fake honeypots) and thereby turn attackers away. The planted clues can be names of known honeypot tools, non-standard system calls in security critical subroutines, reduced number of common files, and appearance of the system being little used. Besides fake honeypots, the deception effects on attackers of ``fake fake honeypots", which refer to real honeypots that pretend to be noticeable fake honeypots, are also investigated. 

Exploitation is another imperative phase in the attacker kill chain. Only after successfully exploiting some vulnerabilities can the adversary gain escalated privilege and be able to move further in the kill chain. In \cite{craneBoobyTrappingSoftware2013}, ``booby trap" codes are inserted into the protected software or system during compilation or program loading. These booby traps remains dormant under normal operation but may be triggered by attackers' exploitation attempts. Once triggered, the booby trap can perform advanced forensics to identify the attack in real time and send attackers deceptive responses. Frederico \emph{et al.} \cite{araujoPatchesHoneyPatchesLightweight2014} propose to use decoy vulnerabilities that have been patched as honeytokens (aka honeypatches). In particular, the vulnerabilities are patched in such a way that attackers' exploitation attempts appear successful but their connections are actually redirected to an ephemeral honeypot with the un-patched version of the system or software. Besides, the honeypot may host a deceptive file system laced with disinformation to further deceive, delay, and misdirect attackers.

\begin{figure}
	\centering
	\includegraphics[width=0.9\linewidth]{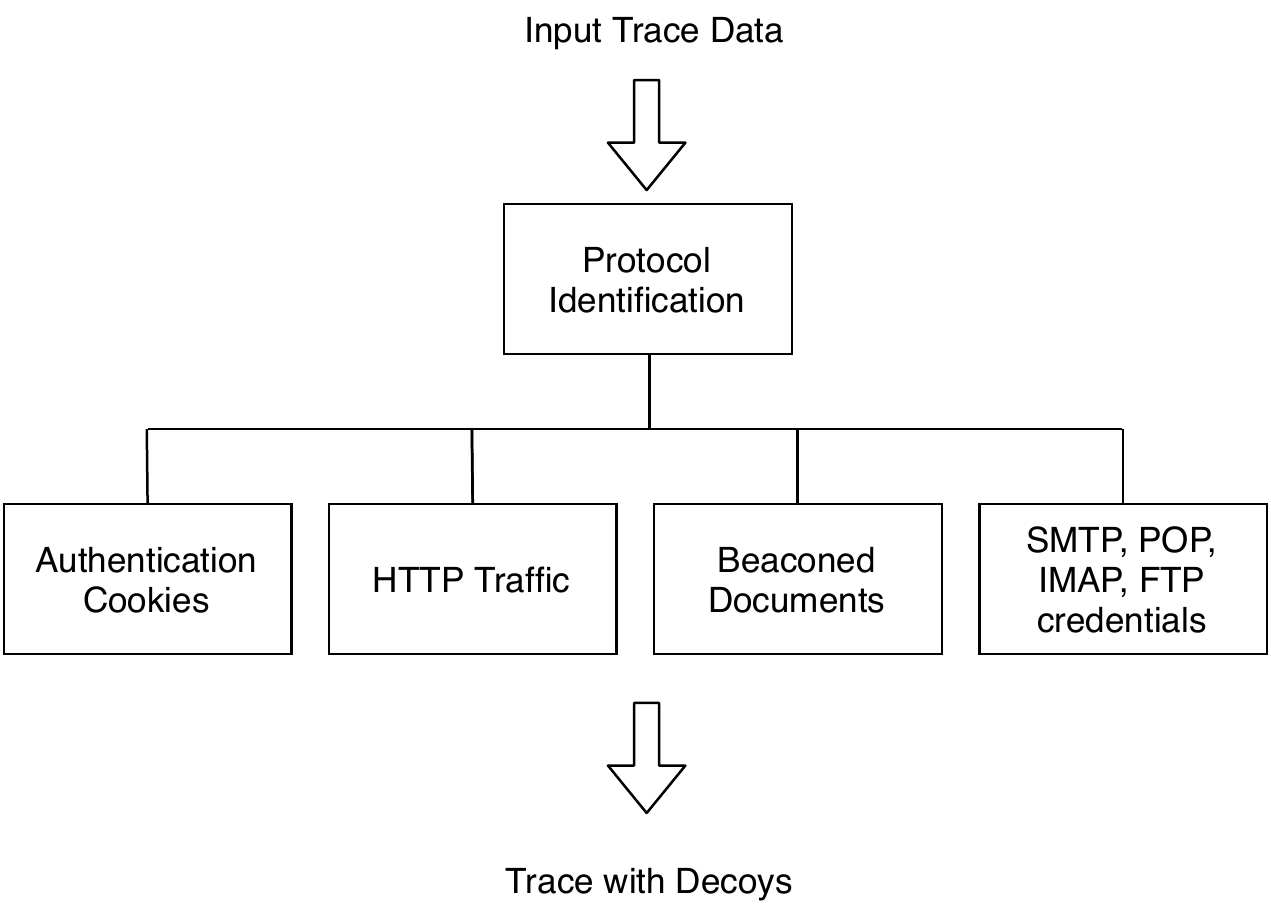}
	\caption[architecture]{The ``record, modify, and replay" process for decoy traffic generation in \cite{bowenSystemGeneratingInjecting2012}} 
	\label{fig:honeyFlow}
\end{figure}

Attackers already inside the network will eavesdrop on the traffic to collect sensitive information and/or use the information to guide their following activities. For example, an attacker may map out systems that do not initiate any network traffic, which are likely to be honeypots, and circumvent them during the lateral movement. Such activity-guided target selection can be disabled by introducing decoy network and user space activities \cite{rrushiHoneypotEvaderActivityguided2019}. Bowen \emph{et al.} \cite{bowenAutomatingInjectionBelievable2010, bowenSystemGeneratingInjecting2012} propose to inject decoy traffic with enticing information that will induce the eavesdropper to take observable actions (e.g., using sniffed credentials to access a decoy account). In particular, to maximize the realism of the decoy traffic, a ``record, modify, and replay" method (see Figure~\ref{fig:honeyFlow}) is used to automatically generate a large amount of decoy traffic; the decoy traffic is also continuously updated to prevent an adversary from recognizing the bait over time. On the other hand, encryption may be used to restrict access to sensitive information in the network traffic. However, the eavesdropper may still reveal the secret through offline brute-force attacks. As decryption with a wrong key will result in random gibberish, the adversary will know that he is successful if the output complies with some expected structure. To mitigate this risk, honey encryption (HE) \cite{juelsHoneyEncryptionSecurity2014} can be used. When the ciphertext generated by HE is decrypted by an incorrect key, a plausible-looking but bogus plaintext will be yielded. The adversary will be confused and may be misdirected to reveal himself if the bogus plaintext is a credential honeytoken. To make the bogus plaintext in HE more deceptive, i.e., contextually correct and domain specific, natural language processing (NLP) based techniques \cite{abiodunReinforcingSecurityInstant2020} and deep learning (DL) based ones\cite{omolaraDeceptionModelRobust2019a} have been used.  

In \cite{juelsBodyguardLiesUse2014, kaghazgaranInsiderThreatDetection2015}, decoy permissions are used to extend role-based access control (RBAC) model for detecting the insider threat. These decoy permissions are not required for the specific roles to handle their tasks, and they are designed to give access to fake versions of sensitive assets. By monitoring attempts to access the fake assets, malicious users can be traced. We think that the decoy permissions are also useful for trapping outside attackers who have managed to infiltrate and reach the credential access attack phase. Legitimate users may know that they are not supposed to use the decoy permissions, but attackers who steal their credentials are not aware of that, leading to their activities being detected. 

The following three categories of honeytoken techniques can be used to disrupt the last three attack phases in the kill chain model, namely the collection, exfiltration, and impact phase. As a result, threat actors may be hampered from achieving their objectives and their malicious activities may be detected.

\textbf{Decoy passwords}: 
Juels and Rivest \cite{juelsHoneywordsMakingPasswordcracking2013} propose to assign multiple false passwords (aka honeywords) along with the real password to each account. This way, even though the adversary manages to crack the passwords from the stolen password hash files, he is still not sure which passwords are real. If the honeywords are used for login, an alarm will be set off. Instead of using multiple fake passwords to protect an account, Almeshekah \emph{et al.} \cite{almeshekahErsatzPasswordsEndingPassword2015} propose to use a machine-dependent function (e.g., a physically unclonable function (PUF) \cite{changRetrospectiveLookForward2017} or a hardware security module (HSM) \cite{PCIHardwareSecurity2009}) at the password server to generate ``ersatzpasswords" from the stored password hashes; the hash of the ersatzpasswords are then stored in place of the original password hashes. This way, without physical access to the target's machine, any offline password cracking attempt will fail. If the attacker is unaware of the scheme and use the recovered ersatzpassword to login, the system administrator will be alerted.

\textbf{Decoy database entries}:
Decoy database objects like \texttt{TABLE CREDIT\_CARDS} or \texttt{VIEW EMPLOYEES\_SALARY} can be inserted into databases to lure attackers. Čenys \emph{et al.}\cite{antanascenysImplementationHoneytokenModule2005} propose to implement modules for Oracle database management system (DBMS), which are responsible for monitoring access to the honeytokens, alerting the DBMS administrator, and logging  malicious activities. To address the challenge of creating realistic decoy entries, HoneyGen \cite{bercovitchHoneyGenAutomatedHoneytokens2011} extrapolates rules describing the data structure, attributes, constraints and logic of real data items, and then automatically generates artificial items that comply with these rules. Padayachee \cite{padayacheeAspectisingHoneytokensContain2014} proposes to leverage aspect-oriented programming (AOP) to seamlessly augment a target DBMS with the basic honeytoken deployment processes, namely honeytoken generation, distribution, management, and detection.   

\textbf{Decoy user/system files}: 
Yuill \emph{et al.} \cite{yuillHoneyfilesDeceptiveFiles2004} propose to use a honeyfile system to generate and monitor baits files; once these files are accessed, alerts will be sent to the system user. To ensure the detectability, Bowen \emph{et al.} \cite{bowenBaitingAttackersUsing2009} propose to embed multiple signals in the decoy files, including a unique watermark that can be detected when the file is loaded in memory or appears in network traffic, a beacon that will signal a remote web site once the file is opened, and bait information such as credential honeytokens that will trigger alerts once used. To maximize the likelihood of an attacker taking the bait (i.e., conspicuousness), Voris \emph{et al.} \cite{vorisBaitSnitchDefending2013,vorisFoxTrapThwarting2015} propose some automated deployment methods which can strategically place the decoy files. With the aim of increasing the enticingness of the decoys, NLP techniques are used in \cite{benwhithamAutomatingGenerationEnticing2017}, where the fake file content is generated based on substitution and transposition of words collected from the target directory and file system. Existing file-based deception techniques mainly focus on decoy user data files, while PhantomFS \cite{leePhantomFSFileBasedDeception2020} proposes to use decoy system files. To prevent false alarms triggered by legitimate activities accessing the decoy system files, a hidden interface is introduced, through which the decoy files are excluded. As an attack has to invoke some system files, this approach can further improve the detection of the adversary, especially for disrupting the impact attack phase.

\section{Moving Target Defense}
\label{sec:mtd}
Sun Tzu once wrote ``just as water remains no constant shape, in warfare there are no constant conditions"  \cite{tzuArtWar2007}. Similarly, in cyber defense, a dynamic, constantly evolving attack surface for the protected network is extremely valuable to retain a resilient security posture. MTD techniques seek to randomize network components to reduce the likelihood of a successful attack, increase network dynamics to reduce the lifetime of an attack, and diversify otherwise homogeneous systems to limit the damage of a large-scale attack \cite{okhraviFindingFocusBlur2014}. In other words, MTD intensifies uncertainty and workload for attackers by making the protected network less static, less deterministic, and less homogeneous. 

Similar to honeytokens, the various MTD techniques are able to disrupt the adversary kill chain through all the four layers of the deception stack. 

In the external reconnaissance phase, attackers have to gain necessary knowledge about the target network before they can move on along the kill chain. This attack phase can be guarded against by obfuscating the following two aspects of network properties:

\textbf{IP obfuscation}:
To prevent attackers from tracing hosts in the target network based on IP addresses, a number of techniques have been proposed. Two early examples are \textit{dynamic network address translation} (DyNAT) \cite{kewleyDynamicApproachesThwart2001} which is a protocol-obfuscation technique that can scramble source and destination IP addresses in packet headers and \textit{network address space randomization} (NASR) \cite{antonatosDefendingHitlistWorms2005} which modifies a DHCP server to have short IP address leases so that host machines' IP addresses are changed frequently. Many recent techniques follow the line of randomly changing IP addresses. OpenFlow Random Host Mutation (OF-RHM) \cite{jafarianOpenflowRandomHost2012} is able to mutate IP addresses with high unpredictability and rate. In particular, OF-RHM frequently assigns each host a random virtual IP~(vIP) address, which will be automatically translated to/from the real IP (rIP) address of the host at the network edge. As a result, IP mutation is transparent to host machines and will not disrupt any active connection. To manage the random host mutation efficiently and minimizes the operational overhead, software-defined networking (SDN)~\cite{kreutzSoftwareDefinedNetworkingComprehensive2015} is utilized, where a centralized approach is realized based on OpenFlow \cite{mckeownOpenFlowEnablingInnovation2008}. A variant of the method, called Random Host Mutation (RHM), is proposed in \cite{al-shaerRandomHostMutation2013}, which changes vIP addresses in a distributed fashion and can be deployed on traditional networks. Due to the IPv4 network's limited unoccupied address space, which reduces the unpredictability of IP address hopping, Dunlop \emph{et al.} propose MT6D \cite{dunlopMT6DMovingTarget2011, dunlopBlindManBluff2012}. By leveraging the immense address space of IPv6, MT6D makes it harder for attackers to locate and subsequently target host machines. Besides, by encapsulating the original packet in a tunnel, MT6D also allows to change the IP address at any time without disrupting ongoing sessions. 

\textbf{OS obfuscation}:
To defend against OS fingerprinting attacks, Kampanakis \emph{et al.} \cite{kampanakisSDNbasedSolutionsMoving2014} propose an SDN  based method, which hides the OS information in the response to detected illicit traffic by randomizing TCP sequence numbers and payload patterns in TCP, UDP, and ICMP protocols. Zhao \emph{et al.} \cite{zhaoSDNBasedFingerprintHopping2017} propose to further model the interaction between the fingerprinting attack and defense as a signaling game \cite{banksEquilibriumSelectionSignaling1987} and develop optimal fingerprint hopping strategies by analyzing the equilibriums of the game. A strategy selection algorithm is also proposed to maximize the defense utility.  

\begin{figure}
	\centering
	\includegraphics[width=\linewidth]{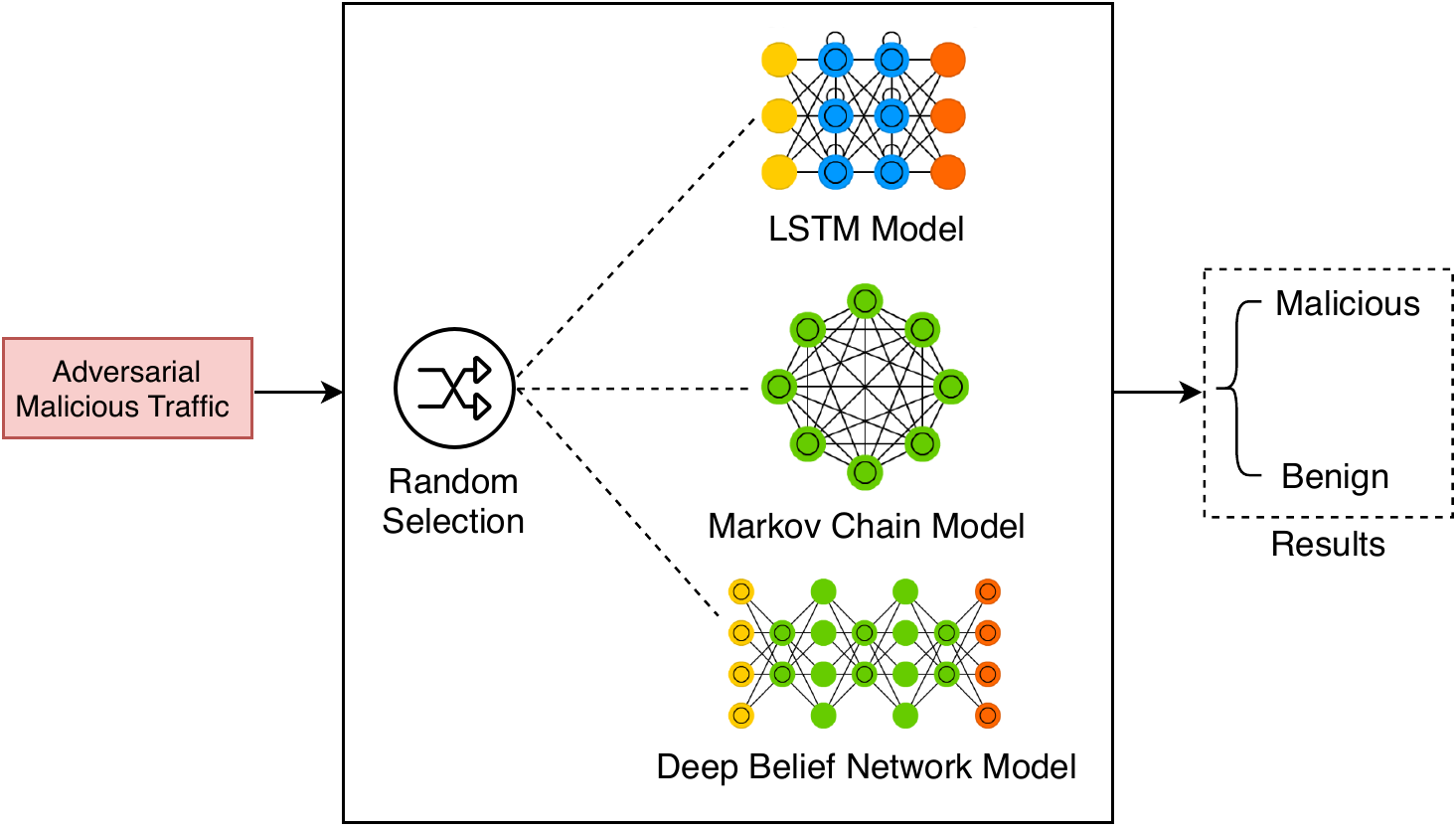}
	\caption[architecture]{Randomized classifiers to mitigate adversarial attacks, where malicious traffic perturbed to evade the long short term memory (LSTM) model will be detected by other models} 
	\label{fig:randomClassifier}
\end{figure}

The defense evasion phase may be disrupted by dynamically and continuously changing the placement of IDS over time. By creating uncertainty about the location of IDS, the likelihood of attackers' actions being detected will be increased. Venkatesan \emph{et al.} \cite{venkatesanMovingTargetDefense2016} analyze the problem of deploying IDS across the network in a resource-constrained environment using a graph-theoretic approach and propose several deployment strategies based on centrality measures \cite{friedkinTheoreticalFoundationsCentrality1991} that capture important properties of the network. Sengupta \emph{et al.}~\cite{senguptaMovingTargetDefense2018} model the same problem as a two-player general-sum Stackelberg game~\cite{korzhykComplexityComputingOptimal2010}. Two scalable algorithms are designed to find the equilibrium of the game, which corresponds to optimal  strategies for switching IDS placement that balance the overall security and usability. On the other hand, as many IDS have been based on artificial intelligence (AI) techniques \cite{sinclairApplicationMachineLearning1999, yinDeepLearningApproach2017, wangHASTIDSLearningHierarchical2018}, there have been several adversarial attacks against the underlying AI models to induce misclassification \cite{huangAdversarialAttacksSDNBased2018, usamaGenerativeAdversarialNetworks2019, linIDSGANGenerativeAdversarial2019}. The AI models may also adopt the moving target concept to improve the resilience against adversarial attacks, e.g., by randomizing the classification schemes \cite{vorobeychikOptimalRandomizedClassification2014, senguptaMTDeepBoostingSecurity2019} as depicted in Figure~\ref{fig:randomClassifier}.   

\begin{figure}
	\centering
	\includegraphics[width=0.8\linewidth]{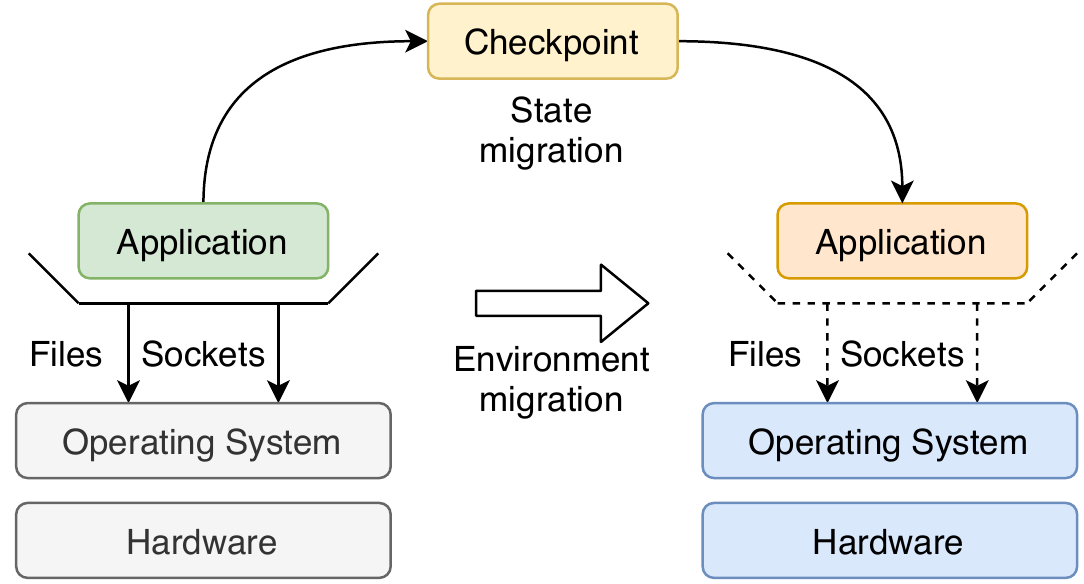}
	\caption[architecture]{The TALENT migration process in \cite{okhraviCreatingCyberMoving2011}} 
	\label{fig:talent}
\end{figure}

The exploitation attack phase may be guarded against by various dynamic system and software techniques, which are also helpful to disrupt the impact attack phase:

\textbf{Dynamic System}: 
Among others, the most commonly used technique for increasing system dynamics is address space layout randomization (ASLR) \cite{PaXAddressSpace2003, liAddressSpaceRandomizationWindows2006}, which hinders the exploitation of memory corruption vulnerabilities by randomizing memory addresses of a loaded software. 
To address code-injection attacks, an instruction set randomization (ISR) technique is proposed in \cite{kcCounteringCodeinjectionAttacks2003}, where an encoded version of software instructions is loaded into the memory and will be decoded by a key before being executed. 
Attackers' exploitation usually depend on vulnerabilities or characteristics of specific OS or CPU architectures. Thompson \emph{et al.} \cite{thompsonMultipleOSRotational2014} propose to enhance the security through a rotation of multiple OSs. Specifically, the method consists of several VMs equipped with different OSs. These VM hosts store shared data in a database and at one time only one of them will be mapped to an external IP address. The periodic rotation of VM hosts is controlled from an administrator machine running a daemon process, and the VM host that was previously in use is analyzed for evidence of intrusion and will be removed from rotation if compromised. Okhravi \emph{et al.} \cite{okhraviCreatingCyberMoving2011} propose a TALENT framework to improve cyber survivability through platform diversity (i.e., different OSs and architectures). In TALENT, as depicted in Figure~\ref{fig:talent}, a running application can be migrated between VMs with different platforms while preserving the state (e.g., the execution state, open files and network connections). A portable checkpoint compiler is used to facilitate the application live migration process. Note that the migration among different platforms must take less time than the time needed for attacking a specific platform. Or else, the migration actually diminishes security because threat actors now have a choice of multiple platforms to attack \cite{okhraviFindingFocusBlur2014}. 

\textbf{Dynamic Software}: There is also a wide range of attacks exploiting software vulnerabilities, which requires precise understanding of the target software. By randomizing the implementation, software diversity introduces uncertainty in the target, increases the cost to attackers, and may provide an effective counter to side-channel attacks \cite{larsenSoKAutomatedSoftware2014}. ChameleonSoft~\cite{azabChameleonSoftMovingTarget2011} proposes to divide a complex software program into smaller tasks, each of which has a set of executable variants that are functionally equivalent but with different quality attributes (e.g., performance, robustness, and mobility). The executable variants can then be shuffled to change the attack surface in accordance with different security situations. To defend against code reuse attacks, such as return-oriented programming (ROP), Gupta \emph{et al.} \cite{guptaMarlinFineGrained2013} propose a fine-grained software diversity approach called Marlin. Marlin breaks a software binary into function blocks and randomly shuffles the order. Such a process can be performed transparently at load time, which ensures every execution instance of the software to be unique. On the other hand, to prevent a software program from being exploited by identified vulnerabilities, Le Goues \emph{et al.} \cite{legouesGenProgGenericMethod2012} propose an automatic software repair method called GenProg. By utilizing an extended form of genetic programming, GenProg is able to evolve a software program with identified vulnerabilities to a functionally equivalent variant that are no longer susceptible to the previous risks. The dynamically patched software can be legacy programs without formal specifications and annotations.  
  
To prevent attackers who are already inside the network from eavesdropping on communication flows, Germano da Silva \emph{et al.} \cite{germanodasilvaCapitalizingSDNbasedSCADA2015} propose a multipath routing strategy, which relies on SDN features to frequently modify communication routes between SCADA devices. As each route transmits only a portion of the packets exchanged during the communication, even though the eavesdropper is well positioned in a strategic point of the network, he will not be able to intercept an entire communication between two devices. As the multipath routing strategy always relies on the shortest path to transmit the acknowledgment (ACK) packets from the receiver, Aseeri \emph{et al.} \cite{aseeriAlleviatingEavesdroppingAttacks2017} found that an attacker can still capture all the packets by eavesdropping on the shortest path and blocking the ACK packet corresponding to the packet sent through other routes until it is retransmitted via the path he is listening to. To address this defect, the SDN controller can be utilized to instruct the receiver to send the ACK packet via the path used by the sender. In the self-shielding dynamic network architecture (SDNA) \cite{yackoskiSelfshieldingDynamicNetwork2011, yackoskiApplyingSelfShieldingDynamics2013}, packets go through one or more intermediate devices before reaching the receiver. The intermediate devices are not simply routers; they also rewrite traffic to conceal the sender and receiver's identities. As a result, the eavesdropping attack can also be thwarted.     

The collection phase in the kill chain may be disrupted by dynamic data approaches. To prevent the cryptographic keys stored in the cloud from being extracted by attackers using cross-VM side-channel attacks \cite{zhangCrossVMSideChannels2012}, Pattuk \emph{et al.} \cite{pattukPreventingCryptographicKey2014} propose to partition the keys into random shares based on the secret sharing and threshold cryptography \cite{beimelSecretSharingSchemesSurvey2011}. The random shares are then stored in different VMs and will be regenerated periodically. As a result, the adversary has to attack multiple VMs to steal the key and the impact of a successful attack will be limited to a certain time period. On the other hand, dynamic data approaches may also impede the impact phase. Smutz and Stavrou \cite{smutzPreventingExploitsMicrosoft2015} propose to randomize the data block order of Microsoft office documents while keeping the visual interpretation intact. As malicious payloads embedded in the documents usually rely on a specific order of internal components, the randomization prevents them from being executed.

\section{Deception for Active Defense}
\label{sec:discussion}
\subsection{Deception in depth}
\label{dind}
Although each of the deception techniques surveyed in Section~\ref{sec:honeypots} to Section~\ref{sec:mtd} is able to disrupt one or several kill chain phases, when used alone, attackers can always find a way to circumvent it. One example is the various honeypot evasion techniques described in Section~\ref{sec_evadeHP}. By contrast, when multiple deception techniques that complement with each other are used together, forming an overall deception fabric covering several or even all layers in the deception stack, a more resilient cyber defense posture can be established. It is believed that such a deception in depth strategy should be leveraged by organizations to achieve comprehensive defense against the onslaught of advanced adversaries and attack techniques~\cite{lawrencepingreeEmergingTechnologyAnalysis2015}. In fact, there have already been commercial products implementing this strategy to create a complete illusion for the adversary \cite{DeceptionDepthCase2017}.        

A number of works have investigated the hybrid use of deception techniques. Through the combination, the deception effect on the adversary can be magnified, leading to the threat actor being deterred, delayed, distracted or detected.   

\begin{figure}
	\centering
	\includegraphics[width=0.93\linewidth]{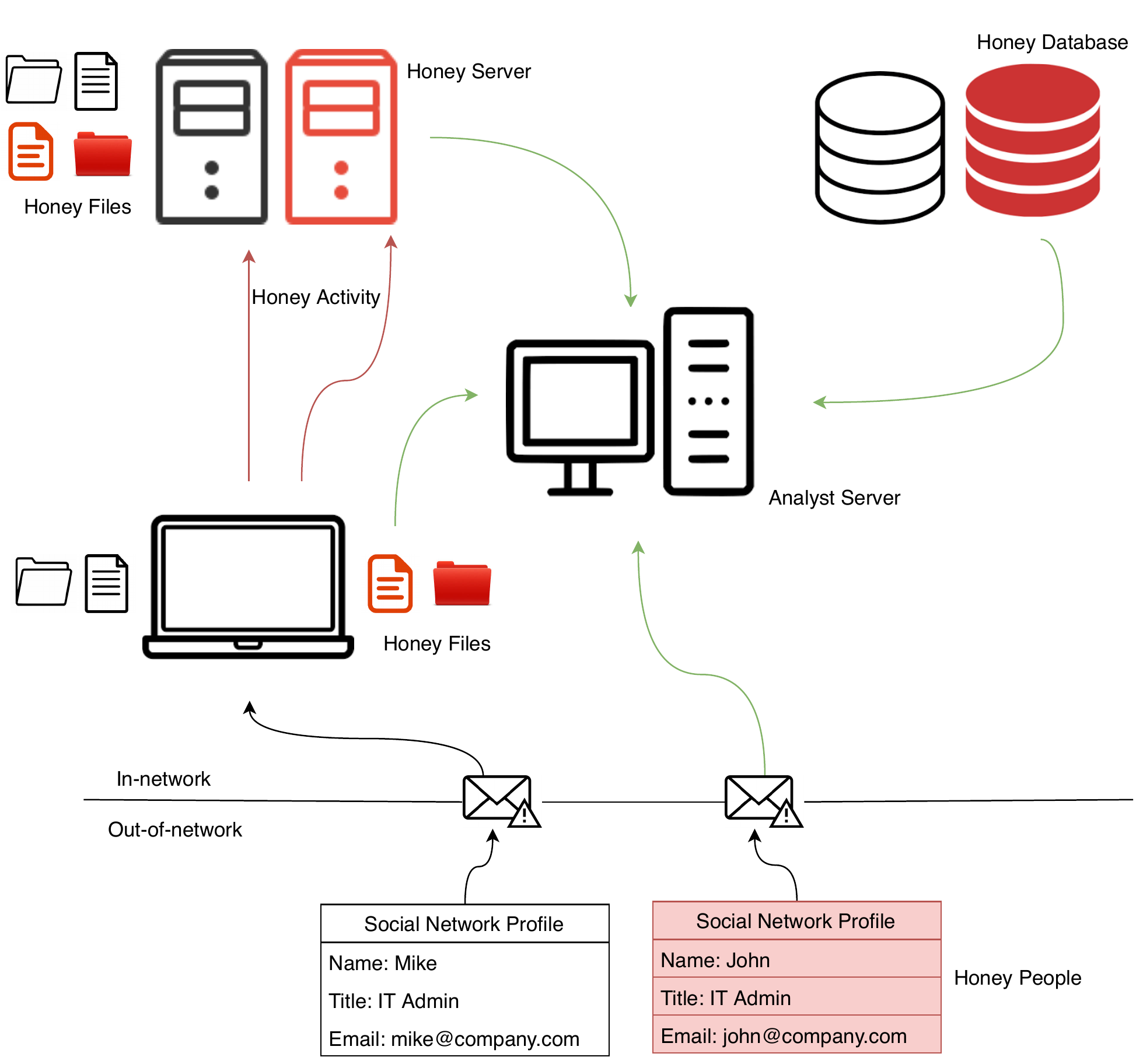}
	\caption[architecture]{The multi-layer deception system in \cite{wangDetectingTargetedAttacks2013}} 
	\label{fig:multilayserDecpt}
\end{figure}

Wang \emph{et al.} \cite{wangDetectingTargetedAttacks2013} propose a multi-layer deception system (see Figure~\ref{fig:multilayserDecpt}), which is composed of honeypot servers and various honeytokens such as honey people, honey files, honey database, and honey activities. The honey people is fake personas created on social network platforms. The honey activities are coupled with honey files and honeypot servers to prevent sophisticated attackers from discerning the bogus resources by observing user behaviors or network traffic. The alerts from all the deception entities are sent to the analyst server, where analysis is performed to confirm or remove the alerts. The analyst server may also correlate different alerts to extract more information of a penetration attempt. For example, if an alert is triggered on a honey file and later on a honey database entry, some correlation analyses may reveal that the two separate alerts correspond to the same espionage campaign. 

\begin{figure}
	\centering
	\includegraphics[width=\linewidth]{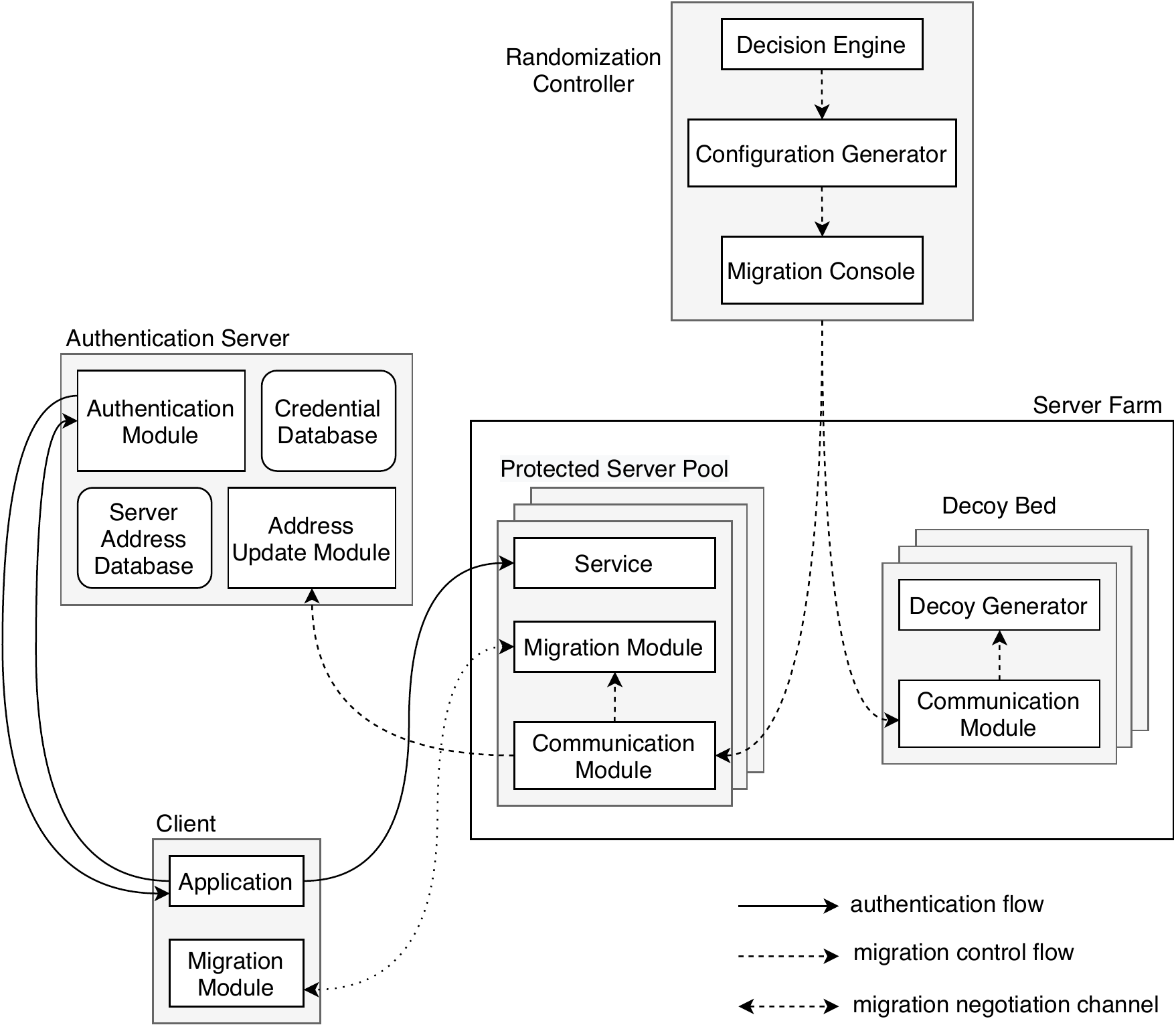}
	\caption[architecture]{The DESIR system in \cite{sunDESIRDecoyenhancedSeamless2016}, which is based on decoy-enhanced network address randomization} 
	\label{fig:desir}
\end{figure}

A decoy-enhanced network address randomization method called DESIR (see Figure~\ref{fig:desir}) is proposed in \cite{sunDESIRDecoyenhancedSeamless2016}, which dynamically mutates the network topology with a number of decoy servers to invalidate attacker's knowledge about the network. DESIR consists of four main components, i.e., an authentication server, a randomization controller, a protected server pool, and a decoy bed. The authentication server is responsible for verifying the client's credential, providing requested server's current IP address upon successful authentication, and updating the IP addresses of servers in the server pool. The randomization controller coordinates the mutation of the network. Its decision module determines the frequency to randomize the network addresses, configuration generator module controls the overall topology of the network, and migration console module distributes the new configurations to the real and decoy servers. Upon receiving new configurations, the decoy generator in the decoy bed will update the decoy network, including the decoy server's IP addresses and MAC addresses as well as the installed or emulated OS and applications. The decoy bed may include both high-interaction and low-interaction honeypots as decoys, which depend on the received configurations. Although the honeypots can be used to attract attackers and learn their TTPs, their main function in DESIR is to further confuse attackers, prolong their network scanning time, and invalidate the knowledge that can be gained. 

\begin{figure}
	\centering
	\begin{subfigure}[b]{0.6\linewidth}
		\includegraphics[width=1\linewidth]{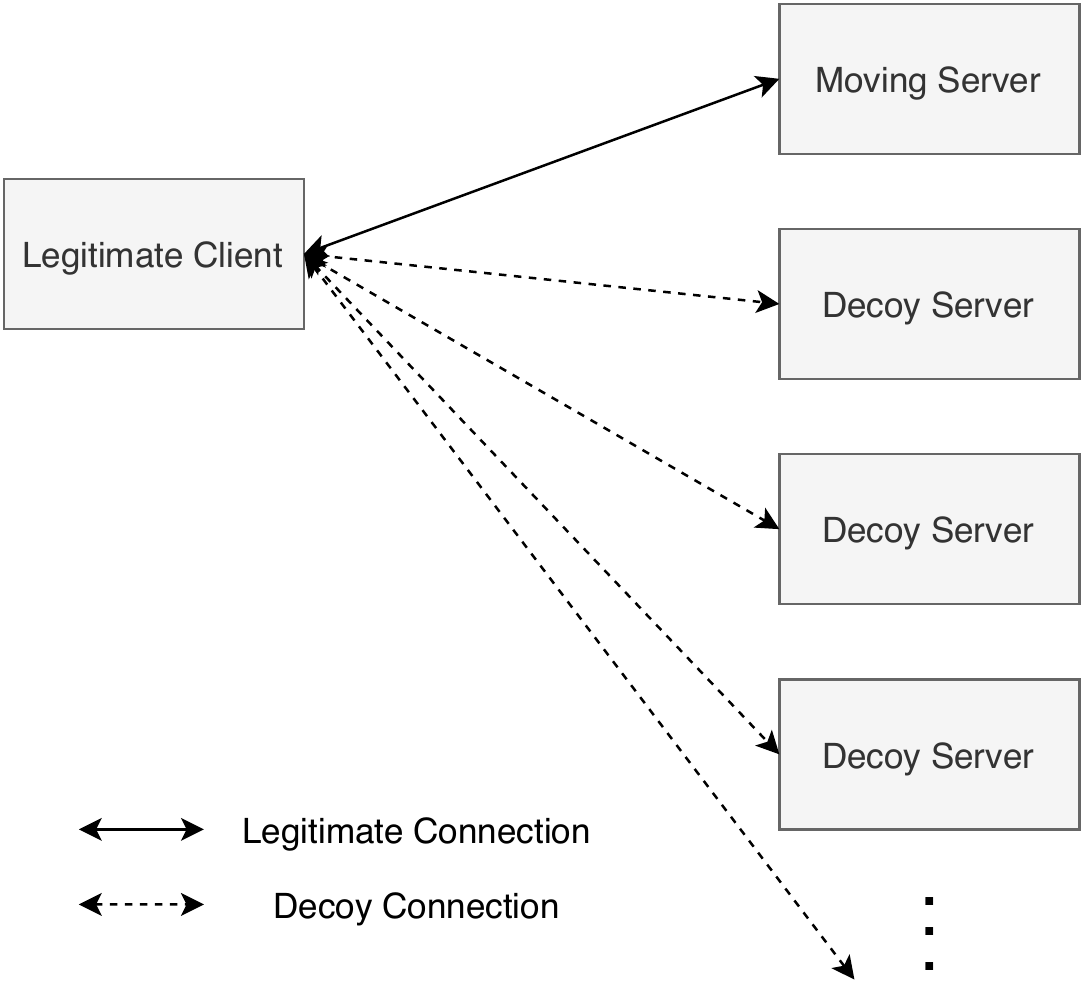}
		\caption{Decoy connections}
		\label{fig:dhm_1}
	\end{subfigure}
	\begin{subfigure}[b]{1\linewidth}
		\includegraphics[width=1\linewidth]{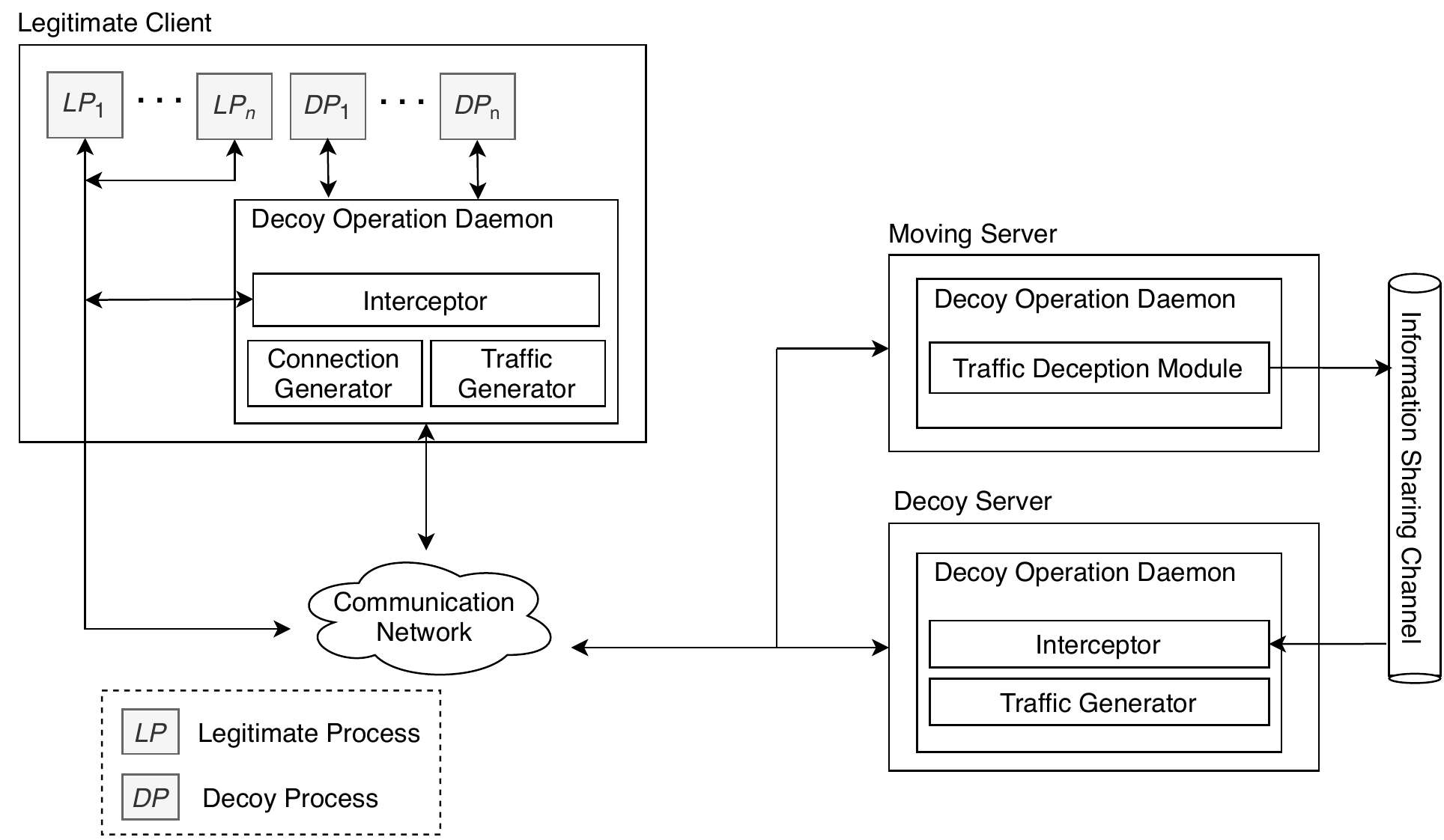}
		\caption{Context-aware decoy traffic injection}
		\label{fig:dhm_2}
	\end{subfigure}
	\caption{The decoy connection and traffic injection in \cite{parkSecureCyberDeception2018}}
	\label{fig:dhm}
\end{figure}

In the DESIR system, as shown in Figure~\ref{fig:desir}, the authenticated client and the moving server is seamlessly connected via the migration module. However, this means that if the client is compromised, it will be easy for the adversary to trace the moving server by analyzing the network traffic. To cope with this threat, Park \emph{et al.} \cite{parkSecureCyberDeception2018} propose to inject decoy connection and traffic with the honeypot servers, as illustrated in Figure~\ref{fig:dhm}. To generate decoy traffic that is even convincing for sophisticated attackers, a context-aware traffic generation mechanism is used (see Figure~\ref{fig:dhm_2}). On the client, the connection generator module of the decoy operation daemon is responsible for creating decoy connections with honeypot servers, and the traffic generator module creates decoy traffic of a similar pattern to the legitimate traffic. The traffic deception module on the moving server shares the characteristics of the outbound traffic, which are imitated by the traffic generator module on the decoy server to generate similar traffic with decoy processes on the client. Besides, similar to \cite{jafarianMultidimensionalHostIdentity2016}, OS fingerprint mutation is also applied on all servers so that the attack surface is further obfuscated.  

\begin{figure}
	\centering
	\includegraphics[width=\linewidth]{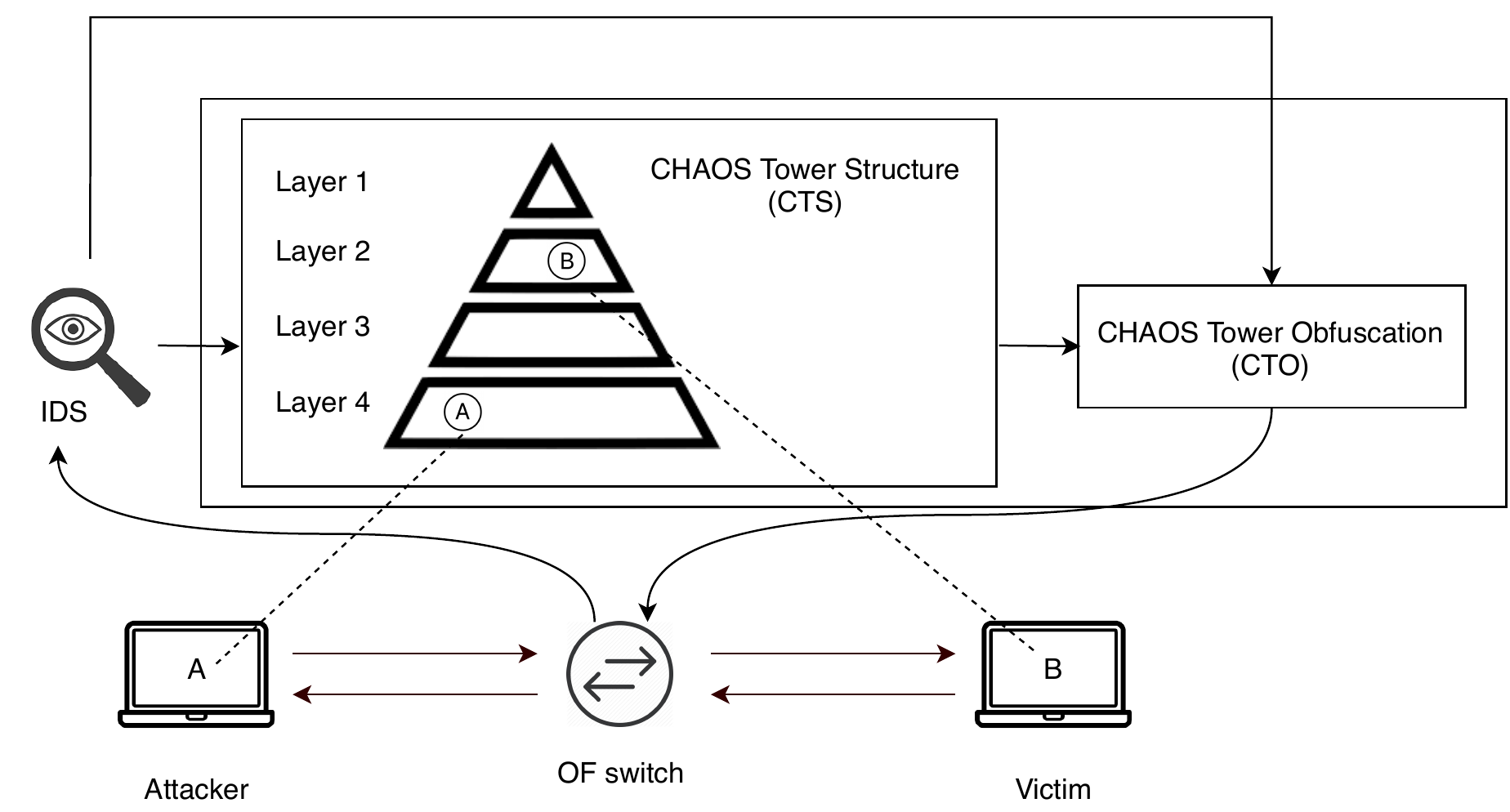}
	\caption[architecture]{The CHAOS system in \cite{shiCHAOSSDNBasedMoving2017}, where suspicious connections identified by IDS or CTS are obfuscated by CTO} 
	\label{fig:chaos}
\end{figure}

The SDN based CHAOS system (see Figure~\ref{fig:chaos}) in \cite{shiCHAOSSDNBasedMoving2017} obfuscates the network attack surface by using honeypot (i.e., decoy servers), honeytoken (i.e, fake response to port scanning), and MTD (i.e, random host mutation) techniques. In particular, host machines in the network is divided into several layers according to their security levels, which forms a CHAOS tower structure (CTS). Communication rules are defined in a CTS module. For example, connection requests from a host machine in lower layers to hosts in higher layers will be deemed as suspicious. The suspicious communications determined by the CTS module and other traffic identified by IDS as malicious will be forwarded to a CHAOS tower obfuscation (CTO) module, where the three types of techniques listed above are implemented. The three obfuscation strategies are applied based on a threshold factor, which can be controlled by the administrator according to the required security level and the structure of the protected network. 

A complete list of the reviewed deception techniques are shown in Table~\ref{table_surveyedMethods}, where they are classified based on the two-dimensional taxonomy, i.e., which attack phases they can disrupt and which deception layer they belong to. Note that some methods, especially the hybrid ones, are able to disrupt multiple attack phases and use techniques from multiple layers. These methods are repeated in the table to fully indicate their characteristics and effects. The reconnaissance phase in Table~\ref{table_surveyedMethods} just refers to the external reconnaissance, while deception techniques that can guard against the internal reconnaissance and target reconnaissance attack phases are the same as those for the lateral movement phase. This separation is aimed to make it easier to understand the different applicabilities and effects of the numerous deception techniques for disrupting attacker reconnaissance.  


\begin{table*}[]
	\caption{Deception techniques classified based on the two-dimensional taxonomy, where HP, HT, and MTD denotes honeypot, honeytoken, and moving target defense, respectively}
	\label{table_surveyedMethods}
	\begin{tabular}{|l|L|L|L|L|}
		\hline
		\multicolumn{1}{|c|}{\multirow{2}{*}{\textbf{Attack Phase}}} & \multicolumn{4}{c|}{\textbf{Deception Layer}}                                                                          \\ \cline{2-5} 
		\multicolumn{1}{|c|}{}                                       & \multicolumn{1}{c|}{Network} & \multicolumn{1}{c|}{System} & \multicolumn{1}{c|}{Software} & \multicolumn{1}{c|}{Data} \\ \hline
		\multirow{3}{*}{Reconnaissance}                              
		& HP:\cite{tomlistonLaBreaStickyHoneypot2001, leslieshingImprovedTarpitNetwork2016, provosVirtualHoneypotFramework2004, bordersOpenFireUsingDeception2007, kumanExperimentUsingIMUNES2017, sunDESIRDecoyenhancedSeamless2016, jafarianMultidimensionalHostIdentity2016, parkSecureCyberDeception2018, shiCHAOSSDNBasedMoving2017, adebayoDeceptorintheMiddleDitMCyber2020}                         
		& HP:\cite{provosVirtualHoneypotFramework2004, dissoPlausibleSolutionSCADA2013, winnConstructingCosteffectiveTargetable2015, lukasristConpot2015, zhaoResearchHighInteractive2017, vetterlHoneypotsAgeUniversal2019, jiangDesignImplementationMachine2020, sunCyberMoatCamouflagingCritical2017}                        
		& HP:\cite{provosVirtualHoneypotFramework2004}                          
		& HP:                      \\ \cline{2-5} 
		& HT:\cite{trassareTechniquePresentingDeceptive2013, parkSecureCyberDeception2018, shiCHAOSSDNBasedMoving2017, sunBelievableDecoySystem2020}                         
		& HT:\cite{murphyApplicationDeceptionCyberspace2010, julianDelayingtypeResponsesUse2002, parkSecureCyberDeception2018, fredcohenMovingTargetDefenses2010, albaneseDeceptionBasedApproach2015, sunBelievableDecoySystem2020}                        
		& HT:\cite{gavrilisFlashCrowdDetection2007, douglasbrewerLinkObfuscationService2010, virvilisChangingGameArt2014, katsinisSecurityMechanismWeb2012,katsinisFrameworkIntrusionDeception2013, hanEvaluationDeceptionBasedWeb2017}                          
		& HT:\cite{bourkeBreachDetectionScale2018, stringhiniDetectingSpammersSocial2010, virvilisChangingGameArt2014, whiteCreatingPersonallyIdentifiable2010, katsinisSecurityMechanismWeb2012,katsinisFrameworkIntrusionDeception2013, wangDetectingTargetedAttacks2013, karunaFakeDocumentGeneration2020}                      \\ \cline{2-5} 
		& MTD:  \cite{kewleyDynamicApproachesThwart2001, antonatosDefendingHitlistWorms2005, jafarianOpenflowRandomHost2012, al-shaerRandomHostMutation2013, dunlopMT6DMovingTarget2011, dunlopBlindManBluff2012, sunDESIRDecoyenhancedSeamless2016, jafarianMultidimensionalHostIdentity2016, parkSecureCyberDeception2018, shiCHAOSSDNBasedMoving2017, fredcohenMovingTargetDefenses2010, sunScalableHighFidelity2019, trassareTechniqueNetworkTopology2013, duanRangeTopologyMutation2020, sunCyberMoatCamouflagingCritical2017}                      
		& MTD:  \cite{kampanakisSDNbasedSolutionsMoving2014, zhaoSDNBasedFingerprintHopping2017, jafarianMultidimensionalHostIdentity2016, parkSecureCyberDeception2018}                     
		& MTD:                         
		& MTD:                     \\ \hline
		\multirow{3}{*}{Delivery}                                    
		& HP:                         
		& HP:\cite{bordersOpenFireUsingDeception2007, biedermannFastDynamicExtracted2012, uriasGatheringThreatIntelligence2016, dissoPlausibleSolutionSCADA2013, winnConstructingCosteffectiveTargetable2015, lukasristConpot2015, zhaoResearchHighInteractive2017, vetterlHoneypotsAgeUniversal2019, jiangDesignImplementationMachine2020}                        
		& HP:\cite{anagnostakisDetectingTargetedAttacks2005}                          
		& HP:                      \\ \cline{2-5} 
		& HT:                         
		& HT:                        
		& HT:                          
		& HT:                      \\ \cline{2-5} 
		& MTD:  \cite{antonatosDefendingHitlistWorms2005, jafarianOpenflowRandomHost2012, al-shaerRandomHostMutation2013, dunlopMT6DMovingTarget2011, dunlopBlindManBluff2012}                      
		& MTD:                       
		& MTD:                         
		& MTD:                     \\ \hline
		\multirow{3}{*}{Defense Evasion}                             
		& HP: \cite{fuRecognizingVirtualHoneypots2006, altUncoveringNetworkTarpits2014}                        
		& HP:                    
		& HP:                          
		& HP:                      \\ \cline{2-5} 
		& HT:                         
		& HT: \cite{rrushiHoneypotEvaderActivityguided2019, roweDefendingCyberspaceFake2007}                       
		& HT:                          
		& HT:                      \\ \cline{2-5} 
		& MTD: \cite{venkatesanMovingTargetDefense2016, senguptaMovingTargetDefense2018}                       
		& MTD: \cite{vorobeychikOptimalRandomizedClassification2014, senguptaMTDeepBoostingSecurity2019}                       
		& MTD:                         
		& MTD:                     \\ \hline
		\multirow{3}{*}{Exploitation}                                
		& HP: \cite{virvilisChangingGameArt2014, rrushiDNICArchitecturalDevelopments2021}                        
		& HP: \cite{bordersOpenFireUsingDeception2007, biedermannFastDynamicExtracted2012, uriasGatheringThreatIntelligence2016, dissoPlausibleSolutionSCADA2013, winnConstructingCosteffectiveTargetable2015, lukasristConpot2015, zhaoResearchHighInteractive2017, wangDetectingTargetedAttacks2013, jafarianMultidimensionalHostIdentity2016, parkSecureCyberDeception2018, vetterlHoneypotsAgeUniversal2019, jiangDesignImplementationMachine2020, sunCyberMoatCamouflagingCritical2017}                       
		& HP:   \cite{anagnostakisDetectingTargetedAttacks2005}                       
		& HP:                      \\ \cline{2-5} 
		& HT:                         
		& HT:  \cite{araujoPatchesHoneyPatchesLightweight2014, leePhantomFSFileBasedDeception2020, choiPhantomFSv2DareYou2020}                      
		& HT:  \cite{craneBoobyTrappingSoftware2013, araujoPatchesHoneyPatchesLightweight2014, virvilisChangingGameArt2014}                        
		& HT:                    \\ \cline{2-5} 
		& MTD:                        
		& MTD: \cite{PaXAddressSpace2003, liAddressSpaceRandomizationWindows2006, kcCounteringCodeinjectionAttacks2003, thompsonMultipleOSRotational2014, okhraviCreatingCyberMoving2011}                      
		& MTD: \cite{azabChameleonSoftMovingTarget2011, guptaMarlinFineGrained2013, legouesGenProgGenericMethod2012}                        
		& MTD:                  \\ \hline
		\multirow{3}{*}{Installation}                                
		& HP:                         
		& HP:  \cite{bordersOpenFireUsingDeception2007, biedermannFastDynamicExtracted2012, uriasGatheringThreatIntelligence2016, dissoPlausibleSolutionSCADA2013, winnConstructingCosteffectiveTargetable2015, lukasristConpot2015, zhaoResearchHighInteractive2017,  vetterlHoneypotsAgeUniversal2019}                        
		& HP:                          
		& HP:                      \\ \cline{2-5} 
		& HT:                         
		& HT:                        
		& HT:                          
		& HT:                      \\ \cline{2-5} 
		& MTD:                        
		& MTD:                       
		& MTD:                         
		& MTD:                     \\ \hline
		\multirow{3}{*}{C2}                                          
		& HP:                         
		& HP: \cite{bordersOpenFireUsingDeception2007, biedermannFastDynamicExtracted2012, uriasGatheringThreatIntelligence2016, dissoPlausibleSolutionSCADA2013, winnConstructingCosteffectiveTargetable2015, lukasristConpot2015, zhaoResearchHighInteractive2017,  vetterlHoneypotsAgeUniversal2019}                      
		& HP: 
		& HP:                      \\ \cline{2-5} 
		& HT:                         
		& HT:                        
		& HT:                          
		& HT:                      \\ \cline{2-5} 
		& MTD:                       
		& MTD:                       
		& MTD:                         
		& MTD:                     \\ \hline
		\multirow{3}{*}{Privilege Escalation}                        
		& HP:                         
		& HP:                        
		& HP:                          
		& HP:                      \\ \cline{2-5} 
		& HT:                         
		& HT: \cite{juelsBodyguardLiesUse2014, kaghazgaranInsiderThreatDetection2015}                       
		& HT:                          
		& HT: 
		                    \\ \cline{2-5} 
		& MTD:                        
		& MTD:                       
		& MTD:                         
		& MTD:                     \\ \hline
		\multirow{3}{*}{Credential Access}                           
		& HP:                         
		& HP:                        
		& HP:                          
		& HP:                      \\ \cline{2-5} 
		& HT:                         
		& HT:  \cite{juelsBodyguardLiesUse2014, kaghazgaranInsiderThreatDetection2015}                      
		& HT:                          
		& HT:  \cite{juelsHoneyEncryptionSecurity2014, abiodunReinforcingSecurityInstant2020, omolaraDeceptionModelRobust2019a, juelsHoneywordsMakingPasswordcracking2013, almeshekahErsatzPasswordsEndingPassword2015, wangDetectingTargetedAttacks2013, karunaFakeDocumentGeneration2020}                    \\ \cline{2-5} 
		& MTD:                        
		& MTD:                       
		& MTD:                         
		& MTD:                     \\ \hline
		\multirow{3}{*}{Lateral Movement}                            
		& HP: \cite{provosVirtualHoneypotFramework2004, bordersOpenFireUsingDeception2007, sunDESIRDecoyenhancedSeamless2016, jafarianMultidimensionalHostIdentity2016, parkSecureCyberDeception2018, shiCHAOSSDNBasedMoving2017}                        
		& HP: \cite{provosVirtualHoneypotFramework2004, bordersOpenFireUsingDeception2007, biedermannFastDynamicExtracted2012, uriasGatheringThreatIntelligence2016, dissoPlausibleSolutionSCADA2013, winnConstructingCosteffectiveTargetable2015, lukasristConpot2015, zhaoResearchHighInteractive2017, wangDetectingTargetedAttacks2013, vetterlHoneypotsAgeUniversal2019, sunCyberMoatCamouflagingCritical2017}                       
		& HP: \cite{provosVirtualHoneypotFramework2004, anagnostakisDetectingTargetedAttacks2005}                         
		& HP:                      \\ \cline{2-5} 
		& HT: \cite{rrushiHoneypotEvaderActivityguided2019, bowenAutomatingInjectionBelievable2010, bowenSystemGeneratingInjecting2012, wangDetectingTargetedAttacks2013, parkSecureCyberDeception2018, shiCHAOSSDNBasedMoving2017,sunBelievableDecoySystem2020}                        
		& HT: \cite{araujoPatchesHoneyPatchesLightweight2014, rrushiHoneypotEvaderActivityguided2019, wangDetectingTargetedAttacks2013, parkSecureCyberDeception2018, fredcohenMovingTargetDefenses2010,sunBelievableDecoySystem2020}                      
		& HT: \cite{araujoPatchesHoneyPatchesLightweight2014}                         
		& HT: \cite{juelsHoneywordsMakingPasswordcracking2013, almeshekahErsatzPasswordsEndingPassword2015,  wangDetectingTargetedAttacks2013, karunaFakeDocumentGeneration2020}                     \\ \cline{2-5} 
		& MTD: \cite{germanodasilvaCapitalizingSDNbasedSCADA2015, aseeriAlleviatingEavesdroppingAttacks2017, yackoskiSelfshieldingDynamicNetwork2011, yackoskiApplyingSelfShieldingDynamics2013, sunDESIRDecoyenhancedSeamless2016, jafarianMultidimensionalHostIdentity2016, parkSecureCyberDeception2018, shiCHAOSSDNBasedMoving2017, fredcohenMovingTargetDefenses2010, sunScalableHighFidelity2019, chiangACyDSAdaptiveCyber2016, sunCyberMoatCamouflagingCritical2017}                       
		& MTD:  \cite{antonatosDefendingHitlistWorms2005, jafarianOpenflowRandomHost2012, al-shaerRandomHostMutation2013, dunlopMT6DMovingTarget2011, dunlopBlindManBluff2012, jafarianMultidimensionalHostIdentity2016, parkSecureCyberDeception2018, kampanakisSDNbasedSolutionsMoving2014, zhaoSDNBasedFingerprintHopping2017}                     
		& MTD:                            
		& MTD:                     \\ \hline
		\multirow{3}{*}{Collection}                                  
		& HP:                         
		& HP: \cite{bordersOpenFireUsingDeception2007, biedermannFastDynamicExtracted2012, uriasGatheringThreatIntelligence2016, dissoPlausibleSolutionSCADA2013, winnConstructingCosteffectiveTargetable2015, lukasristConpot2015, zhaoResearchHighInteractive2017,  vetterlHoneypotsAgeUniversal2019}                      
		& HP: \cite{anagnostakisDetectingTargetedAttacks2005}                         
		& HP:                      \\ \cline{2-5} 
		& HT:                         
		& HT: \cite{juelsBodyguardLiesUse2014, kaghazgaranInsiderThreatDetection2015, yuillHoneyfilesDeceptiveFiles2004, wangDetectingTargetedAttacks2013, virvilisChangingGameArt2014}                       
		& HT: \cite{antanascenysImplementationHoneytokenModule2005, padayacheeAspectisingHoneytokensContain2014}                         
		& HT: \cite{juelsHoneyEncryptionSecurity2014, abiodunReinforcingSecurityInstant2020, omolaraDeceptionModelRobust2019a, juelsHoneywordsMakingPasswordcracking2013, almeshekahErsatzPasswordsEndingPassword2015,  antanascenysImplementationHoneytokenModule2005, bercovitchHoneyGenAutomatedHoneytokens2011, padayacheeAspectisingHoneytokensContain2014, yuillHoneyfilesDeceptiveFiles2004, bowenBaitingAttackersUsing2009, vorisBaitSnitchDefending2013,vorisFoxTrapThwarting2015, benwhithamAutomatingGenerationEnticing2017, wangDetectingTargetedAttacks2013, karunaFakeDocumentGeneration2020}                     \\ \cline{2-5} 
		& MTD: \cite{germanodasilvaCapitalizingSDNbasedSCADA2015, aseeriAlleviatingEavesdroppingAttacks2017, yackoskiSelfshieldingDynamicNetwork2011, yackoskiApplyingSelfShieldingDynamics2013}                       
		& MTD:                       
		& MTD:                         
		& MTD:  \cite{pattukPreventingCryptographicKey2014}                   \\ \hline
		\multirow{3}{*}{Exfiltration}                                
		& HP: \cite{virvilisChangingGameArt2014}                        
		& HP:  \cite{bordersOpenFireUsingDeception2007, biedermannFastDynamicExtracted2012, uriasGatheringThreatIntelligence2016, dissoPlausibleSolutionSCADA2013, winnConstructingCosteffectiveTargetable2015, lukasristConpot2015, zhaoResearchHighInteractive2017,  vetterlHoneypotsAgeUniversal2019}                      
		& HP:                          
		& HP:                      \\ \cline{2-5} 
		& HT:                         
		& HT:                        
		& HT:                       
		& HT:                     \\ \cline{2-5} 
		& MTD:                        
		& MTD:                       
		& MTD:                         
		& MTD:                     \\ \hline
		\multirow{3}{*}{Impact}                                      
		& HP:  \cite{bordersOpenFireUsingDeception2007}                       
		& HP:  \cite{bordersOpenFireUsingDeception2007, biedermannFastDynamicExtracted2012, uriasGatheringThreatIntelligence2016, dissoPlausibleSolutionSCADA2013, winnConstructingCosteffectiveTargetable2015, lukasristConpot2015, zhaoResearchHighInteractive2017, wangDetectingTargetedAttacks2013, jafarianMultidimensionalHostIdentity2016, parkSecureCyberDeception2018, vetterlHoneypotsAgeUniversal2019}                      
		& HP: \cite{anagnostakisDetectingTargetedAttacks2005}                         
		& HP:                      \\ \cline{2-5} 
		& HT:                         
		& HT: \cite{araujoPatchesHoneyPatchesLightweight2014, leePhantomFSFileBasedDeception2020, choiPhantomFSv2DareYou2020}                        
		& HT: \cite{craneBoobyTrappingSoftware2013, araujoPatchesHoneyPatchesLightweight2014}                         
		& HT:                   \\ \cline{2-5} 
		& MTD:                        
		& MTD: \cite{PaXAddressSpace2003, liAddressSpaceRandomizationWindows2006, kcCounteringCodeinjectionAttacks2003, thompsonMultipleOSRotational2014, okhraviCreatingCyberMoving2011}                        
		& MTD: \cite{azabChameleonSoftMovingTarget2011, guptaMarlinFineGrained2013, legouesGenProgGenericMethod2012}                         
		& MTD:  \cite{smutzPreventingExploitsMicrosoft2015}                    \\ \hline
	\end{tabular}
\end{table*}


\subsection{Deception lifecycle}

\begin{figure}
	\centering
	\includegraphics[width=\linewidth]{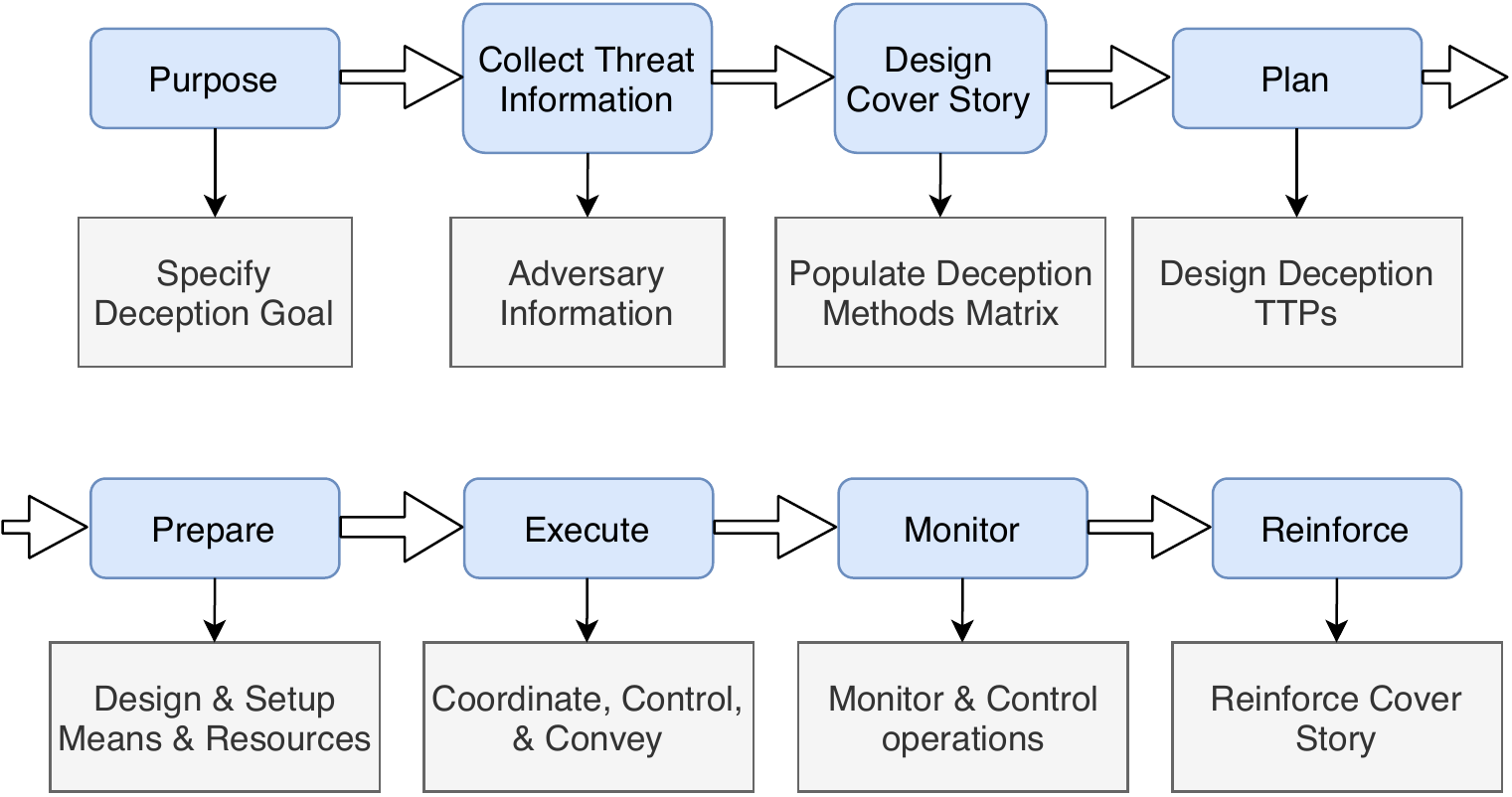}
	\caption[architecture]{The eight-phase cyber deception chain in \cite{heckmanDenialDeceptionCyber2015}}
	\label{fig:cyberDeceptionChain}
\end{figure}

Deception techniques are employed to affect threat actors such that they take action or inaction to the advantage of cyber defenders. To achieve and maintain the desired perceptual and cognitive effects, deception mechanisms have to be properly designed and updated. Almeshekah \cite{almeshekahUsingDeceptionEnhance2015} proposes a deception framework comprising three main phases, i.e., planning, implementing and integrating, and monitoring and evaluating. In the planning phase, the goal of deception is specified, the attacker's bias that can be exploited to achieve desired reactions is analyzed, and the risk that may be introduced by deception techniques is also assessed. De Faveri \emph{et al.} \cite{defaveriGoalDrivenDeceptionTactics2016} propose a goal-driven approach for designing the deception based defense. The approach integrates three phases, i.e., system modeling for specifying the goal, security modeling for specifying security concerns from the attacker perspective, and the deception modeling for specifying the defense (e.g., designing deception stories, monitoring channels, and deception metrics). The first two phases establish the context for modeling the deception. Heckman \emph{et al.} \cite{heckmanDenialDeceptionCyber2015} propose a deception chain for deception operation management from a lifecycle perspective, which is composed of eight phases as depicted in Figure~\ref{fig:cyberDeceptionChain}. Note that its second phase to the fourth phase, i.e., collecting threat information, designing cover story, and planning, can be implemented by leveraging our two-dimensional taxonomy.  

After the coordinated deception tactics are built and executed, they should evolve in response to environment changes and attacker's behavior \cite{defaveriDesigningAdaptiveDeception2016}. Take the honeypot for example. A static honeypot is very likely to be detected by the adversary. By contrast, dynamic honeypots~\cite{budiartoHoneypotsWhyWe2004, zanoramyansiryzakariaReviewDynamicIntelligent2013, wangIntelligentDeploymentPolicy2020}, besides their capability of learning about the network for automated deployment, can continuously monitor the network environment for changes and reconfigure themselves accordingly. Moreover, some dynamic honeypots are able to adapt based on their interactions with attackers. By taking advantage of reinforcement learning, Wagener \emph{et al.} \cite{wagenerAdaptiveSelfconfigurableHoneypots2011, wagenerHelizaTalkingDirty2011} build honeypots that can learn to adopt the best behavior such as blocking or executing commands, returning erroneous messages, and insulting the adversary. The insults act as reverse Turing tests \cite{bairdPessimalPrintReverseTuring2003} and aim to identify whether the opponent is human or an automated tool. Based on the same concept, Pauna and Bica \cite{paunaRASSHReinforcedAdaptive2014} build a self-adaptive honeypot that also emulates a secure shell (SSH) server, where an extra interaction strategy, i.e., delaying the command execution, is added.             

Besides using reinforcement learning to improve the interaction of deception techniques with the adversary, game theory may also be utilized. Carroll and Grosu \cite{carrollGameTheoreticInvestigation2009} model the interaction between the defender and the attacker as a signaling game, which is a non-cooperative two player dynamic game (i.e., the two players take turns to choose actions) of incomplete information. The incomplete information is due to the attacker's uncertainty of the target (e.g., whether the target system is a honeypot). Deceptive equilibrium strategies are then derived to achieve better defense of the network. Rahman \emph{et al.} \cite{rahmanGametheoreticApproachDeceiving2013} model the interaction between OS fingerprinter and the defender as a signaling game and the equilibrium analysis results in a counter-fingerprinting mechanism called DeceiveGame. Unlike many other tools which alter all connections' outgoing packets to deceive fingerprinting and incur significant performance degradation, DeceiveGame can distinguish fingerprinters from benign clients and selectively mystify packets to confuse the fingerprinters, hence minimizing the side effects. Carter \emph{et al.} \cite{carterGameTheoreticApproach2014} model the interaction as a two-palyer Stackelberg game to discover optimal moving target strategies (instead of simple randomization) for dynamic platforms based defense, while Lei \emph{et al.} \cite{leiOptimalStrategySelection2017} model the confrontation in MTD as a Markov game to identify the optimal hopping strategy. In general, game theory makes it possible for cyber defenders to investigate how the adversary's belief evolves and influences his actions, and provides a quantitative framework for optimizing the manipulation of this belief to the benefit of defense \cite{horakManipulatingAdversaryBelief2017}.    

In deception defense, it is critical to continuously monitor the feedback channels to decide whether the desired effects on attackers are achieved. Honeypots may be easily identified, evaded, and even compromised by the adversary, honeytokens may not be enticing, and the hopping frequency in MTD may not be high enough. If the feedback indicates the deception defense is lack of effectiveness, the deception strategies, tactics, and techniques must be immediately adjusted. For this part, game theory may also be helpful. For instance, by building a multi-layer game model, a feedback learning framework is developed in \cite{zhuGameTheoreticApproachFeedbackDriven2013}, which enables the system to monitor its current state and update the defense strategy based on the risk it estimates on the fly.

\section{Reflection and Outlook}
\label{sec:outlook}

In cyber defense, deception techniques exploit attackers' psychological biases and vulnerabilities and have direct impact on their beliefs, decisions, and actions. Even just some clues that the target system's response may be fake will delay or even turn away the adversary \cite{roweModelDeceptionCyberattacks2004}. Compared to conventional attack prevention or detection tools which can only impede the adversary's current actions, deception techniques may have long-term impact on the adversary. Nonetheless, as deception techniques typically involve active adversary engagement, they have to be carefully maintained to stay effective. Especially when addressing APT and insider threats, high-fidelity deception over a long period is necessary. This poses stringent requirements to the deception operator. Although a vast number of deception techniques in various domains have been proposed since their inception in late 1980s, very few of them achieve real-life applications. According to Lance Spitzner, deception techniques were held back not by the concept, but by the technology \cite{antonchuvakinWillDeceptionFizzle2019}. For example, early honeypots require manual customization and management, which is extremely time-consuming and error-prone. Only after recent advancement in virtualization and SDN technology, which simplifies and automates the tedious process, honeypot techniques become scalable in real-life networks. To further enhance the usability of deception techniques, some recent works \cite{islamActiveDeceptionFramework2020, incTrapXIntroducesIndustryFirst2020} propose to provide \textit{deception as a service} through automatically orchestrated deception deployment with minimal human involvement. These efforts will definitely facilitate wide adoption of deception techniques. On the other hand, most of the early deception techniques have the drawback of assuming static network configurations, while recent dynamic techniques leveraging game theory models usually oversimplify the adversary's strategies \cite{yeDifferentiallyPrivateGame2021}. These limitations make the actual deployment less effective and easy to be evaded by the adversary. We think recent efforts in testbeds and experimentation platforms  \cite{acostaCybersecurityDeceptionExperimentation2020, ferguson-walterTularosaStudyExperimental2018} is promising to solve this problem. With deception techniques tested and validated on realistic systems and in realistic settings, not only the possible design flaws can be identified much more easily, but also the effectiveness of different techniques can be compared for easier tradeoff or complementary usage. It has been shown both analytically and experimentally that a single deception technique is not enough to attain highly resilient cyber deception \cite{duanCONCEALStrategyComposition2018}. The testbed and experimentation platforms will be an ideal environment for finding the optimal composition of different deception building blocks. 



The two-dimensional taxonomy, built based on our proposed cyber kill chain model and the four-layer deception stack, facilitates the systematic review of representative approaches from the domains of honeypots, honeytokens, and MTD techniques in a threat-focused manner. To create a holistic deception fabric covering the protected network and form a complete illusion for the adversary, an integrated use of these techniques is believed to be a prerequisite. Our taxonomy may serve as a guide or reference to consolidate and coordinate the different techniques. By adopting the deception in depth strategy and properly managing deception mechanisms throughout their lifecycle, a resilient deception defense will be built, which helps organizations establish the active cyber defense posture.    

For future research directions, we think that there will be more works on effective integration of the deception techniques from different domains. A well-designed deception defense should fully exploit the characteristics of different deception techniques. As these characteristics often complement with each other, the overall deception effect may be magnified and the defense cost may be optimized. For instance, defenders may distribute low-cost honeytokens all over the network to monitor the security status. Based on the indicated threat level, the instances of high-interaction honeypots, the hopping frequencies of MTD techniques, and the density of decoy activities can be dynamically adjusted. Such context awareness will be further enhanced when deception defense is combined with conventional threat detection and response (TDR) solutions. 
On the other hand, to smooth the integration, the deception effects on the adversary and the cost of deception operations should be quantified. In fact, there have been some works in this direction. For the former, Maleki \emph{et al.} \cite{malekiMarkovModelingMoving2016} propose a Markov model based framework for analyzing MTD techniques, where security capacity is defined to measure their strength or effectiveness. For the latter, Wang \emph{et al.} \cite{wangDetectingTargetedAttacks2013} model the design of the multi-layer deception system as an optimization problem to minimize the total expected loss due to system deployment and asset compromise. To better address these two problems, we feel that the quantitative framework offered by game theory will play an important role.   

We may also witness hardware become a more important participant in cyber defense. A lesson from the Spectre and Meltdown attacks \cite{abu-ghazalehHowSpectreMeltdown2019} is that no security is possible if the underlying hardware is vulnerable. Conversely, a more secure hardware may better obfuscate the attack surface and boost the uncertainty. For instance, the Morpheus secure architecture in~\cite{gallagherMorpheusVulnerabilityTolerantSecure2019} implements a hardware based churning mechanism to transparently randomize key program values, which are needed by attackers for crafting successful attacks, at runtime. To enhance the value of the churning mechanism, Morpheus also incorporates an attack detector. Once sensing a potential attack, the detector can immediately trigger an increased churn rate to strengthen the defense and repel the attack. Besides, the ensembles of MTD techniques developed on Morpheus, such as relocating pointers and encrypting code and pointers, can use the hardware support to achieve more randomness at a lower cost. 

The ultimate target of deception defense is the adversary's perception and belief. We think that there will also be more works developed based on better understanding of the human element. Ferguson-Walter \cite{ferguson-walterEmpiricalAssessmentEffectiveness2020} suggests that advances in behavioral science should be leveraged to better influence attacker's target selection and operations. By manipulating threat actors' cognitive biases and cognitive load, it will be made more difficult for them to achieve their objectives.

\bibliographystyle{IEEEtran}

\bibliography{ZLdeception}

\begin{thebibliography}{100}
\providecommand{\url}[1]{#1}
\csname url@samestyle\endcsname
\providecommand{\newblock}{\relax}
\providecommand{\bibinfo}[2]{#2}
\providecommand{\BIBentrySTDinterwordspacing}{\spaceskip=0pt\relax}
\providecommand{\BIBentryALTinterwordstretchfactor}{4}
\providecommand{\BIBentryALTinterwordspacing}{\spaceskip=\fontdimen2\font plus
\BIBentryALTinterwordstretchfactor\fontdimen3\font minus
  \fontdimen4\font\relax}
\providecommand{\BIBforeignlanguage}[2]{{%
\expandafter\ifx\csname l@#1\endcsname\relax
\typeout{** WARNING: IEEEtran.bst: No hyphenation pattern has been}%
\typeout{** loaded for the language `#1'. Using the pattern for}%
\typeout{** the default language instead.}%
\else
\language=\csname l@#1\endcsname
\fi
#2}}
\providecommand{\BIBdecl}{\relax}
\BIBdecl

\bibitem{mitnickArtDeceptionControlling2007}
K.~D. Mitnick, W.~L. Simon, and S.~Wozniak,
  \emph{\BIBforeignlanguage{English}{The {{Art}} of {{Deception}}:
  {{Controlling}} the {{Human Element}} of {{Security}}}}, 1st~ed.\hskip 1em
  plus 0.5em minus 0.4em\relax {Wiley}, Aug. 2007.

\bibitem{2019DataBreacha}
``\BIBforeignlanguage{en}{2019 {{Data Breach Investigations Report}}},''
  https://enterprise.verizon.com/resources/reports/2019-data-breach-investigations-report.pdf.

\bibitem{cimpanuCzechHospitalHit2020}
C.~Cimpanu, ``\BIBforeignlanguage{en}{Czech hospital hit by cyberattack while
  in the midst of a {{COVID}}-19 outbreak},''
  https://www.zdnet.com/article/czech-hospital-hit-by-cyber-attack-while-in-the-midst-of-a-covid-19-outbreak/,
  Mar. 2020.

\bibitem{palmerCoronavirusthemedPhishingAttacks2020}
D.~Palmer, ``\BIBforeignlanguage{en}{Coronavirus-themed phishing attacks and
  hacking campaigns are on the rise},''
  https://www.zdnet.com/article/coronavirus-themed-phishing-attacks-and-hacking-campaigns-are-on-the-rise/,
  Mar. 2020.

\bibitem{cimpanuThereNowCOVID192020}
C.~Cimpanu, ``\BIBforeignlanguage{en}{There's now {{COVID}}-19 malware that
  will wipe your {{PC}} and rewrite your {{MBR}}},''
  https://www.zdnet.com/article/theres-now-covid-19-malware-that-will-wipe-your-pc-and-rewrite-your-mbr/,
  Apr. 2020.

\bibitem{cybenkoCognitiveHackingBattle2002}
G.~Cybenko, A.~Giani, and P.~Thompson, ``Cognitive hacking: A battle for the
  mind,'' \emph{Computer}, vol.~35, no.~8, pp. 50--56, Aug. 2002.

\bibitem{stollCuckooEggTracking1989}
C.~Stoll, \emph{The Cuckoo's Egg: Tracking a Spy through the Maze of Computer
  Espionage}.\hskip 1em plus 0.5em minus 0.4em\relax {USA}: {Doubleday}, 1989.

\bibitem{lynniiiDefendingNewDomain2010a}
W.~J. Lynn~III, ``\BIBforeignlanguage{en}{Defending a {{New Domain}}: {{The
  Pentagon}}'s {{Cyberstrategy}}},'' \emph{\BIBforeignlanguage{en}{Foreign
  Affairs}}, vol.~89, no.~5, pp. 97--108, 2010.

\bibitem{lawrencepingreeEmergingTechnologyAnalysis2015}
{Lawrence Pingree}, ``Emerging {{Technology Analysis}}: {{Deception
  Techniques}} and {{Technologies Create Security Technology Business
  Opportunities}},'' {Gartner, Inc.}, Tech. Rep., Jul. 2015.

\bibitem{ferguson-walterGameTheoryAdaptive2019a}
K.~{Ferguson-Walter}, S.~Fugate, J.~Mauger, and M.~Major,
  ``\BIBforeignlanguage{en}{Game theory for adaptive defensive cyber
  deception},'' in \emph{\BIBforeignlanguage{en}{Proceedings of the 6th
  {{Annual Symposium}} on {{Hot Topics}} in the {{Science}} of {{Security}} -
  {{HotSoS}} '19}}.\hskip 1em plus 0.5em minus 0.4em\relax {Nashville,
  Tennessee}: {ACM Press}, 2019, pp. 1--8.

\bibitem{shadeMoonrakerStudyExperimental2020}
T.~Shade, A.~Rogers, K.~{Ferguson-Walter}, S.~B. Elsen, D.~Fayette, and
  K.~Heckman, ``\BIBforeignlanguage{en}{The {{Moonraker Study}}: {{An
  Experimental Evaluation}} of {{Host}}-{{Based Deception}}},'' in
  \emph{\BIBforeignlanguage{en}{Hawaii {{International Conference}} on {{System
  Sciences}}}}, 2020.

\bibitem{u.s.departmentofhomelandsecurityRecommendedPracticeImproving2016}
{U.S. Department of Homeland Security}, ``Recommended {{Practice}}: {{Improving
  Industrial Control System Cybersecurity}} with {{Defense}}-in-{{Depth
  Strategies}},'' Sep. 2016.

\bibitem{rossDevelopingCyberResilient2019}
R.~Ross, V.~Pillitteri, R.~Graubart, D.~Bodeau, and R.~McQuaid,
  ``\BIBforeignlanguage{en}{Developing cyber resilient systems: A systems
  security engineering approach},'' {National Institute of Standards and
  Technology}, {Gaithersburg, MD}, Tech. Rep. NIST SP 800-160v2, Nov. 2019.

\bibitem{spitznerHoneypotsTrackingHackers2002}
L.~Spitzner, \emph{\BIBforeignlanguage{English}{Honeypots: {{Tracking
  Hackers}}}}.\hskip 1em plus 0.5em minus 0.4em\relax {Boston}: {Addison-Wesley
  Professional}, Sep. 2002.

\bibitem{augustopaesdebarrosIDSRESProtocol2003}
{Augusto Paes de Barros}, ``{{IDS}}: {{RES}}: {{Protocol Anomaly Detection
  IDS}} - {{Honeypots}},'' https://seclists.org/focus-ids/2003/Feb/95, Feb.
  2003.

\bibitem{lancespitznerHoneytokensOtherHoneypot2003}
{Lance Spitzner}, ``Honeytokens: {{The Other Honeypot}},'' Security Focus
  information, Jul. 2003.

\bibitem{yuillHoneyfilesDeceptiveFiles2004}
J.~Yuill, M.~Zappe, D.~Denning, and F.~Feer, ``Honeyfiles: Deceptive files for
  intrusion detection,'' \emph{Proceedings from the Fifth Annual IEEE SMC
  Information Assurance Workshop}, 2004.

\bibitem{lancespitznerProblemsChallengesHoneypots2004}
{Lance Spitzner}, ``Problems and {{Challenges}} with {{Honeypots}},'' Security
  Focus information, Jan. 2004.

\bibitem{NITRDCSIAIWG2010}
``{{NITRD CSIA IWG Cybersecurity Game}}-{{Change Research}} \& {{Development
  Recommendations}},'' 2010.

\bibitem{pawlickGameTheoreticTaxonomySurvey2019}
J.~Pawlick, E.~Colbert, and Q.~Zhu, ``\BIBforeignlanguage{en}{A
  {{Game}}-{{Theoretic Taxonomy}} and {{Survey}} of {{Defensive Deception}} for
  {{Cybersecurity}} and {{Privacy}}},''
  \emph{\BIBforeignlanguage{en}{arXiv:1712.05441 [cs]}}, May 2019.

\bibitem{fredcohenMovingTargetDefenses2010}
{Fred Cohen}, ``Moving target defenses with and without cover deception,''
  http://all.net/Analyst/2010-10.pdf, 2010.

\bibitem{seifertHoneyCLowInteractionClient2007}
C.~Seifert, I.~Welch, and P.~Komisarczuk, ``\BIBforeignlanguage{en}{{{HoneyC}}
  - {{The Low}}-{{Interaction Client Honeypot}}},'' in
  \emph{\BIBforeignlanguage{en}{Proceedings of the 2007 {{NZCSRCS}}}}, Jan.
  2007.

\bibitem{nazarioPhoneyCVirtualClient2009}
J.~Nazario, ``{{PhoneyC}}: {{A Virtual Client Honeypot}},'' in \emph{{{USENIX
  Workshop}} on {{Large}}-Scale {{Exploits}} and {{Emergent Threats}}}, 2009.

\bibitem{christianseifertTaxonomyHoneypots2006}
{Christian Seifert}, {Ian Welch}, and {Peter Komisarczuk}, ``Taxonomy of
  {{Honeypots}},'' Tech. Rep. CS-TR-06-12, Jun. 2006.

\bibitem{nawrockiSurveyHoneypotSoftware2016}
M.~Nawrocki, M.~W{\"a}hlisch, T.~C. Schmidt, C.~Keil, and J.~Sch{\"o}nfelder,
  ``A {{Survey}} on {{Honeypot Software}} and {{Data Analysis}},''
  \emph{arXiv:1608.06249 [cs]}, Aug. 2016.

\bibitem{scottsmithCatchingFliesGuide2016}
{Scott Smith}, ``Catching {{Flies}}: {{A}} guide to the various flavors of
  honeypots,'' {SANS}, Tech. Rep., 2016.

\bibitem{hanDeceptionTechniquesComputer2018}
X.~Han, N.~Kheir, and D.~Balzarotti, ``Deception {{Techniques}} in {{Computer
  Security}}: {{A Research Perspective}},'' \emph{ACM Computing Surveys},
  vol.~51, no.~4, Jul. 2018.

\bibitem{efendiSurveyDeceptionTechniques2019}
A.~M. Efendi, Z.~Ibrahim, M.~A. Zawawi, F.~Abdul~Rahim, N.~M. Pahri, and
  A.~Ismail, ``A {{Survey}} on {{Deception Techniques}} for {{Securing Web
  Application}},'' in \emph{2019 {{IEEE}} 5th {{Intl Conference}} on {{Big Data
  Security}} on {{Cloud}} ({{BigDataSecurity}}), {{IEEE Intl Conference}} on
  {{High Performance}} and {{Smart Computing}}, ({{HPSC}}) and {{IEEE Intl
  Conference}} on {{Intelligent Data}} and {{Security}} ({{IDS}})}, May 2019,
  pp. 328--331.

\bibitem{okhraviSurveyCyberMoving2013}
H.~Okhravi, M.~A. Rabe, T.~J. Mayberry, W.~G. Leonard, T.~R. Hobson,
  D.~Bigelow, and W.~W. Streilein, ``\BIBforeignlanguage{en}{Survey of {{Cyber
  Moving Target Techniques}}},'' {Lincoln Lab, MIT}, Tech. Rep. 1166, Sep.
  2013.

\bibitem{caiMovingTargetDefense2016}
G.-l. Cai, B.-s. Wang, W.~Hu, and T.-z. Wang, ``\BIBforeignlanguage{en}{Moving
  target defense: State of the art and characteristics},''
  \emph{\BIBforeignlanguage{en}{Frontiers of Information Technology \&
  Electronic Engineering}}, vol.~17, no.~11, pp. 1122--1153, Nov. 2016.

\bibitem{b.c.wardSurveyCyberMoving2018}
{B.C. Ward}, {S.R. Gomez}, {R.W. Skowyra}, {D. Bigelow}, {J.N. Martin}, {J.W.
  Landry}, and {H. Okhravi}, ``Survey of {{Cyber Moving Targets}}: 2nd
  {{Edition}},'' {Lincoln Lab, MIT}, Tech. Rep. 1228, Jan. 2018.

\bibitem{leiMovingTargetDefense2018}
C.~Lei, H.-Q. Zhang, J.-L. Tan, Y.-C. Zhang, and X.-H. Liu,
  ``\BIBforeignlanguage{en}{Moving {{Target Defense Techniques}}: {{A
  Survey}}},'' \emph{\BIBforeignlanguage{en}{Security and Communication
  Networks}}, vol. 2018, 2018.

\bibitem{senguptaSurveyMovingTarget2020}
S.~Sengupta, A.~Chowdhary, A.~Sabur, A.~Alshamrani, D.~Huang, and
  S.~Kambhampati, ``\BIBforeignlanguage{en}{A {{Survey}} of {{Moving Target
  Defenses}} for {{Network Security}}},''
  \emph{\BIBforeignlanguage{en}{arXiv:1905.00964 [cs]}}, Mar. 2020.

\bibitem{fraunholzDemystifyingDeceptionTechnology2018a}
D.~Fraunholz, S.~D. Anton, C.~Lipps, D.~Reti, D.~Krohmer, F.~Pohl, M.~Tammen,
  and H.~D. Schotten, ``\BIBforeignlanguage{en}{Demystifying {{Deception
  Technology}}:{{A Survey}}},'' \emph{\BIBforeignlanguage{en}{arXiv:1804.06196
  [cs]}}, Apr. 2018.

\bibitem{hutchinsIntelligenceDrivenComputerNetwork2011}
E.~M. Hutchins, M.~J. Cloppert, and R.~M. Amin,
  ``\BIBforeignlanguage{en}{Intelligence-{{Driven Computer Network Defense
  Informed}} by {{Analysis}} of {{Adversary Campaigns}} and {{Intrusion Kill
  Chains}}},'' \emph{\BIBforeignlanguage{en}{Leading Issues in Information
  Warfare \& Security Research}}, vol.~1, no.~1, 2011.

\bibitem{reidyCombatingInsiderThreat2013}
P.~Reidy, ``\BIBforeignlanguage{en}{Combating the {{Insider Threat}} at the
  {{FBI}}: {{Real World Lessons Learned}}},'' in
  \emph{\BIBforeignlanguage{en}{Black {{Hat}}}}, {USA}, 2013.

\bibitem{gioraengelDeconstructingCyberKill2014}
{Giora Engel}, ``\BIBforeignlanguage{en}{Deconstructing {{The Cyber Kill
  Chain}}},''
  https://www.darkreading.com/attacks-breaches/deconstructing-the-cyber-kill-chain/a/d-id/1317542,
  Nov. 2014.

\bibitem{marclaliberteTwistCyberKill2016}
{Marc Laliberte}, ``\BIBforeignlanguage{en}{A {{Twist On The Cyber Kill
  Chain}}: {{Defending Against A JavaScript Malware Attack}}},''
  https://www.darkreading.com/attacks-breaches/a-twist-on-the-cyber-kill-chain-defending-against-a-javascript-malware-attack/a/d-id/1326952,
  Sep. 2016.

\bibitem{blaked.bryantNovelKillchainFramework2017}
{Blake D. Bryant} and {Hossein Saiedian}, ``\BIBforeignlanguage{en}{A novel
  kill-chain framework for remote security log analysis with {{SIEM}}
  software},'' \emph{\BIBforeignlanguage{en}{Computers \& Security}}, vol.~67,
  pp. 198--210, Jun. 2017.

\bibitem{maloneUsingExpandedCyber2016}
S.~T. Malone, ``\BIBforeignlanguage{en}{Using an {{Expanded Cyber Kill Chain
  Model}} to increase attack resiliency},'' in
  \emph{\BIBforeignlanguage{en}{Black {{Hat}}}}, {USA}, 2016.

\bibitem{polsUnifiedKillChain2017}
P.~Pols, ``The {{Unified Kill Chain}},'' Ph.D. dissertation, Cyber Security
  Academy, Dec. 2017.

\bibitem{kellysheridanDefenseEvasionDominated2020}
{Kelly Sheridan}, ``\BIBforeignlanguage{en}{Defense {{Evasion Dominated}} 2019
  {{Attack Tactics}}},''
  https://www.darkreading.com/vulnerabilities---threats/defense-evasion-dominated-2019-attack-tactics/d/d-id/1337457,
  Mar. 2020.

\bibitem{aminCyberSecurityWater2013}
S.~Amin, X.~Litrico, S.~Sastry, and A.~M. Bayen, ``Cyber {{Security}} of
  {{Water SCADA Systems}}\textemdash{{Part I}}: {{Analysis}} and
  {{Experimentation}} of {{Stealthy Deception Attacks}},'' \emph{IEEE
  Transactions on Control Systems Technology}, vol.~21, no.~5, pp. 1963--1970,
  Sep. 2013.

\bibitem{houRobustPartialNodesBasedState2020}
N.~Hou, Z.~Wang, D.~W.~C. Ho, and H.~Dong, ``Robust
  {{Partial}}-{{Nodes}}-{{Based State Estimation}} for {{Complex Networks Under
  Deception Attacks}},'' \emph{IEEE Transactions on Cybernetics}, vol.~50,
  no.~6, pp. 2793--2802, Jun. 2020.

\bibitem{zhangOptimalStealthyDeception2020}
Q.~Zhang, K.~Liu, Y.~Xia, and A.~Ma, ``Optimal {{Stealthy Deception Attack
  Against Cyber}}-{{Physical Systems}},'' \emph{IEEE Transactions on
  Cybernetics}, vol.~50, no.~9, pp. 3963--3972, Sep. 2020.

\bibitem{meira-goesSynthesisSupervisorsRobust2021}
R.~{Meira-Goes}, S.~Lafortune, and H.~Marchand, ``Synthesis of {{Supervisors
  Robust Against Sensor Deception Attacks}},'' \emph{IEEE Transactions on
  Automatic Control}, 2021.

\bibitem{kitchenhamProceduresPerformingSystematic2004}
B.~Kitchenham, ``\BIBforeignlanguage{en}{Procedures for {{Performing Systematic
  Reviews}}},'' \emph{\BIBforeignlanguage{en}{Keele, UK, Keele University}},
  p.~33, 2004.

\bibitem{lancespitznerValueHoneypotsPart2001}
{Lance Spitzner}, ``The {{Value}} of {{Honeypots}}, {{Part One}}:
  {{Definitions}} and {{Values}} of {{Honeypots}},'' Security Focus
  information, Oct. 2001.

\bibitem{scottbergInternetHoneypotsProtection2002}
B.~Scottberg, W.~Yurcik, and D.~Doss, ``\BIBforeignlanguage{en}{Internet
  honeypots: Protection or entrapment?}'' in
  \emph{\BIBforeignlanguage{en}{{{IEEE International Symposium}} on
  {{Technology}} and {{Society}} ({{ISTAS}}'02)}}.\hskip 1em plus 0.5em minus
  0.4em\relax {Raleigh, NC, USA}: {IEEE}, 2002, pp. 387--391.

\bibitem{fredcohenDeceptionToolKit1998}
{Fred Cohen}, ``Deception {{ToolKit}},'' http://www.all.net/dtk/, Mar. 1998.

\bibitem{tomlistonLaBreaStickyHoneypot2001}
{Tom Liston}, ``{{LaBrea}}: "{{Sticky}}" {{Honeypot}} and {{IDS}},''
  http://labrea.sourceforge.net/labrea-info.html, 2001.

\bibitem{leslieshingImprovedTarpitNetwork2016}
{Leslie Shing}, ``An improved tarpit for network deception,'' Ph.D.
  dissertation, Naval Postgraduate School, 2016.

\bibitem{provosVirtualHoneypotFramework2004}
N.~Provos, ``A virtual honeypot framework,'' in \emph{Proceedings of the 13th
  Conference on {{USENIX Security Symposium}} - {{Volume}} 13}, ser.
  {{SSYM}}'04.\hskip 1em plus 0.5em minus 0.4em\relax {San Diego, CA}: {USENIX
  Association}, Aug. 2004, p.~1.

\bibitem{anagnostakisDetectingTargetedAttacks2005}
K.~G. Anagnostakis, S.~Sidiroglou, P.~Akritidis, K.~Xinidis, E.~Markatos, and
  A.~D. Keromytis, ``\BIBforeignlanguage{en}{Detecting {{Targeted Attacks Using
  Shadow Honeypots}}},'' in \emph{\BIBforeignlanguage{en}{{{USENIX
  Security}}}}, Jan. 2005, p.~16.

\bibitem{bordersOpenFireUsingDeception2007}
K.~Borders, L.~Falk, and A.~Prakash, ``{{OpenFire}}: {{Using}} deception to
  reduce network attacks,'' in \emph{2007 {{Third International Conference}} on
  {{Security}} and {{Privacy}} in {{Communications Networks}} and the
  {{Workshops}} - {{SecureComm}} 2007}, Sep. 2007, pp. 224--233.

\bibitem{biedermannFastDynamicExtracted2012}
S.~Biedermann, M.~Mink, and S.~Katzenbeisser, ``Fast dynamic extracted
  honeypots in cloud computing,'' in \emph{Proceedings of the 2012 {{ACM
  Workshop}} on {{Cloud}} Computing Security Workshop}.\hskip 1em plus 0.5em
  minus 0.4em\relax {Raleigh, North Carolina, USA}: {Association for Computing
  Machinery}, Oct. 2012, pp. 13--18.

\bibitem{uriasGatheringThreatIntelligence2016}
V.~E. Urias, W.~M.~S. Stout, and H.~W. Lin, ``\BIBforeignlanguage{en}{Gathering
  threat intelligence through computer network deception},'' in
  \emph{\BIBforeignlanguage{en}{2016 {{IEEE Symposium}} on {{Technologies}} for
  {{Homeland Security}} ({{HST}})}}.\hskip 1em plus 0.5em minus 0.4em\relax
  {Waltham, MA, USA}: {IEEE}, May 2016, pp. 1--6.

\bibitem{dissoPlausibleSolutionSCADA2013}
J.~P. Disso, K.~Jones, and S.~Bailey, ``A {{Plausible Solution}} to {{SCADA
  Security Honeypot Systems}},'' in \emph{2013 {{Eighth International
  Conference}} on {{Broadband}} and {{Wireless Computing}}, {{Communication}}
  and {{Applications}}}, Oct. 2013, pp. 443--448.

\bibitem{winnConstructingCosteffectiveTargetable2015}
M.~Winn, M.~Rice, S.~Dunlap, J.~Lopez, and B.~Mullins,
  ``\BIBforeignlanguage{en}{Constructing cost-effective and targetable
  industrial control system honeypots for production networks},''
  \emph{\BIBforeignlanguage{en}{International Journal of Critical
  Infrastructure Protection}}, vol.~10, pp. 47--58, Sep. 2015.

\bibitem{lukasristConpot2015}
{Lukas Rist}, {Johnny Vestergaard}, {Daniel Haslinger}, {Andrea Pasquale}, and
  {John Smith}, ``Conpot,'' http://conpot.org/, 2015.

\bibitem{zhaoResearchHighInteractive2017}
C.~Zhao and S.~Qin, ``\BIBforeignlanguage{en}{A research for high interactive
  honepot based on industrial service},'' in \emph{\BIBforeignlanguage{en}{2017
  3rd {{IEEE International Conference}} on {{Computer}} and {{Communications}}
  ({{ICCC}})}}.\hskip 1em plus 0.5em minus 0.4em\relax {Chengdu}: {IEEE}, Dec.
  2017, pp. 2935--2939.

\bibitem{kumanExperimentUsingIMUNES2017}
S.~Kuman, S.~Gro{\v s}, and M.~Mikuc, ``An experiment in using {{IMUNES}} and
  {{Conpot}} to emulate honeypot control networks,'' in \emph{2017 40th
  {{International Convention}} on {{Information}} and {{Communication
  Technology}}, {{Electronics}} and {{Microelectronics}} ({{MIPRO}})}, May
  2017, pp. 1262--1268.

\bibitem{josephcoreyAdvancedHoneyPot2003}
{Joseph Corey}, ``Advanced {{Honey Pot Identification}} and {{Exploitation}},''
  {Phrack Inc.}, Tech. Rep., 2003.

\bibitem{fuRecognizingVirtualHoneypots2006}
X.~Fu, W.~Yu, D.~Cheng, X.~Tan, K.~Streff, and S.~Graham, ``On {{Recognizing
  Virtual Honeypots}} and {{Countermeasures}},'' in \emph{2006 2nd {{IEEE
  International Symposium}} on {{Dependable}}, {{Autonomic}} and {{Secure
  Computing}}}, Sep. 2006, pp. 211--218.

\bibitem{altUncoveringNetworkTarpits2014}
L.~Alt, R.~Beverly, and A.~Dainotti, ``\BIBforeignlanguage{en}{Uncovering
  network tarpits with degreaser},'' in
  \emph{\BIBforeignlanguage{en}{Proceedings of the 30th {{Annual Computer
  Security Applications Conference}} on - {{ACSAC}} '14}}.\hskip 1em plus 0.5em
  minus 0.4em\relax {New Orleans, Louisiana}: {ACM Press}, 2014, pp. 156--165.

\bibitem{vetterlBitterHarvestSystematically2018}
A.~Vetterl and R.~Clayton, ``Bitter harvest: Systematically fingerprinting low-
  and medium-interaction honeypots at internet scale,'' in \emph{Proceedings of
  the 12th {{USENIX Conference}} on {{Offensive Technologies}}}, ser.
  {{WOOT}}'18, {Baltimore, MD, USA}, Aug. 2018.

\bibitem{vetterlHoneypotsAgeUniversal2019}
A.~M. Vetterl, ``\BIBforeignlanguage{en}{Honeypots in the age of universal
  attacks and the {{Internet}} of {{Things}}},'' Ph.D. dissertation, University
  of Cambridge, Nov. 2019.

\bibitem{rrushiHoneypotEvaderActivityguided2019}
J.~L. Rrushi, ``\BIBforeignlanguage{en}{Honeypot {{Evader}}:
  {{Activity}}-guided {{Propagation}} versus {{Counter}}-evasion via {{Decoy OS
  Activity}}},'' in \emph{\BIBforeignlanguage{en}{Proceedings of the 14th
  {{IEEE International Conference}} on {{Malicious}} and {{Unwanted
  Software}}}}, {Nantucket, Massachusetts, USA}, Oct. 2019, p.~11.

\bibitem{wangHoneypotDetectionAdvanced2010}
P.~Wang, L.~Wu, R.~Cunningham, and C.~C. Zou,
  ``\BIBforeignlanguage{en}{Honeypot detection in advanced botnet attacks},''
  \emph{\BIBforeignlanguage{en}{International Journal of Information and
  Computer Security}}, vol.~4, no.~1, p.~30, 2010.

\bibitem{krawetzAntihoneypotTechnology2004}
N.~Krawetz, ``Anti-honeypot technology,'' \emph{IEEE Security Privacy}, vol.~2,
  no.~1, pp. 76--79, Jan. 2004.

\bibitem{hayatleDempsterShaferEvidenceCombining2012}
O.~Hayatle, A.~Youssef, and H.~Otrok,
  ``\BIBforeignlanguage{en}{Dempster-{{Shafer Evidence Combining}} for
  ({{Anti}})-{{Honeypot Technologies}}},''
  \emph{\BIBforeignlanguage{en}{Information Security Journal: A Global
  Perspective}}, vol.~21, no.~6, pp. 306--316, Jan. 2012.

\bibitem{sentzCombinationEvidenceDempsterShafer2002}
K.~Sentz, S.~Ferson, and K.~Sentz, ``Combination of evidence in
  {{Dempster}}-{{Shafer}} theory,'' {Sandia National Laboratories},
  {Albuquerque, New Mexico.}, Tech. Rep., 2002.

\bibitem{huangAutomaticIdentificationHoneypot2019}
C.~Huang, J.~Han, X.~Zhang, and J.~Liu, ``\BIBforeignlanguage{en}{Automatic
  {{Identification}} of {{Honeypot Server Using Machine Learning
  Techniques}}},'' \emph{\BIBforeignlanguage{en}{Security and Communication
  Networks}}, vol. 2019, pp. 1--8, Sep. 2019.

\bibitem{eugeneh.spaffordMorePassiveDefense2011}
{Eugene H. Spafford}, ``\BIBforeignlanguage{en}{More than passive defense},''
  https://www.cerias.purdue.edu/, Jul. 2011.

\bibitem{bourkeBreachDetectionScale2018}
D.~Bourke and D.~Grzelak, ``\BIBforeignlanguage{en}{Breach {{Detection}} at
  {{Scale}} with {{AWS Honey Tokens}}},'' in
  \emph{\BIBforeignlanguage{en}{Black {{Hat Asia}}}}, 2018.

\bibitem{stringhiniDetectingSpammersSocial2010}
G.~Stringhini, C.~Kruegel, and G.~Vigna, ``\BIBforeignlanguage{en}{Detecting
  {{Spammers}} on {{Social Networks}}},'' in
  \emph{\BIBforeignlanguage{en}{Proceedings of the 26th Annual Computer
  Security Applications Conference}}, 2010.

\bibitem{virvilisChangingGameArt2014}
N.~Virvilis, B.~Vanautgaerden, and O.~S. Serrano,
  ``\BIBforeignlanguage{en}{Changing the game: {{The}} art of deceiving
  sophisticated attackers},'' in \emph{\BIBforeignlanguage{en}{2014 6th
  {{International Conference On Cyber Conflict}} ({{CyCon}} 2014)}}.\hskip 1em
  plus 0.5em minus 0.4em\relax {Tallinn, Estonia}: {IEEE}, Jun. 2014, pp.
  87--97.

\bibitem{whiteCreatingPersonallyIdentifiable2010}
J.~White, ``\BIBforeignlanguage{en}{Creating {{Personally Identifiable
  Honeytokens}}},'' in \emph{\BIBforeignlanguage{en}{Innovations and
  {{Advances}} in {{Computer Sciences}} and {{Engineering}}}}, T.~Sobh,
  Ed.\hskip 1em plus 0.5em minus 0.4em\relax {Dordrecht}: {Springer
  Netherlands}, 2010, pp. 227--232.

\bibitem{gavrilisFlashCrowdDetection2007}
D.~Gavrilis, I.~Chatzis, and E.~Dermatas, ``Flash {{Crowd Detection Using Decoy
  Hyperlinks}},'' in \emph{2007 {{IEEE International Conference}} on
  {{Networking}}, {{Sensing}} and {{Control}}}, Apr. 2007, pp. 466--470.

\bibitem{douglasbrewerLinkObfuscationService2010}
{Douglas Brewer}, {Kang Li}, {Laksmish Ramaswamy}, and {Calton Pu}, ``A {{Link
  Obfuscation Service}} to {{Detect Webbots}},'' Jul. 2010.

\bibitem{murphyApplicationDeceptionCyberspace2010}
S.~B. Murphy, J.~T. McDonald, and R.~F. Mills, ``\BIBforeignlanguage{en}{An
  {{Application}} of {{Deception}} in {{Cyberspace}}: {{Operating System
  Obfuscation}}},'' in \emph{\BIBforeignlanguage{en}{International
  {{Conference}} on {{Information Warfare}} and {{Security}}}}, Apr. 2010.

\bibitem{trassareTechniquePresentingDeceptive2013}
S.~T. Trassare, ``\BIBforeignlanguage{en}{A technique for presenting a
  deceptive dynamic network topology},'' Ph.D. dissertation, Naval Postgraduate
  School, 2013.

\bibitem{julianDelayingtypeResponsesUse2002}
D.~P. Julian, ``\BIBforeignlanguage{en}{Delaying-type responses for use by
  software decoys},'' Ph.D. dissertation, 2002.

\bibitem{neilc.roweDeceptionDefenseComputer2007}
{Neil C. Rowe}, ``Deception in defense of computer systems from cyber-attack,''
  in \emph{Cyber {{War}} and {{Cyber Terrorism}}}, {A. Colarik} and {L.
  Janczewski}, Eds., 2007.

\bibitem{katsinisSecurityMechanismWeb2012}
C.~Katsinis and B.~Kumar, ``\BIBforeignlanguage{en}{A {{Security Mechanism}}
  for {{Web Servers Based}} on {{Deception}}},'' in
  \emph{\BIBforeignlanguage{en}{Proceedings on the {{International Conference}}
  on {{Internet Computing}} ({{ICOMP}})}}, 2012, p.~6.

\bibitem{katsinisFrameworkIntrusionDeception2013}
------, ``\BIBforeignlanguage{en}{A {{Framework}} for {{Intrusion Deception}}
  on {{Web Servers}}},'' in \emph{\BIBforeignlanguage{en}{International
  {{Conference}} on {{Internet Computing}}({{ICOMP}})}}, 2013, p.~7.

\bibitem{hanEvaluationDeceptionBasedWeb2017}
X.~Han, N.~Kheir, and D.~Balzarotti, ``\BIBforeignlanguage{en}{Evaluation of
  {{Deception}}-{{Based Web Attacks Detection}}},'' in
  \emph{\BIBforeignlanguage{en}{Proceedings of the 2017 {{Workshop}} on
  {{Moving Target Defense}} - {{MTD}} '17}}.\hskip 1em plus 0.5em minus
  0.4em\relax {Dallas, Texas, USA}: {ACM Press}, 2017, pp. 65--73.

\bibitem{roweDefendingCyberspaceFake2007}
N.~C. Rowe, E.~J. Custy, and B.~T. Duong, ``\BIBforeignlanguage{en}{Defending
  {{Cyberspace}} with {{Fake Honeypots}}},''
  \emph{\BIBforeignlanguage{en}{Journal of Computers}}, vol.~2, no.~2, pp.
  25--36, Apr. 2007.

\bibitem{craneBoobyTrappingSoftware2013}
S.~Crane, P.~Larsen, S.~Brunthaler, and M.~Franz,
  ``\BIBforeignlanguage{en}{Booby trapping software},'' in
  \emph{\BIBforeignlanguage{en}{Proceedings of the 2013 Workshop on {{New}}
  Security Paradigms Workshop - {{NSPW}} '13}}.\hskip 1em plus 0.5em minus
  0.4em\relax {Banff, Alberta, Canada}: {ACM Press}, 2013, pp. 95--106.

\bibitem{araujoPatchesHoneyPatchesLightweight2014}
F.~Araujo, K.~W. Hamlen, S.~Biedermann, and S.~Katzenbeisser, ``From
  {{Patches}} to {{Honey}}-{{Patches}}: {{Lightweight Attacker Misdirection}},
  {{Deception}}, and {{Disinformation}},'' in \emph{Proceedings of the 2014
  {{ACM SIGSAC Conference}} on {{Computer}} and {{Communications Security}}},
  ser. {{CCS}} '14.\hskip 1em plus 0.5em minus 0.4em\relax {Scottsdale,
  Arizona, USA}: {Association for Computing Machinery}, Nov. 2014, pp.
  942--953.

\bibitem{bowenSystemGeneratingInjecting2012}
B.~M. Bowen, V.~P. Kemerlis, P.~Prabhu, A.~D. Keromytis, and S.~J. Stolfo,
  ``\BIBforeignlanguage{en}{A system for generating and injecting
  indistinguishable network decoys},'' \emph{\BIBforeignlanguage{en}{Journal of
  Computer Security}}, vol.~20, no. 2-3, pp. 199--221, Jun. 2012.

\bibitem{bowenAutomatingInjectionBelievable2010}
------, ``\BIBforeignlanguage{en}{Automating the injection of believable decoys
  to detect snooping},'' in \emph{\BIBforeignlanguage{en}{Proceedings of the
  Third {{ACM}} Conference on {{Wireless}} Network Security - {{WiSec}}
  '10}}.\hskip 1em plus 0.5em minus 0.4em\relax {Hoboken, New Jersey, USA}:
  {ACM Press}, 2010, p.~81.

\bibitem{juelsHoneyEncryptionSecurity2014}
A.~Juels and T.~Ristenpart, ``Honey {{Encryption}}: {{Security Beyond}} the
  {{Brute}}-{{Force Bound}},'' Tech. Rep. 155, 2014.

\bibitem{abiodunReinforcingSecurityInstant2020}
E.~O. Abiodun, A.~Jantan, O.~I. Abiodun, and H.~Arshad,
  ``\BIBforeignlanguage{en}{Reinforcing the {{Security}} of {{Instant Messaging
  Systems Using}} an {{Enhanced Honey Encryption Scheme}}: {{The Case}} of
  {{WhatsApp}}},'' \emph{\BIBforeignlanguage{en}{Wireless Personal
  Communications}}, Jan. 2020.

\bibitem{omolaraDeceptionModelRobust2019a}
A.~E. Omolara, A.~Jantan, O.~Isaac~Abiodun, K.~Victoria~Dada, H.~Arshad, and
  E.~Emmanuel, ``A {{Deception Model Robust}} to {{Eavesdropping Over
  Communication}} for {{Social Network Systems}},'' \emph{IEEE Access}, vol.~7,
  pp. 100\,881--100\,898, 2019.

\bibitem{juelsBodyguardLiesUse2014}
A.~Juels, ``\BIBforeignlanguage{en}{A bodyguard of lies: The use of honey
  objects in information security},'' in
  \emph{\BIBforeignlanguage{en}{Proceedings of the 19th {{ACM}} Symposium on
  {{Access}} Control Models and Technologies - {{SACMAT}} '14}}.\hskip 1em plus
  0.5em minus 0.4em\relax {London, Ontario, Canada}: {ACM Press}, 2014, pp.
  1--4.

\bibitem{kaghazgaranInsiderThreatDetection2015}
P.~Kaghazgaran and H.~Takabi, ``Toward an {{Insider Threat Detection Framework
  Using Honey Permissions}},'' \emph{Journal of Internet Services and
  Information Securuirty (JISIS)}, 2015.

\bibitem{juelsHoneywordsMakingPasswordcracking2013}
A.~Juels and R.~L. Rivest, ``\BIBforeignlanguage{en}{Honeywords: Making
  password-cracking detectable},'' in \emph{\BIBforeignlanguage{en}{Proceedings
  of the 2013 {{ACM SIGSAC}} Conference on {{Computer}} \& Communications
  Security - {{CCS}} '13}}.\hskip 1em plus 0.5em minus 0.4em\relax {Berlin,
  Germany}: {ACM Press}, 2013, pp. 145--160.

\bibitem{almeshekahErsatzPasswordsEndingPassword2015}
M.~H. Almeshekah, C.~N. Gutierrez, M.~J. Atallah, and E.~H. Spafford,
  ``\BIBforeignlanguage{en}{{{ErsatzPasswords}}: {{Ending Password Cracking}}
  and {{Detecting Password Leakage}}},'' in
  \emph{\BIBforeignlanguage{en}{Proceedings of the 31st {{Annual Computer
  Security Applications Conference}} on - {{ACSAC}} 2015}}.\hskip 1em plus
  0.5em minus 0.4em\relax {Los Angeles, CA, USA}: {ACM Press}, 2015, pp.
  311--320.

\bibitem{changRetrospectiveLookForward2017}
C.-H. Chang, Y.~Zheng, and L.~Zhang, ``\BIBforeignlanguage{en}{A
  {{Retrospective}} and a {{Look Forward}}: {{Fifteen Years}} of {{Physical
  Unclonable Function Advancement}}},'' \emph{\BIBforeignlanguage{en}{IEEE
  Circuits and Systems Magazine}}, vol.~17, no.~3, pp. 32--62, 2017.

\bibitem{PCIHardwareSecurity2009}
``{{PCI Hardware Security Module Security Requirements}}, {{Version}} 1.0,''
  {PCI Security Standards Council}, Tech. Rep., Apr. 2009.

\bibitem{antanascenysImplementationHoneytokenModule2005}
{Antanas \v{C}enys}, {Darius Rainys}, {Lukas Radvilavi\v{c}ius}, and {Nikolaj
  Goranin}, ``\BIBforeignlanguage{en}{Implementation of {{Honeytoken Module}}
  in {{DBMS Oracle}} 9ir2 {{Enterprise Edition}} for {{Internal Malicious
  Activity Detection}}},'' in \emph{\BIBforeignlanguage{en}{{{IEEE Computer
  Society}}'s {{TC}} on {{Security}} and {{Privacy}}}}, 2005.

\bibitem{bercovitchHoneyGenAutomatedHoneytokens2011}
M.~Bercovitch, M.~Renford, L.~Hasson, A.~Shabtai, L.~Rokach, and {Yuval
  Elovici}, ``{{HoneyGen}}: {{An}} automated honeytokens generator,'' in
  \emph{Proceedings of 2011 {{IEEE International Conference}} on
  {{Intelligence}} and {{Security Informatics}}}, Jul. 2011, pp. 131--136.

\bibitem{padayacheeAspectisingHoneytokensContain2014}
K.~Padayachee, ``\BIBforeignlanguage{en}{Aspectising honeytokens to contain the
  insider threat},'' \emph{\BIBforeignlanguage{en}{IET Information Security}},
  vol.~9, no.~4, pp. 240--247, Dec. 2014.

\bibitem{bowenBaitingAttackersUsing2009}
B.~M. Bowen, S.~Hershkop, A.~D. Keromytis, and S.~J. Stolfo,
  ``\BIBforeignlanguage{en}{Baiting {{Inside Attackers Using Decoy
  Documents}}},'' in \emph{\BIBforeignlanguage{en}{International {{Conference}}
  on {{Security}} and {{Privacy}} in {{Communication Systems}}}}, 2009.

\bibitem{vorisBaitSnitchDefending2013}
J.~Voris, J.~Jermyn, A.~D. Keromytis, and S.~J. Stolfo,
  ``\BIBforeignlanguage{en}{Bait and {{Snitch}}: {{Defending Computer Systems}}
  with {{Decoys}}},'' in \emph{\BIBforeignlanguage{en}{Proceedings of the
  {{Cyber Infrastructure Protection Conference}}, {{Strategic Studies
  Institute}}}}, 2013, p.~25.

\bibitem{vorisFoxTrapThwarting2015}
J.~Voris, J.~Jermyn, N.~Boggs, and S.~Stolfo, ``\BIBforeignlanguage{en}{Fox in
  the trap: Thwarting masqueraders via automated decoy document deployment},''
  in \emph{\BIBforeignlanguage{en}{Proceedings of the {{Eighth European
  Workshop}} on {{System Security}} - {{EuroSec}} '15}}.\hskip 1em plus 0.5em
  minus 0.4em\relax {Bordeaux, France}: {ACM Press}, 2015, pp. 1--7.

\bibitem{benwhithamAutomatingGenerationEnticing2017}
{Ben Whitham}, ``\BIBforeignlanguage{en}{Automating the {{Generation}} of
  {{Enticing Text Content}} for {{High}}-{{Interaction Honeyfiles}}}.''\hskip
  1em plus 0.5em minus 0.4em\relax {IEEE computer society}, 2017.

\bibitem{leePhantomFSFileBasedDeception2020}
J.~Lee, J.~Choi, G.~Lee, S.-W. Shim, and T.~Kim, ``{{PhantomFS}}:
  {{File}}-{{Based Deception Technology}} for {{Thwarting Malicious Users}},''
  \emph{IEEE Access}, vol.~8, pp. 32\,203--32\,214, 2020.

\bibitem{tzuArtWar2007}
S.~Tzu, \emph{\BIBforeignlanguage{English}{The {{Art Of War}}}}, first thus
  edition~ed.\hskip 1em plus 0.5em minus 0.4em\relax {Filiquarian}, Nov. 2007.

\bibitem{okhraviFindingFocusBlur2014}
H.~Okhravi, T.~Hobson, D.~Bigelow, and W.~Streilein,
  ``\BIBforeignlanguage{en}{Finding {{Focus}} in the {{Blur}} of
  {{Moving}}-{{Target Techniques}}},'' \emph{\BIBforeignlanguage{en}{IEEE
  Security \& Privacy}}, vol.~12, no.~2, pp. 16--26, Mar. 2014.

\bibitem{kewleyDynamicApproachesThwart2001}
D.~Kewley, R.~Fink, J.~Lowry, and M.~Dean, ``Dynamic approaches to thwart
  adversary intelligence gathering,'' in \emph{Proceedings {{DARPA Information
  Survivability Conference}} and {{Exposition II}}. {{DISCEX}}'01}, vol.~1,
  Jun. 2001, pp. 176--185 vol.1.

\bibitem{antonatosDefendingHitlistWorms2005}
S.~Antonatos, P.~Akritidis, E.~P. Markatos, and K.~G. Anagnostakis, ``Defending
  against hitlist worms using network address space randomization,'' in
  \emph{Proceedings of the 2005 {{ACM}} Workshop on {{Rapid}} Malcode}, ser.
  {{WORM}} '05.\hskip 1em plus 0.5em minus 0.4em\relax {Fairfax, VA, USA}:
  {Association for Computing Machinery}, Nov. 2005, pp. 30--40.

\bibitem{jafarianOpenflowRandomHost2012}
J.~H. Jafarian, E.~{Al-Shaer}, and Q.~Duan, ``\BIBforeignlanguage{en}{Openflow
  random host mutation: Transparent moving target defense using software
  defined networking},'' in \emph{\BIBforeignlanguage{en}{Proceedings of the
  First Workshop on {{Hot}} Topics in Software Defined Networks - {{HotSDN}}
  '12}}.\hskip 1em plus 0.5em minus 0.4em\relax {Helsinki, Finland}: {ACM
  Press}, 2012, p. 127.

\bibitem{kreutzSoftwareDefinedNetworkingComprehensive2015}
D.~Kreutz, F.~M.~V. Ramos, P.~E. Ver{\'i}ssimo, C.~E. Rothenberg,
  S.~Azodolmolky, and S.~Uhlig, ``Software-{{Defined Networking}}: {{A
  Comprehensive Survey}},'' \emph{Proceedings of the IEEE}, vol. 103, no.~1,
  pp. 14--76, Jan. 2015.

\bibitem{mckeownOpenFlowEnablingInnovation2008}
N.~McKeown, T.~Anderson, H.~Balakrishnan, G.~Parulkar, L.~Peterson, J.~Rexford,
  S.~Shenker, and J.~Turner, ``{{OpenFlow}}: Enabling innovation in campus
  networks,'' \emph{ACM SIGCOMM Computer Communication Review}, vol.~38, no.~2,
  pp. 69--74, Mar. 2008.

\bibitem{al-shaerRandomHostMutation2013}
E.~{Al-Shaer}, Q.~Duan, and J.~H. Jafarian, ``\BIBforeignlanguage{en}{Random
  {{Host Mutation}} for {{Moving Target Defense}}},'' in
  \emph{\BIBforeignlanguage{en}{Security and {{Privacy}} in {{Communication
  Networks}}}}, A.~D. Keromytis and R.~Di~Pietro, Eds.\hskip 1em plus 0.5em
  minus 0.4em\relax {Berlin, Heidelberg}: {Springer Berlin Heidelberg}, 2013,
  vol. 106, pp. 310--327.

\bibitem{dunlopMT6DMovingTarget2011}
M.~Dunlop, S.~Groat, W.~Urbanski, R.~Marchany, and J.~Tront, ``{{MT6D}}: {{A
  Moving Target IPv6 Defense}},'' in \emph{2011 - {{MILCOM}} 2011 {{Military
  Communications Conference}}}, Nov. 2011, pp. 1321--1326.

\bibitem{dunlopBlindManBluff2012}
------, ``The {{Blind Man}}'s {{Bluff Approach}} to {{Security Using IPv6}},''
  \emph{IEEE Security \& Privacy}, vol.~10, no.~4, pp. 35--43, Jul. 2012.

\bibitem{kampanakisSDNbasedSolutionsMoving2014}
P.~Kampanakis, H.~G. Perros, and T.~Beyene, ``{{SDN}}-based solutions for
  {{Moving Target Defense}} network protection,'' \emph{Proceeding of IEEE
  International Symposium on a World of Wireless, Mobile and Multimedia
  Networks}, 2014.

\bibitem{zhaoSDNBasedFingerprintHopping2017}
Z.~Zhao, F.~Liu, and D.~Gong, ``\BIBforeignlanguage{en}{An {{SDN}}-{{Based
  Fingerprint Hopping Method}} to {{Prevent Fingerprinting Attacks}}},''
  \emph{\BIBforeignlanguage{en}{Security and Communication Networks}}, 2017.

\bibitem{banksEquilibriumSelectionSignaling1987}
J.~S. Banks and J.~Sobel, ``\BIBforeignlanguage{en}{Equilibrium {{Selection}}
  in {{Signaling Games}}},'' \emph{\BIBforeignlanguage{en}{Econometrica}},
  vol.~55, no.~3, p. 647, May 1987.

\bibitem{venkatesanMovingTargetDefense2016}
S.~Venkatesan, M.~Albanese, G.~Cybenko, and S.~Jajodia, ``A {{Moving Target
  Defense Approach}} to {{Disrupting Stealthy Botnets}},'' in \emph{Proceedings
  of the 2016 {{ACM Workshop}} on {{Moving Target Defense}}}, ser. {{MTD}}
  '16.\hskip 1em plus 0.5em minus 0.4em\relax {Vienna, Austria}: {Association
  for Computing Machinery}, Oct. 2016, pp. 37--46.

\bibitem{friedkinTheoreticalFoundationsCentrality1991}
N.~E. Friedkin, ``Theoretical {{Foundations}} for {{Centrality Measures}},''
  \emph{American Journal of Sociology}, vol.~96, no.~6, pp. 1478--1504, 1991.

\bibitem{senguptaMovingTargetDefense2018}
S.~Sengupta, A.~Chowdhary, D.~Huang, and S.~Kambhampati,
  ``\BIBforeignlanguage{en}{Moving {{Target Defense}} for the {{Placement}} of
  {{Intrusion Detection Systems}} in the {{Cloud}}},'' in
  \emph{\BIBforeignlanguage{en}{Decision and {{Game Theory}} for
  {{Security}}}}, ser. Lecture {{Notes}} in {{Computer Science}}, L.~Bushnell,
  R.~Poovendran, and T.~Ba{\c s}ar, Eds.\hskip 1em plus 0.5em minus 0.4em\relax
  {Cham}: {Springer International Publishing}, 2018, pp. 326--345.

\bibitem{korzhykComplexityComputingOptimal2010}
D.~Korzhyk, V.~Conitzer, and R.~Parr, ``\BIBforeignlanguage{en}{Complexity of
  {{Computing Optimal Stackelberg Strategies}} in {{Security Resource
  Allocation Games}}},'' \emph{\BIBforeignlanguage{en}{Proceedings of the
  Twenty-Fourth AAAI Conference on Artificial Intelligence (AAAI-10)}}, p.~6,
  2010.

\bibitem{sinclairApplicationMachineLearning1999}
C.~Sinclair, L.~Pierce, and S.~Matzner, ``An application of machine learning to
  network intrusion detection,'' in \emph{Proceedings 15th {{Annual Computer
  Security Applications Conference}} ({{ACSAC}}'99)}, Dec. 1999, pp. 371--377.

\bibitem{yinDeepLearningApproach2017}
C.~Yin, Y.~Zhu, J.~Fei, and X.~He, ``A {{Deep Learning Approach}} for
  {{Intrusion Detection Using Recurrent Neural Networks}},'' \emph{IEEE
  Access}, vol.~5, pp. 21\,954--21\,961, 2017.

\bibitem{wangHASTIDSLearningHierarchical2018}
W.~Wang, Y.~Sheng, J.~Wang, X.~Zeng, X.~Ye, Y.~Huang, and M.~Zhu,
  ``{{HAST}}-{{IDS}}: {{Learning Hierarchical Spatial}}-{{Temporal Features
  Using Deep Neural Networks}} to {{Improve Intrusion Detection}},'' \emph{IEEE
  Access}, vol.~6, pp. 1792--1806, 2018.

\bibitem{huangAdversarialAttacksSDNBased2018}
C.-H. Huang, T.-H. Lee, L.-h. Chang, J.-R. Lin, and G.~Horng,
  ``\BIBforeignlanguage{en}{Adversarial {{Attacks}} on {{SDN}}-{{Based Deep
  Learning IDS System}}},'' in \emph{\BIBforeignlanguage{en}{Mobile and
  {{Wireless Technology}} 2018}}, ser. Lecture {{Notes}} in {{Electrical
  Engineering}}, K.~J. Kim and H.~Kim, Eds.\hskip 1em plus 0.5em minus
  0.4em\relax {Singapore}: {Springer}, 2018, pp. 181--191.

\bibitem{usamaGenerativeAdversarialNetworks2019}
M.~Usama, M.~Asim, S.~Latif, J.~Qadir, and {Ala-Al-Fuqaha}, ``Generative
  {{Adversarial Networks For Launching}} and {{Thwarting Adversarial Attacks}}
  on {{Network Intrusion Detection Systems}},'' in \emph{2019 15th
  {{International Wireless Communications Mobile Computing Conference}}
  ({{IWCMC}})}, Jun. 2019, pp. 78--83.

\bibitem{linIDSGANGenerativeAdversarial2019}
Z.~Lin, Y.~Shi, and Z.~Xue, ``{{IDSGAN}}: {{Generative Adversarial Networks}}
  for {{Attack Generation}} against {{Intrusion Detection}},''
  \emph{arXiv:1809.02077 [cs]}, Jun. 2019.

\bibitem{vorobeychikOptimalRandomizedClassification2014}
Y.~Vorobeychik and B.~Li, ``Optimal randomized classification in adversarial
  settings,'' in \emph{Proceedings of the 2014 International Conference on
  {{Autonomous}} Agents and Multi-Agent Systems}, ser. {{AAMAS}} '14.\hskip 1em
  plus 0.5em minus 0.4em\relax {Paris, France}: {International Foundation for
  Autonomous Agents and Multiagent Systems}, May 2014, pp. 485--492.

\bibitem{senguptaMTDeepBoostingSecurity2019}
S.~Sengupta, T.~Chakraborti, and S.~Kambhampati,
  ``\BIBforeignlanguage{en}{{{MTDeep}}: {{Boosting}} the {{Security}} of {{Deep
  Neural Nets Against Adversarial Attacks}} with {{Moving Target Defense}}},''
  in \emph{\BIBforeignlanguage{en}{Decision and {{Game Theory}} for
  {{Security}}}}.\hskip 1em plus 0.5em minus 0.4em\relax {Cham}: {Springer
  International Publishing}, 2019, pp. 479--491.

\bibitem{PaXAddressSpace2003}
``{{PaX Address Space Layout Randomization}},''
  https://pax.grsecurity.net/docs/aslr.txt, 2003.

\bibitem{liAddressSpaceRandomizationWindows2006}
L.~Li, J.~Just, and R.~Sekar, ``\BIBforeignlanguage{en}{Address-{{Space
  Randomization}} for {{Windows Systems}}},'' in
  \emph{\BIBforeignlanguage{en}{2006 22nd {{Annual Computer Security
  Applications Conference}} ({{ACSAC}}'06)}}.\hskip 1em plus 0.5em minus
  0.4em\relax {Miami Beach, FL, USA}: {IEEE}, Dec. 2006, pp. 329--338.

\bibitem{kcCounteringCodeinjectionAttacks2003}
G.~S. Kc, A.~D. Keromytis, and V.~Prevelakis, ``Countering code-injection
  attacks with instruction-set randomization,'' in \emph{Proceedings of the
  10th {{ACM}} Conference on {{Computer}} and Communications Security}, ser.
  {{CCS}} '03.\hskip 1em plus 0.5em minus 0.4em\relax {Washington D.C., USA}:
  {Association for Computing Machinery}, Oct. 2003, pp. 272--280.

\bibitem{thompsonMultipleOSRotational2014}
M.~Thompson, N.~Evans, and V.~Kisekka, ``Multiple {{OS}} rotational environment
  an implemented {{Moving Target Defense}},'' in \emph{2014 7th {{International
  Symposium}} on {{Resilient Control Systems}} ({{ISRCS}})}, Aug. 2014, pp.
  1--6.

\bibitem{okhraviCreatingCyberMoving2011}
H.~Okhravi, A.~Comella, E.~Robinson, S.~Yannalfo, P.~Michaleas, and J.~Haines,
  ``\BIBforeignlanguage{en}{Creating a {{Cyber Moving Target}} for {{Critical
  Infrastructure Applications}}},'' in
  \emph{\BIBforeignlanguage{en}{International {{Conference}} on {{Critical
  Infrastructure Protection}}}}, ser. {{IFIP Advances}} in {{Information}} and
  {{Communication Technology}}, {Hanover, NH, USA}, 2011, pp. 107--123.

\bibitem{larsenSoKAutomatedSoftware2014}
P.~Larsen, A.~Homescu, S.~Brunthaler, and M.~Franz, ``{{SoK}}: {{Automated
  Software Diversity}},'' in \emph{2014 {{IEEE Symposium}} on {{Security}} and
  {{Privacy}}}, May 2014, pp. 276--291.

\bibitem{azabChameleonSoftMovingTarget2011}
M.~Azab, R.~Hassan, and M.~Eltoweissy, ``{{ChameleonSoft}}: {{A}} moving target
  defense system,'' in \emph{7th {{International Conference}} on
  {{Collaborative Computing}}: {{Networking}}, {{Applications}} and
  {{Worksharing}} ({{CollaborateCom}})}, Oct. 2011, pp. 241--250.

\bibitem{guptaMarlinFineGrained2013}
A.~Gupta, S.~Kerr, M.~S. Kirkpatrick, and E.~Bertino,
  ``\BIBforeignlanguage{en}{Marlin: {{A Fine Grained Randomization Approach}}
  to {{Defend}} against {{ROP Attacks}}},'' in
  \emph{\BIBforeignlanguage{en}{Network and {{System Security}}}}, ser. Lecture
  {{Notes}} in {{Computer Science}}.\hskip 1em plus 0.5em minus 0.4em\relax
  {Berlin, Heidelberg}: {Springer}, 2013, pp. 293--306.

\bibitem{legouesGenProgGenericMethod2012}
C.~Le~Goues, T.~Nguyen, S.~Forrest, and W.~Weimer, ``{{GenProg}}: {{A Generic
  Method}} for {{Automatic Software Repair}},'' \emph{IEEE Transactions on
  Software Engineering}, vol.~38, no.~1, pp. 54--72, Jan. 2012.

\bibitem{germanodasilvaCapitalizingSDNbasedSCADA2015}
E.~{Germano da Silva}, L.~A. Dias~Knob, J.~A. Wickboldt, L.~P. Gaspary, L.~Z.
  Granville, and A.~{Schaeffer-Filho}, ``Capitalizing on {{SDN}}-based
  {{SCADA}} systems: {{An}} anti-eavesdropping case-study,'' in \emph{2015
  {{IFIP}}/{{IEEE International Symposium}} on {{Integrated Network
  Management}} ({{IM}})}, May 2015, pp. 165--173.

\bibitem{aseeriAlleviatingEavesdroppingAttacks2017}
A.~Aseeri, N.~Netjinda, and R.~Hewett, ``Alleviating eavesdropping attacks in
  software-defined networking data plane,'' in \emph{Proceedings of the 12th
  {{Annual Conference}} on {{Cyber}} and {{Information Security Research}}},
  ser. {{CISRC}} '17.\hskip 1em plus 0.5em minus 0.4em\relax {Oak Ridge,
  Tennessee, USA}: {Association for Computing Machinery}, Apr. 2017, pp. 1--8.

\bibitem{yackoskiSelfshieldingDynamicNetwork2011}
J.~Yackoski, P.~Xie, H.~Bullen, J.~Li, and K.~Sun, ``A {{Self}}-shielding
  {{Dynamic Network Architecture}},'' in \emph{Military {{Communications
  Conference}} ({{MILCOM}} 2011)}, Nov. 2011, pp. 1381--1386.

\bibitem{yackoskiApplyingSelfShieldingDynamics2013}
J.~Yackoski, H.~Bullen, X.~Yu, and J.~Li, ``\BIBforeignlanguage{en}{Applying
  {{Self}}-{{Shielding Dynamics}} to the {{Network Architecture}}},'' in
  \emph{\BIBforeignlanguage{en}{Moving {{Target Defense II}}}}, ser. Advances
  in {{Information Security}}.\hskip 1em plus 0.5em minus 0.4em\relax {New
  York, NY}: {Springer}, 2013, pp. 97--115.

\bibitem{zhangCrossVMSideChannels2012}
Y.~Zhang, A.~Juels, M.~K. Reiter, and T.~Ristenpart, ``Cross-{{VM}} side
  channels and their use to extract private keys,'' in \emph{Proceedings of the
  2012 {{ACM}} Conference on {{Computer}} and Communications Security}, ser.
  {{CCS}} '12.\hskip 1em plus 0.5em minus 0.4em\relax {Raleigh, North Carolina,
  USA}: {Association for Computing Machinery}, Oct. 2012, pp. 305--316.

\bibitem{pattukPreventingCryptographicKey2014}
E.~Pattuk, M.~Kantarcioglu, Z.~Lin, and H.~Ulusoy, ``Preventing cryptographic
  key leakage in cloud virtual machines,'' in \emph{Proceedings of the 23rd
  {{USENIX}} Conference on {{Security Symposium}}}, ser. {{SEC}}'14.\hskip 1em
  plus 0.5em minus 0.4em\relax {San Diego, CA}: {USENIX Association}, Aug.
  2014, pp. 703--718.

\bibitem{beimelSecretSharingSchemesSurvey2011}
A.~Beimel, ``\BIBforeignlanguage{en}{Secret-{{Sharing Schemes}}: {{A
  Survey}}},'' in \emph{\BIBforeignlanguage{en}{Coding and
  {{Cryptology}}}}.\hskip 1em plus 0.5em minus 0.4em\relax {Berlin,
  Heidelberg}: {Springer}, 2011, pp. 11--46.

\bibitem{smutzPreventingExploitsMicrosoft2015}
C.~Smutz and A.~Stavrou, ``Preventing {{Exploits}} in {{Microsoft Office
  Documents Through Content Randomization}},'' in \emph{{{RAID}}}, 2015.

\bibitem{DeceptionDepthCase2017}
``\BIBforeignlanguage{en-US}{Deception in {{Depth}} \textendash{} {{The Case}}
  for a {{Full}}-{{Stack Architecture}}},''
  https://trapx.com/deception-in-depth-the-case-for-a-full-stack-architecture/,
  Mar. 2017.

\bibitem{wangDetectingTargetedAttacks2013}
W.~Wang, {Jeffrey Bickford}, {Ilona Murynets}, {Ramesh Subbaraman}, and {Gokul
  Singaraju}, ``\BIBforeignlanguage{en}{Detecting {{Targeted Attacks By
  Multilayer Deception}}},'' \emph{\BIBforeignlanguage{en}{Journal of Cyber
  Security and Mobility}}, vol.~2, no.~2, pp. 175--199, 2013.

\bibitem{sunDESIRDecoyenhancedSeamless2016}
J.~Sun and K.~Sun, ``{{DESIR}}: {{Decoy}}-enhanced seamless {{IP}}
  randomization,'' in \emph{{{IEEE INFOCOM}} 2016 - {{The}} 35th {{Annual IEEE
  International Conference}} on {{Computer Communications}}}, Apr. 2016, pp.
  1--9.

\bibitem{parkSecureCyberDeception2018}
K.~Park, S.~Woo, D.~Moon, and H.~Choi, ``\BIBforeignlanguage{en}{Secure {{Cyber
  Deception Architecture}} and {{Decoy Injection}} to {{Mitigate}} the
  {{Insider Threat}}},'' \emph{\BIBforeignlanguage{en}{Symmetry}}, vol.~10,
  no.~1, p.~14, Jan. 2018.

\bibitem{jafarianMultidimensionalHostIdentity2016}
J.~H. Jafarian, A.~Niakanlahiji, E.~{Al-Shaer}, and Q.~Duan,
  ``\BIBforeignlanguage{en}{Multi-dimensional {{Host Identity Anonymization}}
  for {{Defeating Skilled Attackers}}},'' in
  \emph{\BIBforeignlanguage{en}{Proceedings of the 2016 {{ACM Workshop}} on
  {{Moving Target Defense}} - {{MTD}}'16}}.\hskip 1em plus 0.5em minus
  0.4em\relax {Vienna, Austria}: {ACM Press}, 2016, pp. 47--58.

\bibitem{shiCHAOSSDNBasedMoving2017}
Y.~Shi, H.~Zhang, J.~Wang, F.~Xiao, J.~Huang, D.~Zha, H.~Hu, F.~Yan, and
  B.~Zhao, ``\BIBforeignlanguage{en}{{{CHAOS}}: {{An SDN}}-{{Based Moving
  Target Defense System}}},'' \emph{\BIBforeignlanguage{en}{Security and
  Communication Networks}}, 2017.

\bibitem{adebayoDeceptorintheMiddleDitMCyber2020}
A.~Adebayo and D.~B. Rawat, ``Deceptor-in-the-{{Middle}} ({{DitM}}): {{Cyber
  Deception}} for {{Security}} in {{Wireless Network Virtualization}},'' in
  \emph{2020 {{IEEE}} 17th {{Annual Consumer Communications Networking
  Conference}} ({{CCNC}})}, Jan. 2020, pp. 1--6.

\bibitem{jiangDesignImplementationMachine2020}
K.~Jiang and H.~Zheng, ``Design and {{Implementation}} of {{A Machine Learning
  Enhanced Web Honeypot System}},'' in \emph{2020 13th {{International
  Congress}} on {{Image}} and {{Signal Processing}}, {{BioMedical Engineering}}
  and {{Informatics}} ({{CISP}}-{{BMEI}})}, Oct. 2020, pp. 957--961.

\bibitem{sunCyberMoatCamouflagingCritical2017}
J.~Sun, K.~Sun, and Q.~Li, ``{{CyberMoat}}: {{Camouflaging}} critical server
  infrastructures with large scale decoy farms,'' in \emph{2017 {{IEEE
  Conference}} on {{Communications}} and {{Network Security}} ({{CNS}})}, Oct.
  2017, pp. 1--9.

\bibitem{sunBelievableDecoySystem2020}
------, ``Towards a {{Believable Decoy System}}: {{Replaying Network
  Activities}} from {{Real System}},'' in \emph{2020 {{IEEE Conference}} on
  {{Communications}} and {{Network Security}} ({{CNS}})}, Jun. 2020, pp. 1--9.

\bibitem{albaneseDeceptionBasedApproach2015}
M.~Albanese, E.~Battista, and S.~Jajodia, ``A deception based approach for
  defeating {{OS}} and service fingerprinting,'' in \emph{2015 {{IEEE
  Conference}} on {{Communications}} and {{Network Security}} ({{CNS}})}, Sep.
  2015, pp. 317--325.

\bibitem{karunaFakeDocumentGeneration2020}
P.~Karuna, H.~Purohit, S.~Jajodia, R.~Ganesan, and O.~Uzuner, ``Fake {{Document
  Generation}} for {{Cyber Deception}} by {{Manipulating Text
  Comprehensibility}},'' \emph{IEEE Systems Journal}, pp. 1--11, 2020.

\bibitem{sunScalableHighFidelity2019}
J.~Sun, S.~Liu, and K.~Sun, ``A {{Scalable High Fidelity Decoy Framework}}
  against {{Sophisticated Cyber Attacks}},'' in \emph{Proceedings of the 6th
  {{ACM Workshop}} on {{Moving Target Defense}}}, ser. {{MTD}}'19.\hskip 1em
  plus 0.5em minus 0.4em\relax {New York, NY, USA}: {Association for Computing
  Machinery}, Nov. 2019, pp. 37--46.

\bibitem{trassareTechniqueNetworkTopology2013}
S.~T. Trassare, R.~Beverly, and D.~Alderson, ``A {{Technique}} for {{Network
  Topology Deception}},'' in \emph{{{MILCOM}} 2013 - 2013 {{IEEE Military
  Communications Conference}}}, Nov. 2013, pp. 1795--1800.

\bibitem{duanRangeTopologyMutation2020}
Q.~Duan, E.~{Al-Shaer}, and J.~Xie, ``Range and {{Topology Mutation Based
  Wireless Agility}},'' in \emph{Proceedings of the 7th {{ACM Workshop}} on
  {{Moving Target Defense}}}, ser. {{MTD}}'20.\hskip 1em plus 0.5em minus
  0.4em\relax {New York, NY, USA}: {Association for Computing Machinery}, Nov.
  2020, pp. 59--67.

\bibitem{rrushiDNICArchitecturalDevelopments2021}
J.~L. Rrushi, ``{{DNIC Architectural Developments}} for 0-{{Knowledge
  Detection}} of {{OPC Malware}},'' \emph{IEEE Transactions on Dependable and
  Secure Computing}, vol.~18, no.~1, pp. 30--44, Jan. 2021.

\bibitem{choiPhantomFSv2DareYou2020}
J.~Choi, H.~Lee, Y.~Park, H.~K. Kim, J.~Lee, Y.~Kim, G.~Lee, S.-W. Shim, and
  T.~Kim, ``{{PhantomFS}}-v2: {{Dare You}} to {{Avoid This Trap}},'' \emph{IEEE
  Access}, vol.~8, pp. 198\,285--198\,300, Oct. 2020.

\bibitem{chiangACyDSAdaptiveCyber2016}
C.~J. Chiang, Y.~M. Gottlieb, {Shridatt James Sugrim}, R.~Chadha, C.~Serban,
  A.~Poylisher, L.~M. Marvel, and J.~Santos, ``{{ACyDS}}: {{An}} adaptive cyber
  deception system,'' in \emph{{{MILCOM}} 2016 - 2016 {{IEEE Military
  Communications Conference}}}, Nov. 2016, pp. 800--805.

\bibitem{almeshekahUsingDeceptionEnhance2015}
M.~H. Almeshekah, ``Using {{Deception}} to {{Enhance Security}}: {{A
  Taxonomy}}, {{Model}}, and {{Novel Uses}},'' Ph.D. dissertation, Jan. 2015.

\bibitem{defaveriGoalDrivenDeceptionTactics2016}
C.~De~Faveri, A.~Moreira, and V.~Amaral,
  ``\BIBforeignlanguage{en}{Goal-{{Driven Deception Tactics Design}}},'' in
  \emph{\BIBforeignlanguage{en}{2016 {{IEEE}} 27th {{International Symposium}}
  on {{Software Reliability Engineering}} ({{ISSRE}})}}.\hskip 1em plus 0.5em
  minus 0.4em\relax {Ottawa, ON, Canada}: {IEEE}, Oct. 2016, pp. 264--275.

\bibitem{heckmanDenialDeceptionCyber2015}
K.~E. Heckman, F.~J. Stech, B.~S. Schmoker, and R.~K. Thomas, ``Denial and
  {{Deception}} in {{Cyber Defense}},'' \emph{IEEE Computer}, vol.~48, no.~4,
  pp. 36--44, Apr. 2015.

\bibitem{defaveriDesigningAdaptiveDeception2016}
C.~De~Faveri and A.~Moreira, ``Designing {{Adaptive Deception Strategies}},''
  in \emph{2016 {{IEEE International Conference}} on {{Software Quality}},
  {{Reliability}} and {{Security Companion}} ({{QRS}}-{{C}})}, Aug. 2016, pp.
  77--84.

\bibitem{budiartoHoneypotsWhyWe2004}
R.~Budiarto, A.~Samsudin, C.~Heong, and S.~Noori, ``Honeypots: Why we need a
  dynamics honeypots?'' in \emph{Proceedings of {{International Conference}} on
  {{Information}} and {{Communication Technologies}}: {{From Theory}} to
  {{Applications}}}, Apr. 2004, pp. 565--566.

\bibitem{zanoramyansiryzakariaReviewDynamicIntelligent2013}
W.~Zanoramy Ansiry~Zakaria and M.~Laiha Mat~Kiah, ``\BIBforeignlanguage{en}{A
  review of dynamic and intelligent honeypots},''
  \emph{\BIBforeignlanguage{en}{ScienceAsia}}, vol. 39S, 2013.

\bibitem{wangIntelligentDeploymentPolicy2020}
S.~Wang, Q.~Pei, J.~Wang, G.~Tang, Y.~Zhang, and X.~Liu, ``An {{Intelligent
  Deployment Policy}} for {{Deception Resources Based}} on {{Reinforcement
  Learning}},'' \emph{IEEE Access}, vol.~8, pp. 35\,792--35\,804, 2020.

\bibitem{wagenerAdaptiveSelfconfigurableHoneypots2011}
G.~Wagener, R.~State, T.~Engel, and A.~Dulaunoy, ``Adaptive and
  self-configurable honeypots,'' in \emph{12th {{IFIP}}/{{IEEE International
  Symposium}} on {{Integrated Network Management}} ({{IM}} 2011) and
  {{Workshops}}}, May 2011, pp. 345--352.

\bibitem{wagenerHelizaTalkingDirty2011}
G.~Wagener, R.~State, A.~Dulaunoy, and T.~Engel,
  ``\BIBforeignlanguage{en}{Heliza: Talking dirty to the attackers},''
  \emph{\BIBforeignlanguage{en}{Journal in Computer Virology}}, vol.~7, no.~3,
  pp. 221--232, Aug. 2011.

\bibitem{bairdPessimalPrintReverseTuring2003}
H.~S. Baird, A.~L. Coates, and R.~J. Fateman,
  ``\BIBforeignlanguage{en}{{{PessimalPrint}}: A reverse {{Turing}} test},''
  \emph{\BIBforeignlanguage{en}{International Journal on Document Analysis and
  Recognition}}, vol.~5, no.~2, pp. 158--163, Apr. 2003.

\bibitem{paunaRASSHReinforcedAdaptive2014}
A.~Pauna and I.~Bica, ``{{RASSH}} - {{Reinforced}} adaptive {{SSH}} honeypot,''
  in \emph{Proceedings of the 10th {{International Conference}} on
  {{Communications}} ({{COMM}})}, May 2014, pp. 1--6.

\bibitem{carrollGameTheoreticInvestigation2009}
T.~E. Carroll and D.~Grosu, ``\BIBforeignlanguage{en}{A {{Game Theoretic
  Investigation}} of {{Deception}} in {{Network Security}}},'' in
  \emph{\BIBforeignlanguage{en}{Proceedings of 18th {{International
  Conference}} on {{Computer Communications}} and {{Networks}}}}, 2009, p.~6.

\bibitem{rahmanGametheoreticApproachDeceiving2013}
M.~A. Rahman, M.~H. Manshaei, and E.~{Al-Shaer}, ``A game-theoretic approach
  for deceiving {{Remote Operating System Fingerprinting}},'' in
  \emph{Proceedings of {{IEEE Conference}} on {{Communications}} and {{Network
  Security}} ({{CNS}})}, Oct. 2013, pp. 73--81.

\bibitem{carterGameTheoreticApproach2014}
K.~M. Carter, J.~F. Riordan, and H.~Okhravi, ``A {{Game Theoretic Approach}} to
  {{Strategy Determination}} for {{Dynamic Platform Defenses}},'' in
  \emph{Proceedings of the {{First ACM Workshop}} on {{Moving Target
  Defense}}}, ser. {{MTD}} '14.\hskip 1em plus 0.5em minus 0.4em\relax
  {Scottsdale, Arizona, USA}: {Association for Computing Machinery}, Nov. 2014,
  pp. 21--30.

\bibitem{leiOptimalStrategySelection2017}
C.~Lei, D.-H. Ma, and H.-Q. Zhang, ``Optimal {{Strategy Selection}} for
  {{Moving Target Defense Based}} on {{Markov Game}},'' \emph{IEEE Access},
  vol.~5, pp. 156--169, 2017.

\bibitem{horakManipulatingAdversaryBelief2017}
K.~Hor{\'a}k, Q.~Zhu, and B.~Bo{\v s}ansk{\'y},
  ``\BIBforeignlanguage{en}{Manipulating {{Adversary}}'s {{Belief}}: {{A
  Dynamic Game Approach}} to {{Deception}} by {{Design}} for {{Proactive
  Network Security}}},'' in \emph{\BIBforeignlanguage{en}{International
  {{Conference}} on {{Decision}} and {{Game Theory}} for {{Security}}}}, vol.
  10575, 2017, pp. 273--294.

\bibitem{zhuGameTheoreticApproachFeedbackDriven2013}
Q.~Zhu and T.~Ba{\c s}ar, ``\BIBforeignlanguage{en}{Game-{{Theoretic Approach}}
  to {{Feedback}}-{{Driven Multi}}-stage {{Moving Target Defense}}},'' in
  \emph{\BIBforeignlanguage{en}{Decision and {{Game Theory}} for
  {{Security}}}}, ser. Lecture {{Notes}} in {{Computer Science}}, 2013, pp.
  246--263.

\bibitem{roweModelDeceptionCyberattacks2004}
N.~C. Rowe, ``A model of deception during cyber-attacks on information
  systems,'' in \emph{{{IEEE First Symposium onMulti}}-{{Agent Security}} and
  {{Survivability}}, 2004}, Aug. 2004, pp. 21--30.

\bibitem{antonchuvakinWillDeceptionFizzle2019}
{Anton Chuvakin}, ``\BIBforeignlanguage{en}{Will {{Deception Fizzle}} ...
  {{Again}}?}'' Mar. 2019.

\bibitem{islamActiveDeceptionFramework2020}
M.~M. Islam and E.~{Al-Shaer}, ``Active {{Deception Framework}}: {{An
  Extensible Development Environment}} for {{Adaptive Cyber Deception}},'' in
  \emph{2020 {{IEEE Secure Development}} ({{SecDev}})}, Sep. 2020, pp. 41--48.

\bibitem{incTrapXIntroducesIndustryFirst2020}
T.~S. Inc, ``\BIBforeignlanguage{en}{{{TrapX Introduces Industry}}-{{First
  Deception}}-{{As}}-{{A}}-{{Service Solution}}, {{TrapX
  Flex}}\texttrademark},''
  https://www.prnewswire.com/news-releases/trapx-introduces-industry-first-deception-as-a-service-solution-trapx-flex-301162363.html,
  Oct. 2020.

\bibitem{yeDifferentiallyPrivateGame2021}
D.~Ye, T.~Zhu, S.~Shen, and W.~Zhou, ``A {{Differentially Private Game
  Theoretic Approach}} for {{Deceiving Cyber Adversaries}},'' \emph{IEEE
  Transactions on Information Forensics and Security}, vol.~16, pp. 569--584,
  2021.

\bibitem{acostaCybersecurityDeceptionExperimentation2020}
J.~C. Acosta, A.~Basak, C.~Kiekintveld, N.~Leslie, and C.~Kamhoua,
  ``Cybersecurity {{Deception Experimentation System}},'' in \emph{2020 {{IEEE
  Secure Development}} ({{SecDev}})}, Sep. 2020, pp. 34--40.

\bibitem{ferguson-walterTularosaStudyExperimental2018}
K.~{Ferguson-Walter}, T.~Shade, A.~Rogers, M.~C.~S. Trumbo, K.~S. Nauer, K.~M.
  Divis, A.~Jones, A.~Combs, and R.~G. Abbott,
  ``\BIBforeignlanguage{English}{The {{Tularosa Study}}: {{An Experimental
  Design}} and {{Implementation}} to {{Quantify}} the {{Effectiveness}} of
  {{Cyber Deception}}.}'' {Sandia National Lab. (SNL-NM), Albuquerque, NM
  (United States)}, Tech. Rep. SAND2018-5870C, May 2018.

\bibitem{duanCONCEALStrategyComposition2018}
Q.~Duan, E.~{Al-Shaer}, M.~Islam, and H.~Jafarian, ``{{CONCEAL}}: {{A Strategy
  Composition}} for {{Resilient Cyber Deception}}-{{Framework}}, {{Metrics}}
  and {{Deployment}},'' in \emph{2018 {{IEEE Conference}} on {{Communications}}
  and {{Network Security}} ({{CNS}})}, May 2018, pp. 1--9.

\bibitem{malekiMarkovModelingMoving2016}
H.~Maleki, S.~Valizadeh, W.~Koch, A.~Bestavros, and M.~{van Dijk},
  ``\BIBforeignlanguage{en}{Markov {{Modeling}} of {{Moving Target Defense
  Games}}},'' in \emph{\BIBforeignlanguage{en}{Proceedings of the 2016 {{ACM
  Workshop}} on {{Moving Target Defense}} - {{MTD}}'16}}.\hskip 1em plus 0.5em
  minus 0.4em\relax {Vienna, Austria}: {ACM Press}, 2016, pp. 81--92.

\bibitem{abu-ghazalehHowSpectreMeltdown2019}
N.~{Abu-Ghazaleh}, D.~Ponomarev, and D.~Evtyushkin, ``How the spectre and
  meltdown hacks really worked,'' \emph{IEEE Spectrum}, vol.~56, no.~3, pp.
  42--49, Mar. 2019.

\bibitem{gallagherMorpheusVulnerabilityTolerantSecure2019}
M.~Gallagher, L.~Biernacki, S.~Chen, Z.~B. Aweke, S.~F. Yitbarek, M.~T. Aga,
  A.~Harris, Z.~Xu, B.~Kasikci, V.~Bertacco, S.~Malik, M.~Tiwari, and
  T.~Austin, ``\BIBforeignlanguage{en}{Morpheus: {{A Vulnerability}}-{{Tolerant
  Secure Architecture Based}} on {{Ensembles}} of {{Moving Target Defenses}}
  with {{Churn}}},'' in \emph{\BIBforeignlanguage{en}{Proceedings of the
  {{Twenty}}-{{Fourth International Conference}} on {{Architectural Support}}
  for {{Programming Languages}} and {{Operating Systems}}}}.\hskip 1em plus
  0.5em minus 0.4em\relax {Providence RI USA}: {ACM}, Apr. 2019, pp. 469--484.

\bibitem{ferguson-walterEmpiricalAssessmentEffectiveness2020}
K.~J. {Ferguson-Walter}, ``\BIBforeignlanguage{en}{An {{Empirical Assessment}}
  of the {{Effectiveness}} of {{Deception}} for {{Cyber Defense}}},'' Ph.D.
  dissertation, University of Massachusetts Amherst, Feb. 2020.

\end{thebibliography}

%

\end{document}